\documentclass[aps,prd,twocolumn]{revtex4-1}

\usepackage{amssymb}

\usepackage{amsmath}
\usepackage{latexsym}
\usepackage{amsfonts}
\usepackage{soul}
\usepackage{array}
\usepackage{graphicx}
\usepackage{ragged2e}
\usepackage[svgnames,table]{xcolor}
\usepackage{changepage}
\usepackage{setspace}
\usepackage{hhline}
\usepackage{float}
\usepackage{multirow}
\usepackage[
	hidelinks
	]{hyperref}
\newcommand{\sectionsubtitle}[1]{\begin{quotation}\centering\sffamily\vspace{10pt}{#1}\vspace{10pt}\end{quotation}}
\usepackage{booktabs}
\newcommand{\pL}{p_{\mathrm{L}}}
\newcommand{\pP}{p_{\mathrm{P}}}
\newcommand{\pth}{p_{\mathrm{th}}}
\newcommand{\fQEC}{f_{\mathrm{QEC}}}
\newcommand{\R}{\mathbb{R}}
\newcommand{\tp}{\top}

\begin{document}




\title{Forecasting timelines of quantum computing}


\author{Jaime Sevilla}
\email{j.sevilla.20@abdn.ac.uk}
\affiliation{Aberdeen University, Aberdeen, UK}
\affiliation{Center for the Study of Existential Risk, Cambridge, UK}

\author{C. Jess Riedel}
\email{jessriedel@gmail.com}
\affiliation{NTT Research, Inc., Physics \& Informatics Laboratories, Sunnyvale, CA, USA}

\begin{abstract}
We consider how to forecast progress in the domain of quantum computing. 
For this purpose we collect a dataset of quantum computer systems to date, scored on their physical qubits and gate error rate, and we define an index combining both metrics, the generalized logical qubit. 
We study the relationship between physical qubits and gate error rate, and tentatively conclude that they are positively correlated (albeit with some room for doubt), indicating a frontier of development that trades-off between them. 
We also apply a log-linear regression on the metrics to provide a tentative upper bound on how much progress can be expected over time.  Within the (generally optimistic) assumptions of our model, including the key assumption that exponential progress in qubit count and gate fidelity will continue, we estimate that that proof-of-concept fault-tolerant computation based on superconductor technology is unlikely ($<$5\% confidence) to be exhibited before 2026, and that quantum devices capable of factoring RSA-2048 are unlikely ($<$5\% confidence) to exist before 2039.  It is of course possible that these milestones will in fact be reached earlier, but that this would require faster progress than has yet been seen.

\end{abstract}











\maketitle

In this article we consider the problem of anticipating progress on Quantum Computing.  The following is an outline of the contents.

Section \ref{introduction} contextualizes this article as part of our ongoing investigation of the transformative impact of quantum computing.

\begin{itemize}
\item In Section~\ref{previous-work} we discuss previous related work on anticipating quantum computing timelines and technological forecasting.
\item In Section~\ref{limitations} we discuss shortcomings of our methods.
\end{itemize}

Section \ref{operationalizating progress} describes how this article quantifies progress in quantum computing.  

\begin{itemize}
\item In Section~\ref{types-of-quantum-computation} we introduce the distinction between quantum annealers, Noisy Intermediate Scale Quantum (NISQ) computers, and fault-tolerant quantum computers (FTQC).

\item In Section~\ref{milestones} we discuss what concrete milestones we intend to forecast. Drawing from \citep{grumbling2019quantum}, we establish the milestone of the first large-scale fault-tolerant quantum computer capable of breaking the modern cryptographic scheme RSA 2048 as the question of interest in this report.

\item In Section~\ref{metrics}, we discuss metrics for tracking progress so far. We establish the number of physical qubits and the average two-qubit error rate as our metrics of interest. We combine these into a single metric, the number of ``generalized logical qubits,'' which estimates how many logical qubits a fault-tolerant computer with a given number of physical qubits and two-qubit gate errors could support, but which is defined even for current-generation devices which are yet too noisy to support a single (true) logical qubit.

\item In Section~\ref{operationalization}, we relate our forecasting question to our metrics of interest. As a precise operationalization of our question of interest, we choose to forecast when we will be able to showcase proof-of-concept quantum error correction --- specifically when a quantum computer will be powerful enough to break the commonly used RSA-2048 cryptographic scheme. This is operationalized as a function of the number of generalized logical qubits, concretely at 4100 logical qubits.
\end{itemize}

Section \ref{current-state-of-the-art} reviews progress so far in the metrics of interest. Accompanying this section we present a publicly available database of major contributions in quantum computing hardware we have curated.


Section \ref{modelling-future-progress} presents our main model and results.  

\begin{itemize}
\item In Section~\ref{a-technological-development-frontier}, we apply a multivariate log linear regression model to study the relationship between the number of physical qubits and the error rate in a system. When conditioning on time we find a positive correlation between the metrics, with more qubits being associated with higher error rates, suggesting a frontier of development that trades-off one against another.

\item In Section~\ref{extrapolating-current-trends}, we \emph{assume} exponential progress of QC and apply a log linear multivariate model to provide an upper bound of likely progress of QC based on superconductors. We find that it is 95$\%$  likely that proof-of-concept FTQL will be not be developed before 2026, and that RSA-2048 will be safe from quantum computation by 2039 with the same confidence. Note however that our model is very limited in data and makes strong assumptions about the statistical nature of the underlying metrics.

\item In Section~\ref{model-robustness} we explore the robustness of our extrapolation model in the previous section, discussing alternative modelling choices and how they would affect our conclusions.

\item In Section~\ref{model-validation} we provide minimal validation of the extrapolation model we propose.
\end{itemize}

Section \ref{conclusions-and-open-questions} collects our conclusions and suggests possible improvements and open questions for future work.

The \href{https://docs.google.com/spreadsheets/d/1pwb4gf0FxlxgfVhtXTaqEGS9b7FwsstsJ0v7Zb1naQ0/}{\textcolor[HTML]{1155CC}{\ul{data}}} \footnote{Full URL: \texttt{https://\allowbreak{}docs.\allowbreak{}google.\allowbreak{}com/\allowbreak{}spreadsheets/\allowbreak{}d/\allowbreak{}1pwb4gf0Fxl\allowbreak{}xgfVhtXTaq\allowbreak{}EGS9b7FwsstsJ\allowbreak{}0v7Zb1naQ0/}.} we collected and the \href{https://\allowbreak{}colab.\allowbreak{}research.\allowbreak{}google.\allowbreak{}com/\allowbreak{}drive\allowbreak{}/1XkWcs\allowbreak{}Uy-CiN\allowbreak{}DDJffPC3dN\allowbreak{}JbgvR6iDAqh}{\textcolor[HTML]{1155CC}{\ul{code for our models}}} \footnote{Full URL: \texttt{https://\allowbreak{}colab.\allowbreak{}research.google.\allowbreak{}com/\allowbreak{}drive/\allowbreak{}1XkWcsUy-CiND\allowbreak{}DJffPC3d\allowbreak NJbgvR6iDAqh}.} is freely available online.

To aid the reader in quickly navigating this article, each section begin with the central question it intends to answer.

\section{Introduction}
\label{introduction}
\addcontentsline{toc}{section}{1. Introduction}
\sectionsubtitle{Why bother thinking about progress in quantum computing, and what's already known?}

In this article we engage with the task of predicting the arrival of major milestones in quantum computing.  We will not describe the basics of quantum computing here, but instead will refer the reader to introductions at the popular \citep{aaronson2008limits,aaronson2011quantum}, semi-technical \citep{aaronson2013quantum,perry2019quantum,rudolph2017q,grumbling2019quantum}, and technical levels \citep{kitaev2002classical,matuschak2019quantum,mermin2007quantum,nielsen2000quantum,ozols2018quantum}.

Instead, we mention only these brief ideas: Like traditional (classical) computers, quantum computers are information-processing devices that physically implement mathematical computations. Unlike classical computers, quantum computers harness physical phenomena associated with quantum mechanics which generally can only be observed on microscopic scales.  Although very difficult to build, such devices will be able to solve some specific math problems which are intractable to solve on classical computers.  

However, even when quantum computers are very advanced they are not expected to fully replace classical computers because, for most mathematical problems, quantum effects do not offer a computational speed-up; it is only useful on certain kinds of tasks. Indeed, within realistic architecture proposals, the core quantum processing device relies on pre- and post-processing by classical computers for its basic operation.  At the \emph{theoretical} level, quantum computers represent a true paradigm shift in our basic understanding of physical computation, but at the \emph{practical} level quantum computers are likely to be analogous to existing specialized computational modules like graphics-processing units and tensor-processing units, i.e., they will be dedicated physical devices optimized to solve a particular subset of computational problems.

The most anticipated applications of quantum computing include constrained optimization, efficient simulation of chemical reactions and other large interacting quantum mechanical systems, and cryptanalysis. (Improvements to machine-learning computations is also possible, although currently more speculative.) It is as of yet unclear how important or far reaching these applications will be.  Some further reading about applications can be found in \citep{grumbling2019quantum,dewolf2017potential,majot2015global,moeller2017impact}.

This article is part of a broader investigation we are conducting, trying to anticipate the transformative impact that quantum computing will have on society. Understanding better the major milestones of this technology will give us better insight into what quantum computing is capable of, by when and how many warning signs and room-for-maneuver will society have to adapt to unexpected developments.

\subsection{Previous work}
\label{previous-work}
\addcontentsline{toc}{subsection}{1.1 Previous work}

To our knowledge, the most comprehensive assessment of future progress in quantum computing is the recent report by the National Academies of Sciences, Engineering, and Medicine \citep{grumbling2019quantum}. They propose several milestones and metrics to track progress in Quantum Computing \footnote{These milestones include (1) experimental quantum annealers, (2) small (tens of qubits) computers, (3) gate-based quantum computers that demonstrates quantum supremacy, (4) quantum annealers that demonstrate quantum supremacy, (5) implementation of QEC for improved qubit quality, (6) commercially useful quantum computers and (7) large (\textgreater 1,000 qubits), fault-tolerant, modular quantum computers.}. They find that RSA 2048 cryptography schemes will be safe during the next decade \footnote{``Given the current state of quantum computing and recent rates of progress, it is highly unexpected that a quantum computer that can compromise RSA 2048 or comparable discrete logarithm-based public key cryptosystems will be built within the next decade.''  \citep{grumbling2019quantum}.}. We borrow from this work to set our milestones and metrics of interest in Section~\ref{operationalizating progress} \footnote{As one interesting point, the Academies report highlights Rock's law and contextualizes it as a warning that quantum computing may fail to achieve exponential progress unless it secures increasing funding: ``Moore's law is the result of a virtuous cycle, where improvements in integrated circuit manufacturing allow the manufacturer to reduce the price of their product, which in turn causes them to sell more products and increase their sales and profits. This increased revenue then enables them to improve the manufacturing process again, which is harder this time, since the easier changes have already been made.''}.

Ref.~\citep{piani2019quantum} survey experts in quantum computing about their projected timelines. They find that 22.7$\%$  of the experts they surveyed think it is likely or highly likely that quantum computers will be able to crack RSA-2048 keys by 2030, and 50$\%$  think that is likely or highly likely that we will be able to crack RSA-2048 keys by 2035. We use their work as a baseline for comparison in Section~\ref{extrapolating-current-trends}.

Ref.~\citep{farmer2016how} conducted a quantitative study of progress in several technologies, finding evidence of exponential progress in the cost of many technologies such as transistors, genomic sequencing, DRAMs, etc. Their findings lend support to our modelling assumption of exponential progress in quantum computing, which is detailed in Section~\ref{modelling-future-progress}.

On the topic of trend robustness, \citep{grace2020discontinuous} examines 37 technological trends, and finds a robust discontinuity in 32$\%$  of these trends, with a base yearly rate of 0.1$\%$  chances of a discontinuity per year for each trend. This result is indicative of the base chance for modelling error through Section~\ref{modelling-future-progress}.

Ref.~\citep{kott2019toward} also grapples with the question of how to combine different aspects of technology to find regularities in progress, focusing on military tech as a case study. His approach is conceptually similar to our overall analysis, with two marked differences: a) Kott's figure of merit is automatically derived, while ours is motivated by expert knowledge; b) Kott's approach is closer to machine learning, in that he fits his models according to a loss function while we use classical statistical analysis.

\subsection{Limitations}
\label{limitations}
\addcontentsline{toc}{subsection}{1.2 Limitations}
Young technologies have meager track records, and predictions about the future rely heavily on educated guesses by experts (sometimes supplemented by more reliable arguments, e.g., constraints from basic physics).  As a technology matures, data accumulates and predictions become more strongly driven by the extrapolation of past trends. 

Quantum computing is still a very young field.  Experimentation with quantum control of information has been occurring for decades, but devices advanced enough to be interpreted as non-trivial ``computers'' with quantifiable and comparable parameters are quite new.  Because of this paucity of evidence, essentially all existing predictions about the future are based on expert wisdom (with all of its known flaws).

We consider this work to be merely a first attempt at systematically gathering and extrapolating data.  At best, our results should be considered one piece of relevant evidence that can supplement expert opinion, and perhaps a reason to somewhat decrease one's credence in more extreme predictions.

Our data sources are opportunistic.  We include industry blog announcements, not just academic journal articles.  We are exposed to significant noise, e.g., from what numbers researchers choose to report, and what numbers we (as non-experts) are able to interpret.  Some of this noise is random and will average out as the dataset accumulates, but much of it is biased and will not \footnote{For example, one may expect that papers may only report the metrics they were explicitly trying to improve.}.

We consider our modeling assumptions to be optimistic in at least three ways. First, we are extrapolating from the \emph{best} reported qubit numbers and error rates and ignore other values that have been reported. Second, we ignore both qubit connectivity and the trade-off between the number of physical qubits and the error rate. Third, we presume that constant-rate exponential growth will continue in the long term, which is disputed, especially for the average two-qubit gate error rate. These assumptions are discussed in more detail later.

For these reasons, the confidence intervals predicted by our model are quite different from the credences one might get from a wise and holistic Bayesian analysis. Still, we do expect the former to be useful input for the latter.

We are partially motivated by aggressive predictions often made informally and in the media about when quantum computers will have important real-world effects (e.g., by threatening the security of cryptography systems).  We hope our work will advance the discussion about the future of this field by compelling commentators to acknowledge that aggressive predictions require making assumptions about the rate of progress increasing above the current trend, when in fact the expert wisdom seems to point, if anything, toward future progress falling below the current trend, especially for gate fidelity.

\section{Operationalizing progress }
\label{operationalizating progress}
\addcontentsline{toc}{section}{2 Operationalizing progress }
\sectionsubtitle{How should we measure progress in quantum computing?}

In this section we define the concrete milestone we are trying to predict, and relate it to metrics of current quantum computing systems.

For this purpose we define our own index, the \textit{generalized logical qubits} (GLQs), which estimates the number of qubits that will be available for computation after accounting for error-correction overheads, and which extends to fractions less than 1 for machines (e.g. all extant ones) that fall short of the important technical threshold of fault tolerance, described further below.

We then explain why the thresholds of 1 and 4100 GLQs roughly correspond to the very important milestones of showcasing fault tolerance and compromising the widely used online security standard RSA 2048.

Readers uninterested in the motivations of our metric definitions and subsequent operationalization can just take these as given and skip to Section~\ref{current-state-of-the-art}.

\subsection{Types of quantum computation}
	\label{types-of-quantum-computation}
	\addcontentsline{toc}{subsection}{2.1 Types of quantum computation}
	\sectionsubtitle{What approaches are there towards large-scale quantum computing?}
	
	Very roughly, approaches to computing (both quantum and classical) can be divided into analog and digital devices.  Within classical computing these categories are now strongly associated with representing information using continuous and discrete variables, respectively, but they are also deeply connected with the degree to which the dynamics of the physical system implementing the computation are abstracted away from the mathematical form of the computation itself.  In short: analog computers take advantage of natural physical dynamics that ``look like''  the computation one wants to perform, and are often \textit{limited} to such computations, while digital computers are \textit{universal} --- in the sense that they are capable of performing any reasonable computation given enough resources --- but require greater technological sophistication.  
	
	The analog-digital distinction is not a perfect dichotomy, but it has key implications for the use of error correction and, more broadly, fault tolerance \footnote{There are important technical distinctions between error correction and fault tolerance, but we will not address them in this work.}. In general, analog devices are less suitable for error correction.  In the long run this leaves them vulnerable to noise as that problem becomes progressively worse for longer computations on larger devices.
	
	Historically, for both quantum and classical computers, rudimentary analog devices without error correction became available before digital devices.  In the case of classical computers, digital devices ultimately won out as technological capabilities improved.  Experts disagree whether quantum analog computers will enjoy a time period where they are more practically useful than both quantum digital devices and classical devices. By far the most advanced form of analog quantum computing are so-called quantum annealers \citep{das2008colloquium} (though alternative approaches are being pursued, e.g., \citep{yamamoto2017coherent}).
	
	For these reasons, one can usefully identify three broad categories of quantum computing \citep{grumbling2019quantum}: 
	\begin{itemize}
		\item 
			Analog \textbf{quantum annealers};
		\item 
			\textbf{Noisy Intermediate Scale Quantum }(\textbf{NISQ}) computers; and 
		\item 
			Scalable, gate-based, digital \textbf{fault-tolerant quantum computers} (\textbf{FTQC}).
	\end{itemize}
	Strictly speaking, the category of NISQ computers includes both analog and digital devices, but it can be useful, if crude, to simply think of them as digital devices that are not fault tolerant, and hence are limited in the scale of the computations they can perform by noise. Most experts agree that FTQC devices will dominate in the long-term \citep{grumbling2019quantum} (though there are notable dissenters), and the timescale on which this comes about is very uncertain.

\subsection{Milestones}
	\label{milestones}
	\addcontentsline{toc}{subsection}{2.2 Milestones}
	\sectionsubtitle{What are the most important technical milestones on the road to scalable quantum computing?}
	
	For each of the three categories in the previous subsection we can identify three relevant milestones beyond the theoretical.
	\begin{enumerate}
		\item 
			\textbf{Prototype}: A proof-of-concept implementation of the most basic machine operations.
	
		\item 
			\textbf{Quantum supremacy} (or ``quantum advantage''~\footnote{ Unfortunately, terminology is not completely standardized.  Some authors take ``quantum supremacy''  and ``quantum advantage''  to be synonyms, while others distinguish them based on whether the problem could be solved by \emph{existing} classical (super)computers vs. \emph{any} feasible classical computer, or based on the economic value of the problem.}): 
				The resolution of a mathematical problem by the quantum computer that would be unfeasible to solve classically \footnote{It is worth emphasizing that the mathematical problem selected for demonstrating quantum supremacy is invariably contrived to be as easy as possible for the quantum devices and as hard as possible for the classical devices, i.e., the ``quantum home-field advantage''  has been maximized.  In particular, quantum supremacy does not mean the quantum computer is better at all computational problems, or even a large class of them.}.

		\item 
			\textbf{Practical demonstration}: Application to problems that are of interest \textit{independent} from quantum computing.
	\end{enumerate}
	
	Progress along these miles stones is summarized in Table~\ref{tab:milestones}. Proof-of-concept machines showcasing analog quantum annealers \citep{karimi2011investigating} and NISQ computers \citep{kelly2015state} have existed for several years. Google recently claimed the quantum supremacy milestone \citep{arute2019quantum} with a NISQ computer, showcasing a quantum processor capable of sampling in $0.02$ seconds a probability distribution that would take the current fastest supercomputer (IBM's Summit as of this writing) $2.5$ days to sample \citep{pednault2019quantum}.
	
	However, the highest-quality quantum computing experiments can apply only of order 1,000 gate operations before an error is likely to occur \footnote{This is quantified by the average gate error rate, that we will explain in Section~\ref{metrics}. Google's average error rate is of the order of 1e3 \citep{arute2019quantum}.}. 
	In order to perform usefully long \textit{logical} computations, it will be necessary to use error-correction on the \textit{physical} qubits.  Fault-tolerant quantum computers will be able to drive logical errors down arbitrarily low, with overhead requirements that scale logarithmically with the desired logical error rate (which is inversely proportional to the average length of computation that can be performed without error).  The theory behind error correction and FTQC is well-developed \citep{preskill1998quantum} but no functional experimental prototype exists as of the writing of this paper.

Our question of interest is when either of these technologies will be good enough to solve problems inaccessible to classical computers.

\textbf{Question of interest:} When will we have quantum computers capable of solving problems which cannot be feasibly be solved classically and which have noticeable real-world implications?

This prompts the related question, which of the three candidate technologies will first achieve this milestone?

Analog quantum annealers are not seen as a practical solution for large scale quantum computing due to the lack of a practical error correction scheme to enable computations of arbitrary length \citep{grumbling2019quantum}.

Similarly, while some small applications on certified randomness have been proposed using NISQ \footnote{For a popular-level description, see \citep{aaronson2017quantum}.} \citep{acin2016certified}, no major applications are envisioned yet \citep{grumbling2019quantum}.

Thus for the purposes of this report we will choose to study progress in FTQC.

\textbf{Proxy for question of interest:} When will we have large scale, fault-tolerant quantum computing?

\onecolumngrid

\begin{table}[H]
	\centering
	\def\arraystretch{1.3}
\begin{tabular}{|>{\centering}m{0.15\textwidth}|>{\centering}m{0.24\textwidth}|>{\centering}m{0.26\textwidth}|>{\centering\arraybackslash}m{0.21\textwidth}|}
	\hline
	Technology & \textbf{Analog Quantum Annealers} & \textbf{NISQ Computer} & \textbf{FTQC}\\ \hline
	Stage      & Prototype                                                                            & Quantum supremacy                                                                   & Theoretical                                                                                              \\ \hline
	Example    & D-Wave's 2000Q \citep{king2018observation} & Google's Scamore \citep{arute2019quantum} & Fault-tolerant quantum computation \citep{preskill1998quantum} \\ \hline
\end{tabular}
%
\caption{A map of milestones in quantum computing. In this article we focus on forecasting progress in FTQC.}
\label{tab:milestones}
 \end{table}
\twocolumngrid


\subsection{Metrics}
	\label{metrics}
	\addcontentsline{toc}{subsection}{2.3 Metrics}
	\sectionsubtitle{How can we quantitatively track progress toward the relevant milestones?}
	
	The performance of current systems can be monitored via several useful quantities. Some of the most relevant are \citep{grumbling2019quantum}:
	
	\begin{itemize}
		\item 
			\textbf{Number of physical qubits in a system. }This is the simplest measure we have. It is the number of two-state physical subsystems in the computer in which coherent quantum information can be stored, such as the orientation (up or down) of a particle, or the direction (clockwise or counterclockwise) that a current in a superconducting wire is flowing.
	
		\item 
			\textbf{Average two-qubit-gate error rate. }This is measured as the probability that when a quantum logic gate \footnote{Here we use ``logic gate''  simply to distinguish this machine operation from other sorts of gates, like the ones that keep in horses. We are still referring to a gate operation on physical qubits. (In this paper, we will not need to discuss logical gates, which are a counterpart to logical qubits.) }
			(similar to a classical AND and OR gate) is applied in the system it fails to produce the correct output.  An analogous metric exists for one-qubit gates, but the single-qubit error rate is typically much lower, so the overall error rate per operation is dominated by the two-qubit-gate error rate. In this work we will often simply refer to this as the error rate. 
	
		\item 
			\textbf{Coherence time. }The computational power of quantum computers relies on the physical qubits remaining highly isolated from their surrounding environment.  When the qubits interact with their environment, information about the state of the qubits ``leaks out'' , a process known as \textit{decoherence}, inhibiting the necessary quantum coherent effects. Longer coherence times allow more gates to be performed before this happens, allowing for more complicated quantum computation.  
	
		\item 
			\textbf{Qubit connectivity. }Two physical qubits are \emph{connected} when a gate operation can be achieved by inducing the qubits to physically interact with each other, which often is only possible if they are a short distance apart.  The connectivity of a computer is the (graph) structure of these connections.  To apply a gate to unconnected qubits, multiple costly interactions between intermediate qubits in a chain must be used.  One quantitative measure of a computer's connectivity is the average number of connections it takes to link two qubits in the system. 
	
		\item 
			\textbf{Number of logical qubits in a system}. A quantum computer with error correction and a given number of noisy \emph{physical} qubits can emulate a noiseless computation on a smaller number of \emph{logical} qubits.  As the noise in a device gets greater, it can generally support fewer logical qubits for the same number of physical qubits.  If the noise is greater than some threshold, \emph{zero} logical qubits can be supported.  Very few extant devices have achieved noise levels below this threshold, and none of them have created a single logical qubit (in part because of insufficient, and insufficiently connected, physical qubits).
	\end{itemize}
	
	In this report we choose to focus on (1) the physical number of qubits in a system and (2) the average two-qubit error rate, because they are of high importance and are widely reported in the literature \footnote{We emphasize that qubit connectivity is essentially for fault tolerance and useful computations.  (Trivially, $N$ separate and non-interacting quantum devices, each consisting of $2$ physical qubits interacting through a gate with error rate $r$, could be construed as a quantum ``computer''  with arbitrarily high physical qubits ($2N$) and arbitrarily low error rate ($r$).) The reasonableness of our model is based on the premise that, in practice, connectivity will to some extent track qubit counts and error rates within the field overall.}.
	
	Beyond these immediate metrics, we can try to combine them to produce a metric that captures progress more accurately.  IBM opts for this approach and tracks the \textbf{quantum volume} \citep{cross2019validating}. 
	
	In a similar spirit, we construct our own figure of merit, the number of \textbf{generalized logical qubits}.  This figure approximates the number of logical qubits that a device theoretically would be able to simulate using quantum error correction at a given average error rate with a given number of physical qubits using the surface code \citep{barends2014superconducting,fowler2011surface,fowler2012surface}, the leading choice error correction scheme for systems based on superconducting qubits.  Unlike the actual number of logical qubits, this metric extends smoothly and sensibly to fractional qubits.  It is well-defined for devices that have achieved two-qubit gate error rates below the fault-tolerance threshold mentioned above.
	
	Formally, we define the number of generalized logical qubits to be $N_L = N_P \fQEC$, where $N_P$ is the number of physical qubits, $\pP$ is the two-qubit gate error rate, 
	\begin{align}
	\fQEC = \left[4 \frac{\log \left(\sqrt{10}\, \pP/\pL\right)}{\log\left( \pth/\pP\right)} + 1 \right]^{-2}
	\end{align}
	is an overhead factor ($0 < \fQEC < 1$) accounting for error correction, $\pth \approx 10^{-2}$ is the approximate threshold error under which fault-tolerance becomes possible for the surface code \citep{fowler2011surface,fowler2012surface,javadi-abhari2017towards}, and $\pL=10^{-18}$ is the acceptable logical error rate.  A contour plot of this function is given in Figure \ref{fig:countour_plot}, and additional detail can be found in appendix \ref{appendix:generalized-logical-qubits}.


\onecolumngrid

\begin{figure}[H]
	\centering
		\includegraphics[width=\textwidth]{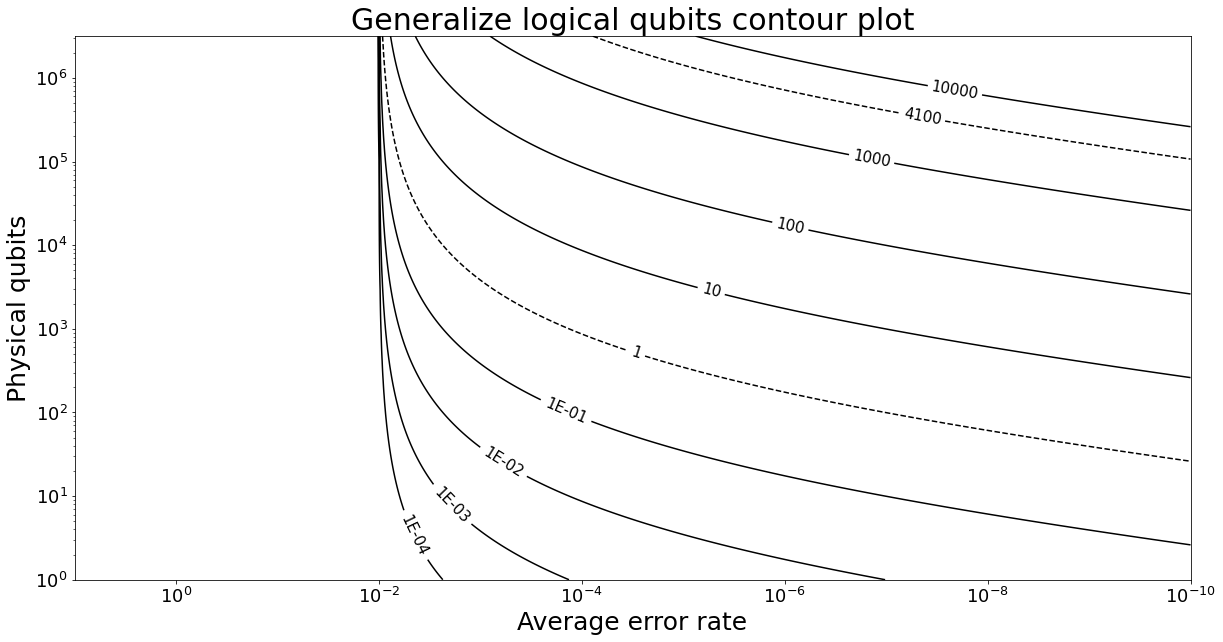}
		\label{fig:countour_plot}
		\caption{Contour plot of the generalized logical qubit map. Our primary thresholds of interest, 1 and 4100, are delineated with dotted curves. The contours asymptote at the fault tolerance threshold $\pth\approx 10^{-2}$. Note that our definition depends on the underlying error correction scheme, in this case we are assuming the surface code.}
\end{figure}



\begin{table}[H]
	 \setlength\extrarowheight{2pt}
\centering
\begin{tabular}{|>{\centering}p{0.45\textwidth}|>{\centering\arraybackslash}p{0.45\textwidth}|} 
	\hline
	\textbf{Milestone}      & \textbf{Operationalization}     \\ \hline
	Quantum fault tolerance & 1 generalized logical qubits    \\ \hline
	RSA 2048 quantum attack & 4100 generalized logical qubits \\ \hline
\end{tabular}
%
\caption{Milestones of interest operationalized in terms of generalized logical qubits.}
\label{tab:operationalization}
 \end{table}
\twocolumngrid

 
\subsection{Operationalization}
\label{operationalization}
\addcontentsline{toc}{subsection}{2.4 Operationalization}
\sectionsubtitle{How can we convert our primary forecasting questions into statements about generalized logical qubits?}

In this subsubsection, we develop our preferred operationalization of progress milestones, as summarized in Table~\ref{tab:operationalization}.
 
\textbf{Question of interest:} When will we have fault-tolerant quantum computing?

As discussed in Section~\ref{types-of-quantum-computation}, the first step to that achievement is proof-of-concept quantum error correction. One reasonable operationalization of this intermediate milestone based on quantum computing is straightforward.

\textbf{Operationalization:} When will the number of generalized logical qubits exceed 1?

The next step would be to scale up fault-tolerant quantum computers to a size where they can perform useful computation. As a landmark of disruptive application we chose the development of a quantum computer capable of threatening a commonly used cryptographic algorithm, RSA 2048.

\textbf{Question of interest: }When will a large fault-tolerant quantum computer that can run Shor's algorithm to break RSA 2048 be developed?

This operationalization is based on one of the most anticipated uses of a quantum computer. However, this operationalization does not allow a ready estimation based on short term metrics. We address that now:

\onecolumngrid

\begin{figure}[H]
	\centering
		\includegraphics[width=0.8\textwidth]{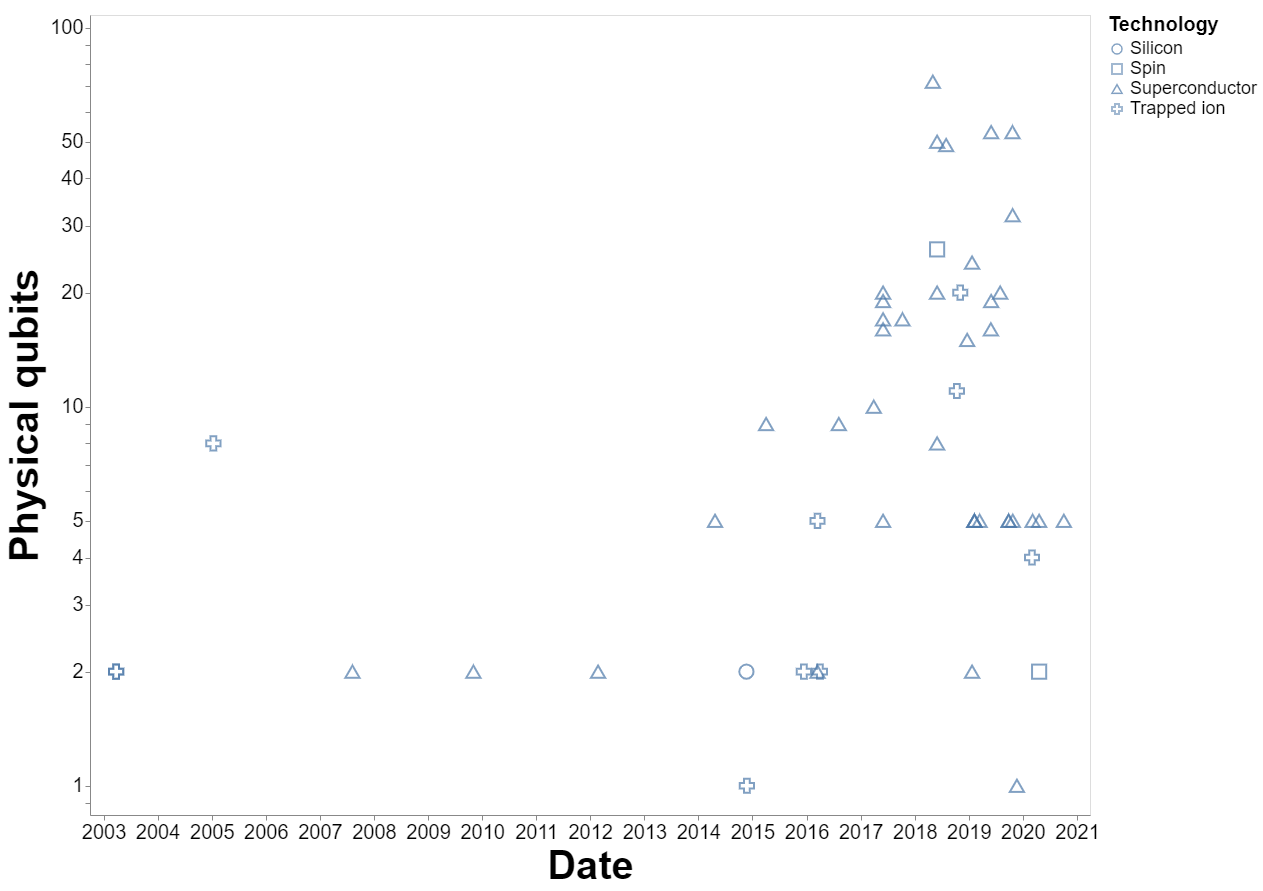}
		\caption{Reported number of physical qubits of systems between 2003 and 2020. $n=52$ data points.}
		\label{fig:historical_physical_qubits}
\end{figure}




\begin{figure}[H]
    \centering
		\includegraphics[width=0.8\textwidth]{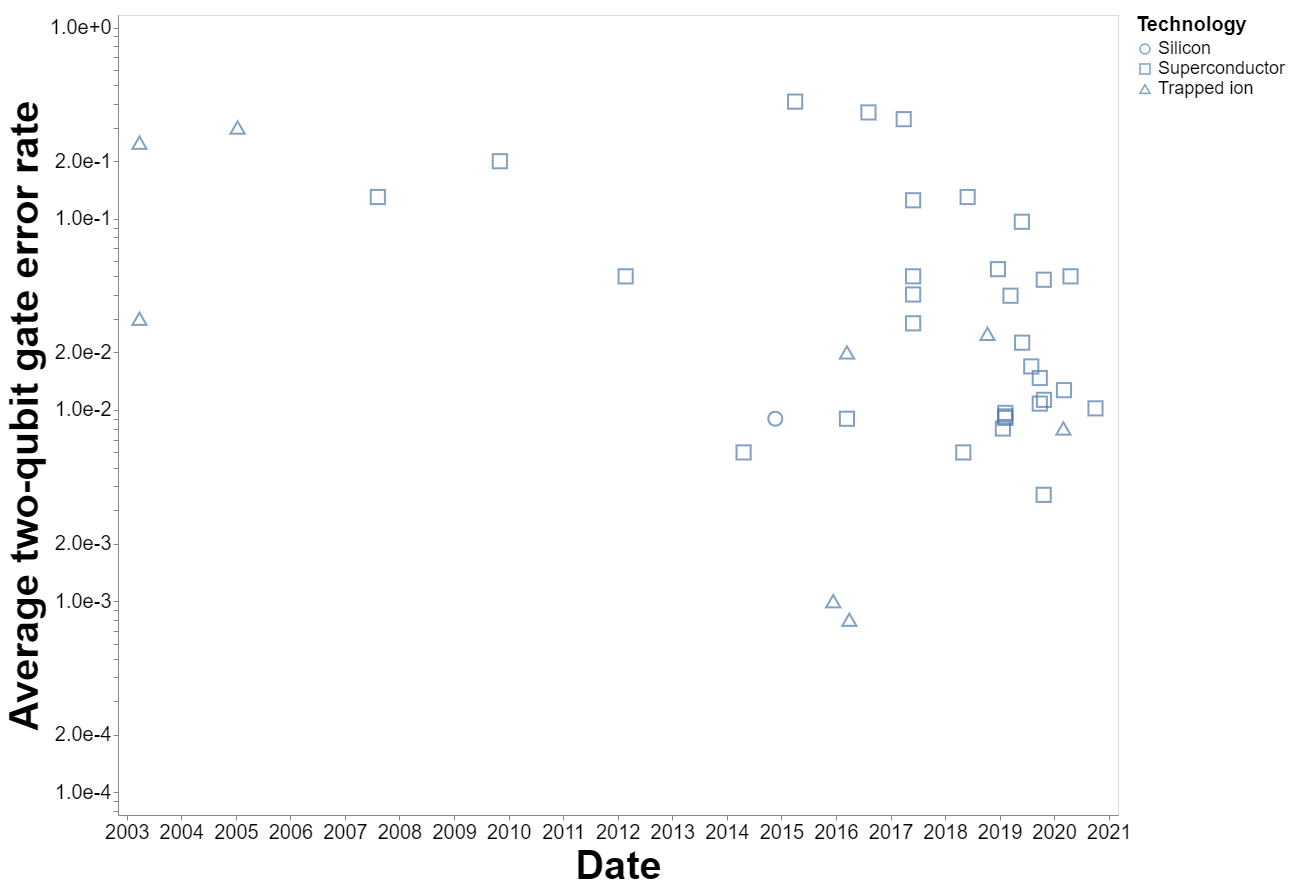}
		\caption{Reported average two-qubit gate error rate of systems between 2003 and 2020. $n=40$ data points. (This is smaller than Fig.~\ref{fig:historical_physical_qubits} because fewer papers reported qubit counts than error rates.)}
		\label{fig:historical_error_rate}
\end{figure}
\twocolumngrid


First, note that running Shor's algorithm to break RSA 2048 will almost certainly require error correction \footnote{``Without QEC, it is unlikely that a complex quantum program, such as one that implements Shor's algorithm, would ever run correctly on a quantum computer.''  \citep{grumbling2019quantum}.}; the computation is of sufficient length that many physical errors will reliably occur in any hardware available in the foreseeable future. Next we observe that the number of logical qubits needed to run Shor's algorithm to break RSA 2048 has been estimated at $ \sim $4098 logical qubits \citep{gidney2019how,haener2016factoring,roetteler2017quantum} \footnote{Häner et al. describe a procedure for using $2(N+2)$ qubits to factor an $N$-bit number \citep{haener2016factoring}. Recently, Gidney \&  Ekerå have described a procedure that uses $ \sim $50$\%$  more qubits, $3N + 0.002 N \log_2 N$, but runs two orders of magnitude faster for these size numbers \citep{gidney2019how}.}. We can thus use our generalized logical qubits metric to operationalize this question.

\textbf{Operationalization: }When will a quantum computer with more than $\sim$4098 generalized logical qubits be built?


\section{Current state of the art}
\label{current-state-of-the-art}
\addcontentsline{toc}{section}{3. Current state of the art}
\sectionsubtitle{How do current systems perform on the metrics of interest?}


Using Refs.~\citep{grumbling2019quantum,finke2020scorecards} as a helpful starting point, we compiled a large historical dataset available \href{https://\allowbreak{}docs.\allowbreak{}google.\allowbreak{}com/\allowbreak{}spreadsheets/\allowbreak{}d/\allowbreak{}1pwb4gf\allowbreak{}0FxlxgfVh\allowbreak{}tXTaqEGS\allowbreak{}9b7Fwsst\allowbreak{}sJ0v7Zb1naQ0/\allowbreak{}edit#\allowbreak{}gid=0}{\textcolor[HTML]{1155CC}{\ul{here}}} \footnote{Full URL: \texttt{https://\allowbreak{}docs.\allowbreak{}google.\allowbreak{}com/\allowbreak{}spreadsheets/\allowbreak{}d/\allowbreak{}1pwb4gf0\allowbreak{}FxlxgfV\allowbreak{}htXTaq\allowbreak{}EGS9b7\allowbreak{}FwsstsJ0v7Z\allowbreak{}b1naQ0/\allowbreak{}edit\#\allowbreak{}gid=0}.}.  Altogether, it is based on data points reported by Refs.~\citep{intel2018, kelly2018preview, gambetta2019cramming, nay2019ibm, ibmquantumexperience2020ibm, intel2017intel, amazon2019amazon, rigetticomputing2018quantum, arute2019quantum, ballance2016high-fidelity, barends2014superconducting, barends2016digitized, chow2012complete, rigetticomputing2018rigetti, debnath2016demonstration, dicarlo2009demonstration, finke2020qubit, friis2018observation, gaebler2016high-fidelity, haeffner2005scalable, hong2020demonstration, kelly2015state, leibfried2003experimental, nersisyan2019manufacturing, reagor2018demonstration, schmidt-kaler2003realization, sheldon2016procedure, song2017-qubit, steffen2006measurement, watson2017ibm, veldhorst2015two, wright2019benchmarking}.

\begin{figure}[H]
    \centering
		\includegraphics[width=0.8\textwidth]{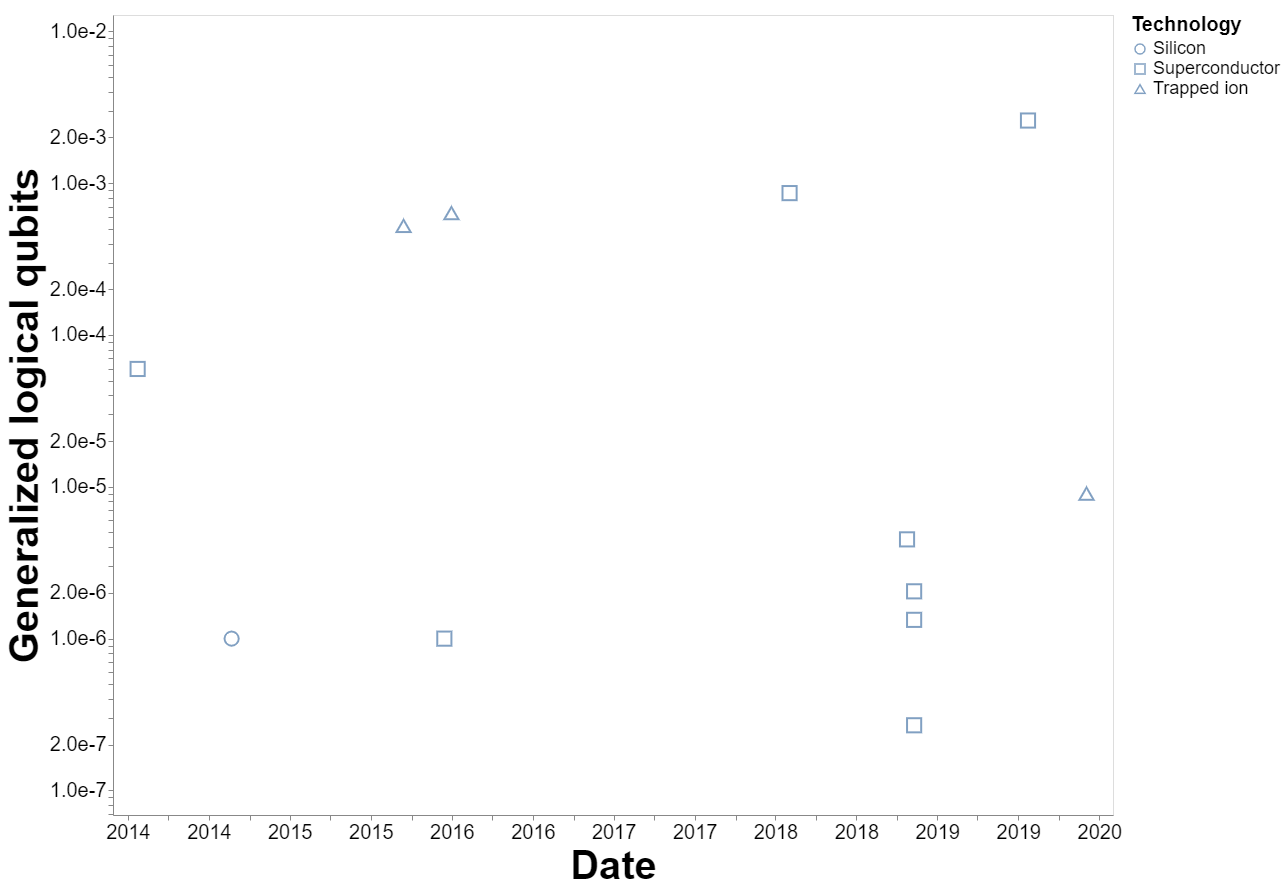}
		\caption{Calculated GLQs for systems between 2003 and 2020. No data points are shown prior to 2014 because the metric is not defined for devices with gate error rates that do not satisfy the fault-tolerance threshold. $n=12$ data points.}
		\label{fig:historical_glqs}
\end{figure}

The metrics we are plotting in Figs. \ref{fig:historical_physical_qubits}, \ref{fig:historical_error_rate}, \ref{fig:historical_glqs} are the physical qubits, average two-qubit gate error rate and our own index, the generalized logical qubits. Explanation of these metrics can be found in Section~\ref{metrics}. They are plotted against the date when the system specifications were made public.\footnote{When we only know the year of publication we imputed the date as 1 June that year.} Our data runs from the year 2003, the date of the earliest source we could find satisfying our requirements, to the first half of 2020, when we froze our dataset for the purpose of our analysis. 

In total we have $n=52$ data points for which we know the number of physical qubits and the date, $n=40$ where we know the physical qubits, the error rate and the date, $n=12$ where we know the physical qubits, the error rate, the date and the GLQs are well defined.

The datapoints collect systems using different quantum computing technologies, including superconductorts, trapped ion qubits, spin qubits and silicon qubits.

\section{Modelling future progress}
\label{modelling-future-progress}
\addcontentsline{toc}{section}{4. Modelling future progress}
\sectionsubtitle{When will various quantum computing abilities be achieved, and how certain are we? }

In previous sections we have identified the key metrics that indicate progress (physical qubits and gate error rate), constructed an index that combines the two (the generalized logical qubit, abbreviated as GLQ) and casted the milestones we are interested in terms of our index (1 GLQ for the proof-of-concept error correction and 4100 GLQs for a quantum Shor attack on RSA 2048).

Here we statistically analyze the historic trends on these metrics. 

First in Section~\ref{a-technological-development-frontier} we show that quantum computer designs so far exhibit a trade off between a large number of physical qubits and low gate error rate.

Then, specializing to superconducting qubit devices, in Section~\ref{extrapolating-current-trends} we extrapolate the current trends in the best performance on these metrics to predict the trend in GLQs and make predictions about our milestones of interest. Section \ref{model-robustness} and Section \ref{model-validation} explore our extrapolation model in depth, respectively looking at alternate modelling choices and a rolling validation of the model.

Through this section we are assuming a continued trend of exponential progress for physical qubits and average gate error. 

Ref.~\citep{grace2020discontinuous} examines 37 technological trends, and finds a robust discontinuity in $32\%$  of these trends, with a base yearly rate of $0.1\%$  chances of a discontinuity per year for each trend. This work suggests a substantial chance of modelling error that we are not explicitly accounting for.

\subsection{A technological development frontier}
\label{a-technological-development-frontier}
\addcontentsline{toc}{subsection}{4.1 A technological development frontier}
 
\begin{adjustwidth}{0.5in}{0.5in}
\sectionsubtitle{How do the physical qubits and average two-qubit gate error rate relate to each other?}

\end{adjustwidth}

To answer this question, we will model how papers score in both metrics as a multivariate log linear model \footnote{We could instead have used multiple linear regression where, for example, we predict the log error rate based on the year and log physical qubits. However this would require us to assume a linear dependency between the log error rate and log physical qubits, whereas the multivariate model is agnostic about the relation of the two metrics.} that takes as input a date and outputs a distribution for the combination of metrics that papers around that date are likely to produce. In particular, we are not just looking at the total correlation between these two metrics over our full dataset, since steady but independent progress in each would induce a positive correlation; rather we are looking at the correlation between them \emph{conditional} on the year of publication to see whether there is a trade-off between these two metrics at any given level of technological maturity.

For a quick and informal introduction to multivariate models, see \citep{vega2020multivariate}. For a more formal treatment, see 
Chapter 6 of Ref.~\citep{izenman2008modern}.

Our model has the form:
\begin{align}
	Y=X B+ \Xi ;\qquad   \Xi \sim N\left( 0,  \Sigma   \right)
\end{align}
where each column in $Y \in  \R^{n\times 2}$ corresponds to the logarithm of our metrics of interest (the physical qubits and the average two-qubit gate error rate) for each of the papers in our dataset, and $X \in \R^{n\times 2}$ is the date as a fractional year stacked with a constant intercept 1.


\onecolumngrid

\begin{figure}[H]
    \centering
		\includegraphics[width=0.8\textwidth]{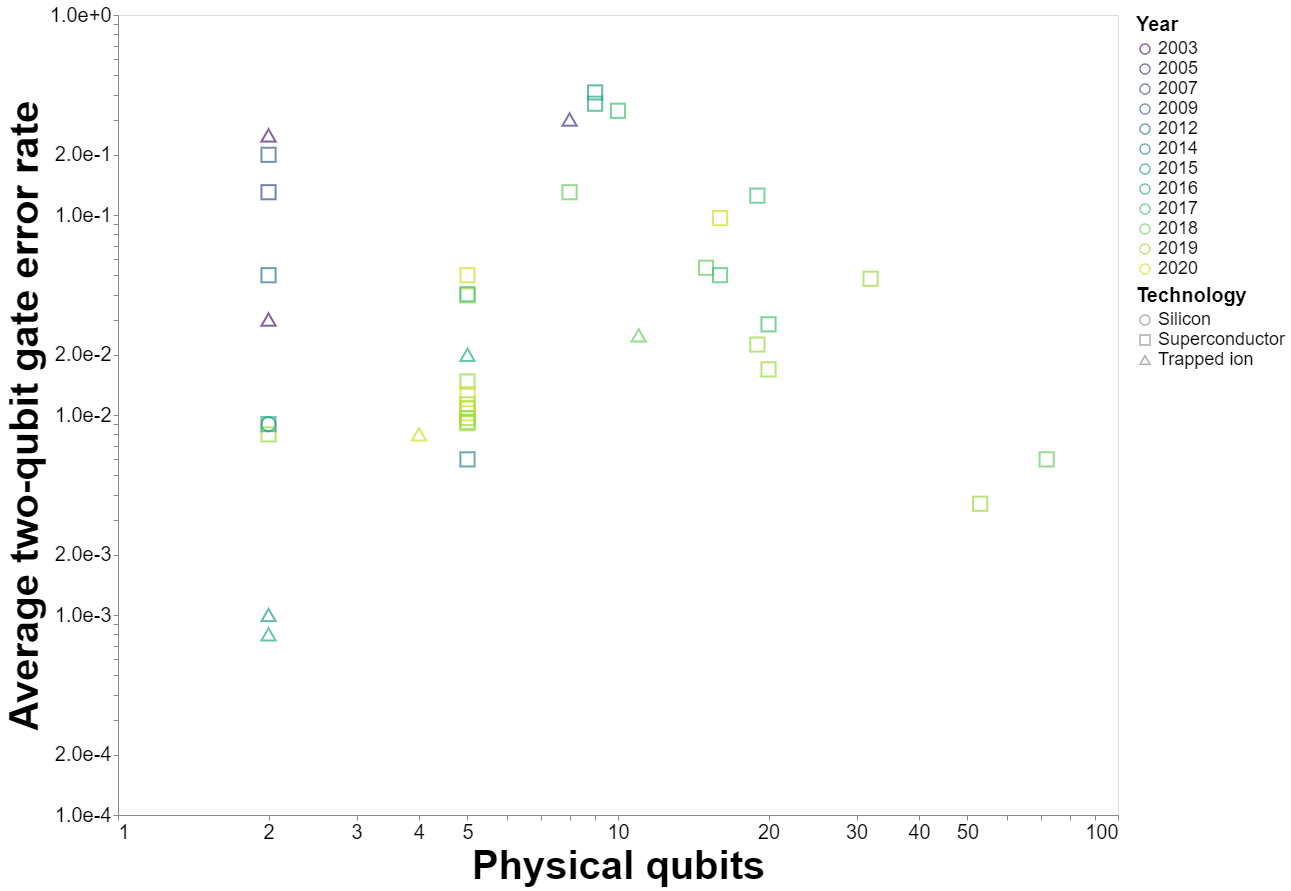}
		\caption{Physical qubits plotted against gate error rate for systems between 2003 and 2020. $n=40$ data points.}
		\label{fig:historical_error_rate_vs_physical_qubits}
\end{figure}
\twocolumngrid


The model is characterized by a matrix $B \in \R^{2\times 2}$ of drift parameters and a covariant noise matrix $  \Sigma \in \R^{2\times2}$. We can estimate their values with the max likelihood estimators  $\hat B$ and  $\hat \Sigma$, calculated as:

\begin{align}
	\hat B &= \left( X^{\tp}X \right)^{-1}X^{\tp}Y\\
	\Sigma  &=\frac{1}{n-2} \left( Y-XB \right)^{\tp} \left( Y-XB \right)
\end{align}

Using all data from 2003 onwards the maximum likelihood estimation of the parameters are displayed in Table~\ref{tab:b_and_sigma}.


The off-diagonal parameter of the symmetric matrix  $\hat \Sigma$  estimates the covariance between the two metrics.

We can provide a confidence interval for the covariance using the naive bootstrap procedure \footnote{For those unfamiliar with bootstrapping, we repeatedly sample n papers with replacements from our dataset and compute the estimator of the covariance $\sigma^2_{12} = \hat \Sigma_{1,0} $  over the resample. The quantiles 0.05 and 0.95 of this sampling procedure converge to a $90\%$  confidence interval for the covariance.} 
(see Chapter 11 of Ref.~\citep{efron2016computer}).

Using all data from 2003 onward ($n=40$ data points) we estimate that the covariance is positive with $98.8\%$  confidence, with a $90\%$  CI of (0.13, 0.76). We note that the physical qubits metric is positively oriented while the error rate metric is negatively oriented, so the positivity of the covariance suggests the existence of a robust trade-off between both metrics.

\onecolumngrid

\begin{table}[!htbp]
\def\arraystretch{1.4} 
\begin{tabular}{ccc}\hline 
	 & $\hat{B}$ & $\hat \Sigma$\\ \hline
	\begin{tabular}{c} 
		{}		 \\
		Intercept\\
		Slope 
	\end{tabular}& 
	\begin{tabular}{cc} 
		 Phys. qubits & Gate error \\
		 -181 & 256\\
		 0.090 & -0.13
	\end{tabular}& 
	\begin{tabular}{cc}
		Phys. qubits & Gate error \\
		 0.76 & 0.44\\
		 0.44 & 2.02
	\end{tabular} \\\hline
\end{tabular}
\caption{Point estimates of the parameters of the multivariate log linear model for data between 2003 and 2020, $n=40$ data points. The first row of $\hat B$  are the intercepts, while the second row corresponds to the yearly log slope. Similarly, the first column of $\hat B$  corresponds to the parameters associated with the physical qubits, while the second corresponds to the parameters associated with the average two-qubit gate error rate. To cast these values in a more amenable way, these parameters indicate that the median value by the year 2020 are 9.22 physical qubits and 0.02 error rate. The first diagonal entry of  $\hat \Sigma$  is the estimated variance for the log physical qubits, while the second diagonal entry is the variance of the log error rate, and the off-diagonal symmetrical entries are the estimated covariance between the two metrics.}
\label{tab:b_and_sigma}
\end{table}




\begin{figure}[H]
	\centering
		\includegraphics[width=0.8\textwidth]{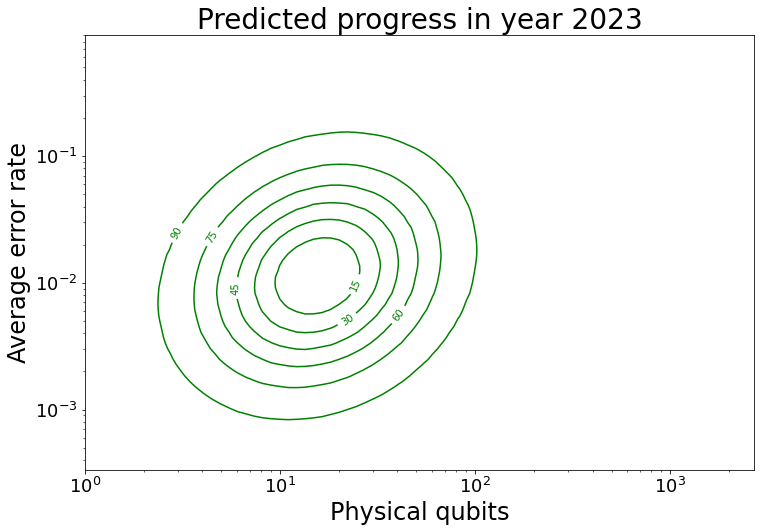}
		\caption{Visualization of the Gaussian pdf of the model with the parameterization in Table~\ref{tab:b_and_sigma} conditioned on the year 2023. That is, the model predicts that a certain percentile of papers published in that year will report metrics inside the respective ellipse.  (Note that this is a prediction about the the distribution of \textit{all} sources satisfying our selection criteria, \textit{not} a prediction about the best values obtained or the technological frontier.) The size and orientation of the ellipse represents the covariance of the metrics. Each curve delineates an $x\%$  probability region, centered around the median.}
\end{figure}
\twocolumngrid


\onecolumngrid

\begin{figure}[H]
	\centering
		\includegraphics[width=0.8\textwidth]{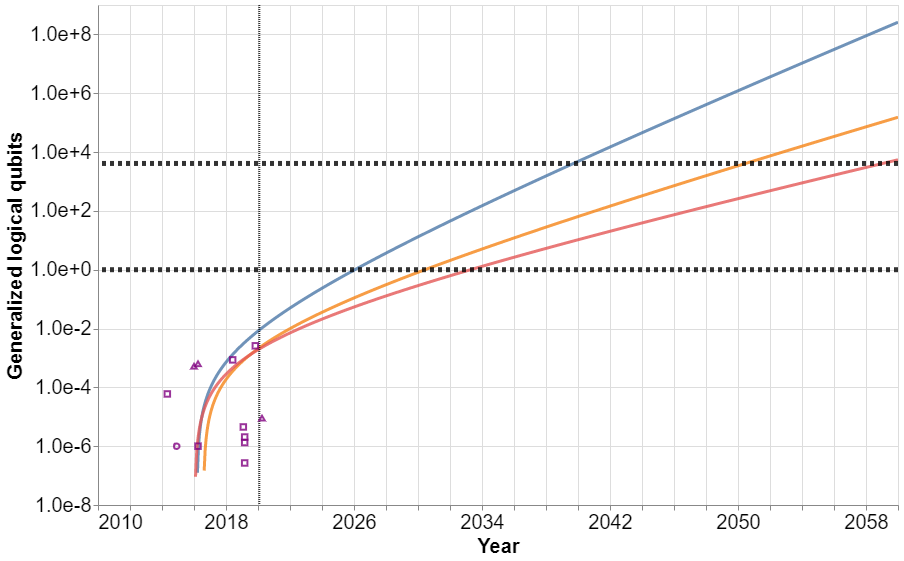}
		\label{fig:extrapolation_main_GLQ}
		\caption{Extrapolated progress of generalized logical qubits at $5\%$, $50\%$, and $95\%$ bootstrapping quantiles (red, orange, and blue, respectively). The trend lines are based on $n=39$ data points corresponding to superconducting quantum computers developed between 2007 to 2020 and constitute the main result of our paper. Also shown are the 12 data points from our dataset for which the number of GLQs are defined. Directly extrapolating these data points would be very unreliable because of their low number and position near the divergence at the fault-tolerance threshold.}
\end{figure}
\twocolumngrid


In plain English, this suggests that quantum computer designers face a trade-off between trying to optimize for quantum computers with many physical qubits and quantum computers with very low gate error rate.

If we focus on data on the most promising substrate so far, superconducting qubits, from 2003 onward ($n=31$ data points) we find that the correlation weakens, with a $90\%$  CI of (-0.11, 0.48) and a confidence of $84.5\%$  that the covariance is positive. The significantly lower confidence indicates that our finding is not robust.

\subsection{Extrapolating current trends}
\label{extrapolating-current-trends}
\addcontentsline{toc}{subsection}{4.2 Extrapolating current trends}
\sectionsubtitle{When will we reach the relevant quantum computing milestones?}

Now, to extrapolate current trends of growth, we assume continued exponential progress on the best physical qubit and error rate metrics achieved so far.  For reasons described below, we consider this to provide a soft \textit{upper} \textit{bound} on the likely rate of quantum computing progress.

We have chosen to focus on predicting the development of superconductor-based quantum computers for two reasons: First it is generally regarded as the single most promising substrate for large-scale FTQC. Second, and more practically, there are substantially more papers in the literature quoting physical qubit numbers and gate error rates in the superconducting category.

We now describe our main modeling method. We begin by considering our two primary metrics --- the number of physical qubits and the error rate --- separately. For each metric, we select only the subset of data points corresponding to papers reporting a new best achieved value of that metric (as of the year of publication).  We then take the two distinct dataset and, to each, we fit a log-linear minimum square error model with log-Gaussian noise. 

In other words, for each of our two metrics we consider just the set of new ``world records''  and perform traditional linear regression on that data in log space.\footnote{During bootstrapping, the median of ``record setting''  physical qubits data points was 11, and the median of ``record setting''  average two-qubit gate error rate data points was 6.} Although this maximization procedure loses some information before feeding it to our model, it is necessary in order to correctly track our topic of interest: the bleeding-edge capabilities of the field. Notably, this means our model resists being influenced by a glut of papers describing devices that are not intended to be competitive on our chosen metrics.

The median trajectory of each model can be extrapolated forward in time, and we combine these to produce a prediction for the progress in generalized logical qubits.

Here is a summary of our model:
\begin{align}
    Y_{1}&=XB_{1}+ \Xi_{1}; \qquad \Xi _{1}\sim N \left( 0, \sigma _{1}^{2} \right)\\
    Y_{2}&=XB_{2}+ \Xi_{2}; \qquad \Xi _{2}\sim N \left( 0, \sigma _{2}^{2} \right) \\
    Z&=f \left( \exp \left( Y_{1} \right) ,\exp \left( Y_{2} \right)  \right)
\end{align}
$X = (1\,\, t)$ stacks an intercept 1 and the date $t$. $Y_1$ is the maximum log physical qubits at date $t$. $Y_2$ is the minimum log average gate error rate at date $t$. $Z$ is the number of generalized logical qubits given the log physical qubit metric $Y_1$ and log gate error rate $Y_2$.

To produce our confidence intervals we are going to use a bootstrapping procedure 
(see Chapter 11 of Ref.~\citep{efron2016computer}).

Our dataset of $n=39$ papers is resampled with repetition $B=1000$ times and log scaled. 

Then the data for each of the metrics is aggregated via a sparse maximum, where only the years where an actual improvement is made are registered. 

This produces two derivative datasets $(X_1,Y_1)$ and $(X_2,Y_2)$ for the trend of max log physical qubits and the trend of min log error rate respectively. 

We use these datasets to compute the maximum likelihood estimates for the parameters (intercept, slope, and noise variance in log space) of the log-linear model.
\begin{align}
B_{i}&= \left( X_{i}^{\tp}X_{i} \right)^{-1}X_{i}^{\tp}Y_{i}\\
\sigma _{i}^{2}&=\frac{1}{n-1} \left( Y_{i}-X_{i}B_{i} \right)^{\tp} \left( Y_{i}-X_{i}B_{i} \right)  
\end{align}

Note that unlike in Section~\ref{a-technological-development-frontier}  here we are fitting two separate log linear models instead of a multivariate one. Furthemore, we emphasize that we are fitting only the subset of data corresponding to new best metric values instead of all reported values like in Section~\ref{a-technological-development-frontier}.

We use these estimates to estimate $T$, the date when the critical threshold of 4100 qubits is crossed by the median trajectory. 
\begin{align}
  T&=\min  \{ t : Z > 4100 \}  \\
  &\approx \min  \Bigg\{ t : f \Bigg( \exp \left( \hat B_{1} \begin{pmatrix}1\\t\end{pmatrix}  \right) , \\
  &\qquad\qquad\qquad\qquad \exp \left( \hat B_{2} \begin{pmatrix}1\\t\end{pmatrix} \right)  \Bigg)  > 4100 \Bigg\}   
\end{align}  

The result is three representative trajectories corresponding to the quantiles 0.05, 0.5 and 0.95 of $T$. For more details on our calculations, please refer to \href{https://\allowbreak{}colab.\allowbreak{}research.\allowbreak{}google.\allowbreak{}com/\allowbreak{}drive/\allowbreak{}1XkWcs\allowbreak{}Uy-Ci\allowbreak{}NDDJffPC3d\allowbreak{}NJbgvR\allowbreak{}6iDAqh#\allowbreak{}scrollTo=\allowbreak{}jV29mJw\allowbreak{}Oiz9W&\allowbreak{}uniqifier=3}{\textcolor[HTML]{1155CC}{\ul{our code}}} \footnote{Full URL: \texttt{https://\allowbreak{}colab.\allowbreak{}research.\allowbreak{}google.com/\allowbreak{}drive/\allowbreak{}1XkWcs\allowbreak{}Uy-Ci\allowbreak{}NDDJffPC3d\allowbreak{}NJbgv\allowbreak{}R6iDAqh\#\allowbreak{}scrollTo=\allowbreak{}jV29mJw\allowbreak{}Oiz9W\&uniqifier\allowbreak{}=3}.}.

We consider this model optimistic in three ways. First, it is being extrapolated over the maximum data instead of the whole dataset. Second, it ignores the trade-off between both metrics we uncovered in Section~\ref{a-technological-development-frontier}. Third, it presumes that the regime of exponential growth will continue in the long term, which is a dubious assumption --- especially for the average two-qubit gate error rate \footnote{There are informal arguments for why error rates may fail to sustain exponential progress in the future even if the number of physical qubits grows exponentially.  Once the fault-tolerant threshold is achieved, there do not seem to be fundamental physical barriers to adding more physical qubits given sufficient effort and resources, but error rates refer to a single gate, and these will plausibly hit strongly diminishing returns with respect to how perfectly they can be engineered. Even if lower error rates are experimentally possible, there are reasons to think devoting engineering resources to more physical qubits may have a higher payoff in terms of the number of logical qubits. Unlike for traditional CPUs, superconducting qubits currently use parallel control channels.  Adding arbitrary numbers of physical qubits may require partially serializing this control, through shared control lines, which hasn't yet been attempted. We thank Daniel Sank for discussion on these points.}.

This model predicts that proof-of-concept FTQL will be developed between early 2026 and early 2033 with $90\%$  confidence with the median in early 2030, and that RSA-2048 Shor attacks will become feasible between mid 2039 and mid 2058 with $90\%$  confidence with the median in early 2050.

Because we are not completely confident in the model, these do not exactly coincide with our Bayesian credences for when these events will take place, but we do find them a strong starting point for further deliberation.

These results are more pessimistic but broadly comparable to those produced through the survey of experts in \citep{piani2019quantum}. We emphasize that $22.7\%$  of the experts they surveyed think it is likely or highly likely that quantum computers will be able to crack RSA-2048 keys by 2030, and $50\%$  think that is likely or highly likely that we will be able to crack RSA-2048 keys by 2035.

Note that our models have two levels of modelled uncertainty: the uncertainty introduced by bootstrapping and the gaussian noise of each estimated model. 

However, the gaussian noise is negligible compared to the bootstrapping uncertainty. This is evident when we plot the quantiles of gaussian noise for each bootstrapped mode, see figure~\ref{fig:extrapolation_gaussian_GLQs}. Thus we ignore the gaussian uncertainty in our analysis.

\subsection{Model robustness}
\label{model-robustness}
\addcontentsline{toc}{subsection}{4.3 Model robustness}
\sectionsubtitle{How robust are the conclusions of our extrapolation model to various alternative choices we might have made?}

As we analyzed the data, we also considered the following variations on our model.  Although our main model is substantially better justified than these variations, we will examine how they would affect things.

Choices made:

\begin{itemize}
	\item Using a max procedure vs. using a best fit line for all data points

	\item All data available vs using data of quantum computers with more than two qubits
\end{itemize}

\onecolumngrid

\begin{figure}[H]
	\centering
		\includegraphics[width=0.8\textwidth]{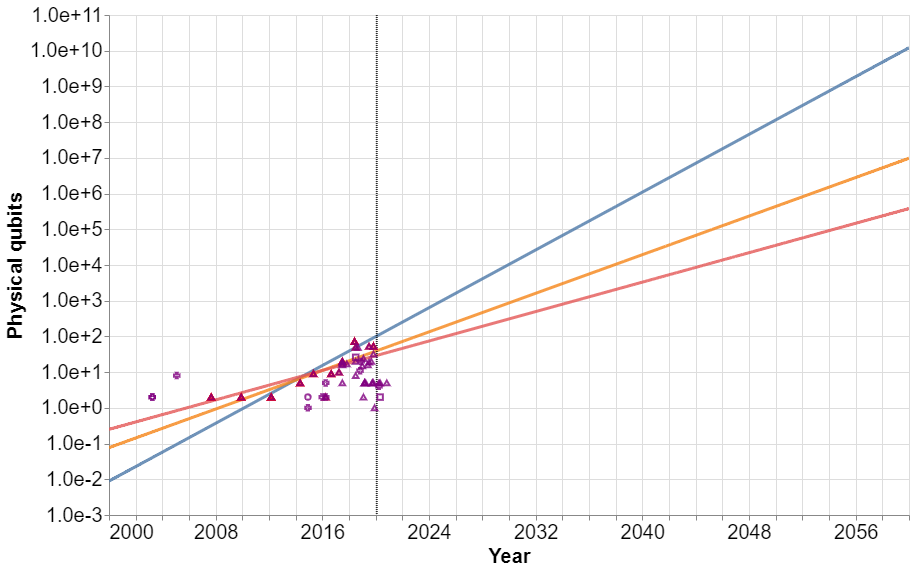}
		\includegraphics[width=0.8\textwidth]{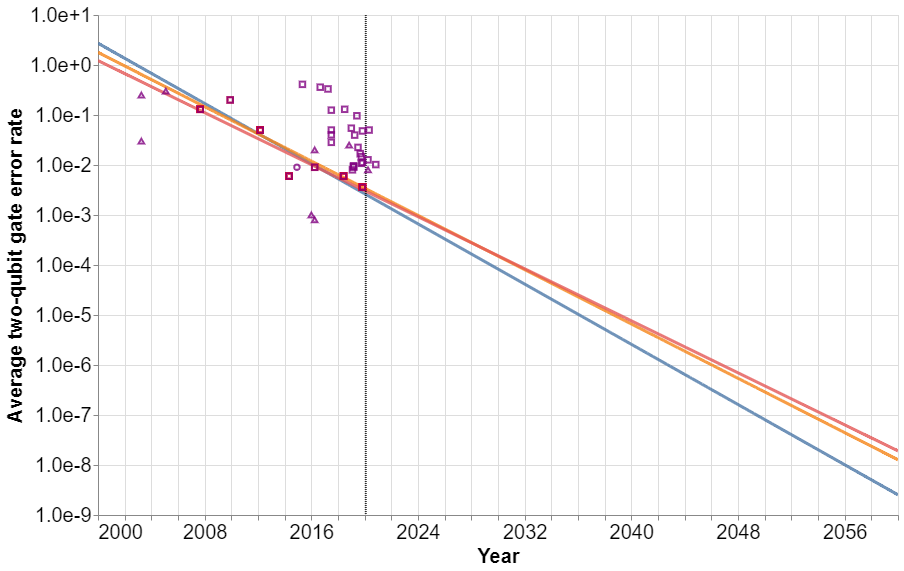}
		\caption{Extrapolated progress of physical qubits (upper plot) and average two-qubit gate error rate (lower plot), from $n=39$ data points corresponding to superconducting quantum computers developed between 2007 to 2020. There are compelling expert assessments that qubit error rates will not and need not fall as dramatically as depicted here (though note that these error rates are still far larger than the rates experienced by modern classical computers).}
\end{figure}
\twocolumngrid


If we do not aggregate the data through a maximum at all we unsurprisingly get far more pessimistic predictions, see Fig \ref{fig:extrapolation_no_max_GLQs}. Worth noting that in this case we are not estimating the \textit{best} results each year, but rather the \textit{typical} results each year. 

Since many papers will report numbers of physical qubits and gate error rates without attempting to push the technological frontier on those particular metrics, we believe our (sparse) max procedure allows our model to better track the hypothetical technological level of the field as a whole. (In particular, the max is not affected by whether lots of non-frontier papers are published.) A constraint imposed by using a max procedure is that it greatly reduces the effective size of our dataset.

\onecolumngrid

\begin{figure}[H]
	\centering
		\includegraphics[width=0.8\textwidth]{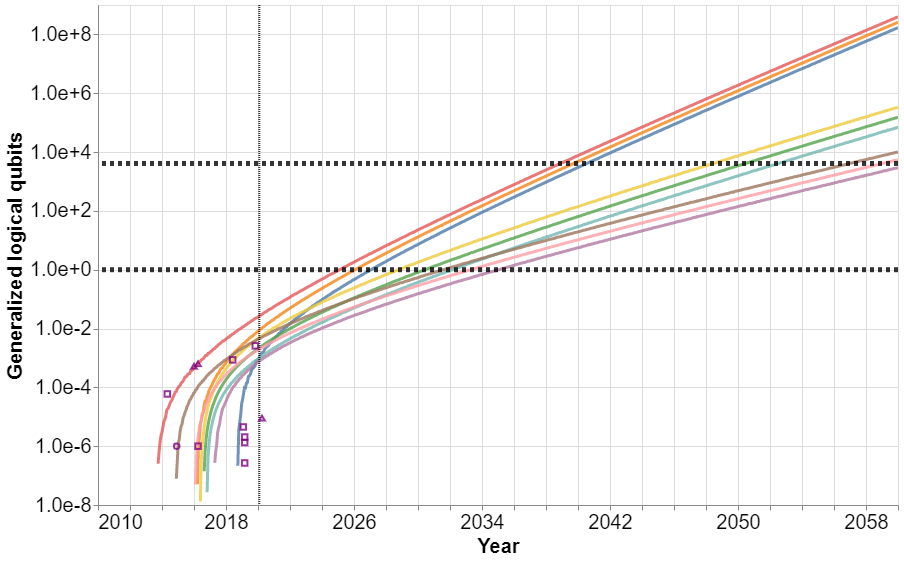}
		\caption{An optimistic extrapolation of current progress on GLQs, based on  $n=39$ data points corresponding to superconducting quantum computers developed between 2007 to 2020, where we have plotted the $5\%$, $50\%$ and $95\%$  quantiles of the gaussian noise credence intervals for the $5\%$, $50\%$ and $95\%$ quantile bootstrapped trajectories.}
		\label{fig:extrapolation_gaussian_GLQs}
\end{figure}
\twocolumngrid


\onecolumngrid

\begin{figure}[H]
	\centering
		\includegraphics[width=0.8\textwidth]{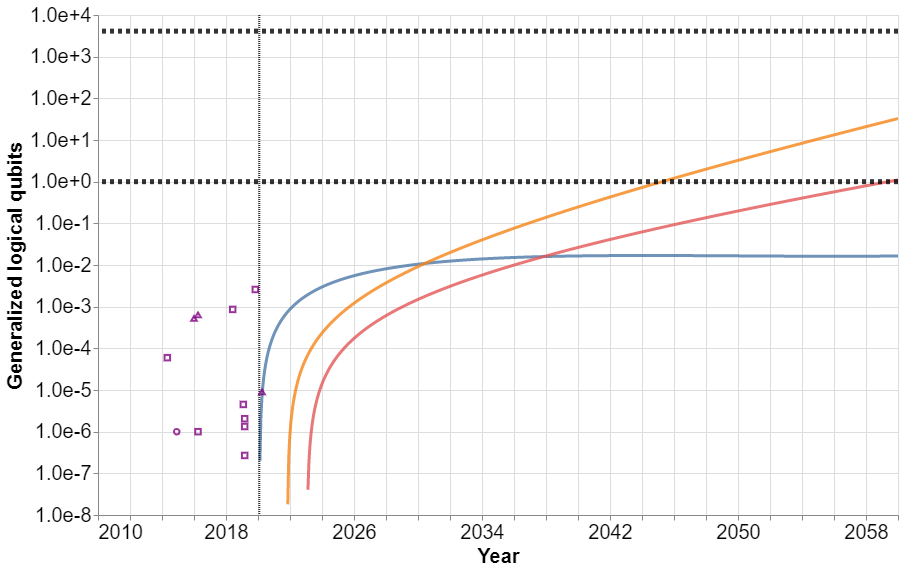}
		\caption{Extrapolated progress of generalized logical qubits based on $n=39$ data points corresponding to superconducting quantum computers developed between 2007 to 2020 and fitted to the raw data, no maximum.}
		\label{fig:extrapolation_no_max_GLQs}
\end{figure}
\twocolumngrid

\begin{figure}[H]
	\centering
		\includegraphics[width=0.8\textwidth]{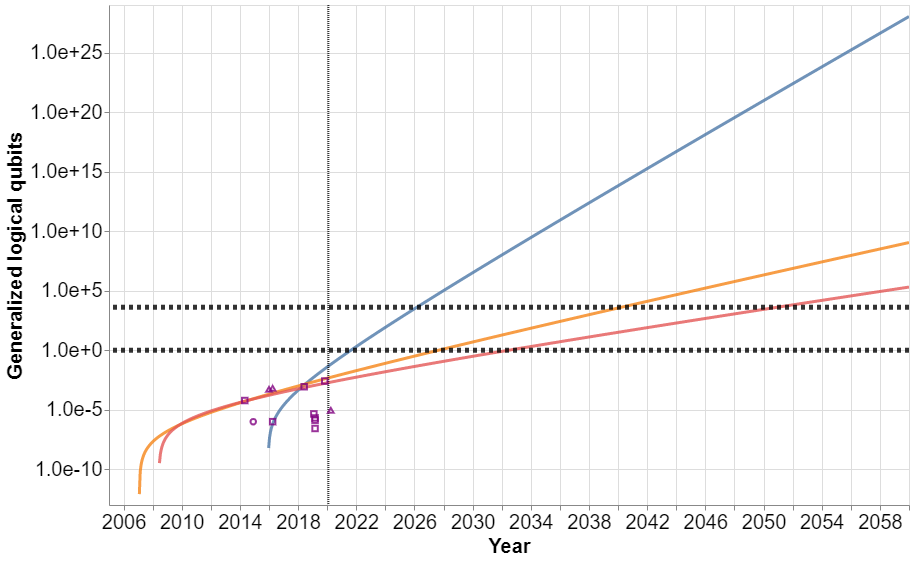}
		\caption{Extrapolated progress of generalized logical qubits based on $n=33$ data points corresponding to superconducting quantum computers developed between 2014 to 2020.}
		\label{fig:extrapolation_2014_GLQs}
\end{figure}

Our main model utilizes data from 2007 to 2020. If we include data from 2014 onwards the resulting predictions are substantially more optimistic, see fig \ref{fig:extrapolation_2014_GLQs}. We have selected this time threshold as 2014 marks the appearance of superconducting quantum computers with more than two qubits, which are arguably more representative of smooth progress.

Our model is quite simple. A more sophisticated approach would likely rely on modelling progress with a geometric drift model as in \citep{farmer2016how}. Unfortunately, this approach was not workable because (1) we have insufficient GLQ data points to apply the geometric drift model directly to univariate GLQ data and (2) the geometric drift model has not yet be generalized to the multivariate setting, which would require significant modifications \footnote{Private communication with Farmer and Lafond.}.  Therefore, we have decided to use our simpler approach instead.

On the other hand, we could have applied an even simpler model by just fitting the raw GLQs. However, the GLQ metric is not well defined for most of the dataset, as the gate error rate of most quantum computers is below the 1e-2 threshold. Thus our dataset is quite limited and the model very noisy. See figure \ref{fig:extrapolation_simple_GLQs} below.

Overall, we observe that the modelling choice makes a significant difference to the predicted time when quantum computing will reach the milestone of being able to crack RSA 2048, which reduces our confidence on the main model. 

In particular, the much faster timelines for the 5\% quantile of the model fitted to data from 2014 to 2020 suggests that the trend of development may have sped up in the last decade. However, given the paucity of data and the dramatic differences with the 50\% and 95\% quantile trajectories in this model this might be just the result of noise.


\subsection{Model validation}
\label{model-validation}
\addcontentsline{toc}{subsection}{4.4 Model validation}
\sectionsubtitle{Does our extrapolation model perform well on historic data?}

In this subsection, we attempt to validate our model by applying it to a subset of our data and using it to predict the rest.

In figure \ref{fig:validation_2018} we plot the predictions when using data from 2007 to 2018, and check the predictions for 2019 against the actual values. In Table~\ref{tab:validation} we expand this procedure to a rolling validation that encompasses different years used as data. The  results are consistent with a well-calibrated model, with 1 value out of 10 falling outside of the predicted 90\% confidence intervals.

\section{Conclusions and open questions}
\label{conclusions-and-open-questions}
\addcontentsline{toc}{section}{5. Conclusions and open questions}
\subsection{Summary}
\label{summary}
\addcontentsline{toc}{subsection}{5.1 Summary}
We began this work by briefly summarizing some of the past literature on technological progress, and argued that exponential growth in hardware technology is a reasonable (though by no means assured) expectation. 

We then compiled a quantitative database of achievements in quantum computing research in the last two decades, focusing especially on two metrics: the highest physical qubit count and lowest average two-qubit gate error rates.

When conditioning on year, we found an (anti\nobreakdash-)correlation between the highest qubit counts and the lowest error rates across all technologies, in line with expectations that these metrics are in tension at a given level of technological development. However, when looking at superconducting data specifically our confidence weakens significantly, indicating that this finding might be spurious.


\onecolumngrid

\begin{figure}[H]
	\centering
		\includegraphics[width=0.8\textwidth]{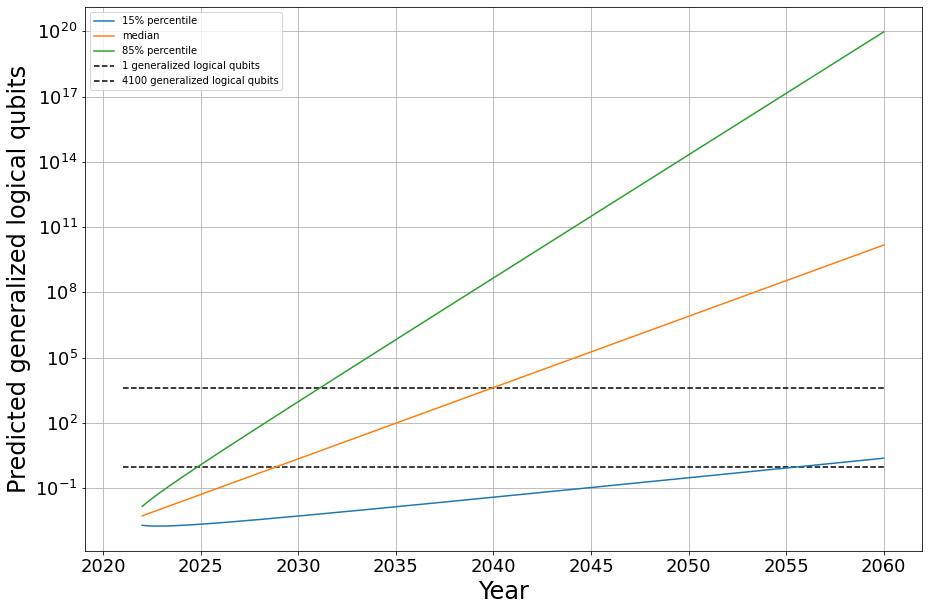}
		\label{fig:geometric_drift}
		\caption{The fit and extrapolation of the GLQ data ($m=5$ years of data, $n=10$ data points corresponding to superconducting quantum computers developed between 2015 to 2020) to a geometric drift model.}
\end{figure}
\twocolumngrid


\onecolumngrid

\begin{figure}[H]
	\centering
		\includegraphics[width=0.8\textwidth]{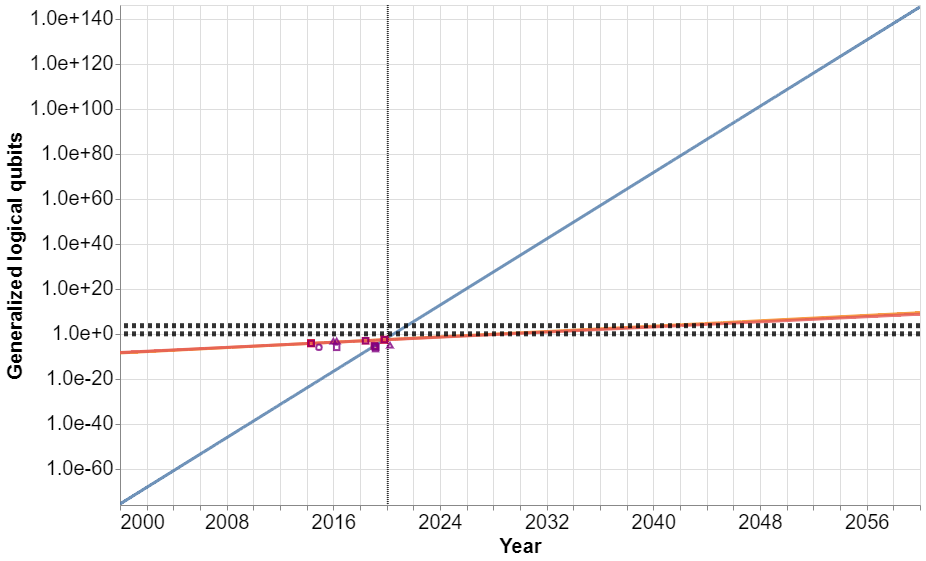}
		\caption{Simple log-linear extrapolation of the GLQs of superconducting quantum computers between 2007 and 2020, $n=12$.}
		\label{fig:extrapolation_simple_GLQs}
\end{figure}
\twocolumngrid


\onecolumngrid

\begin{figure}[H]
	\centering
		\includegraphics[width=0.8\textwidth]{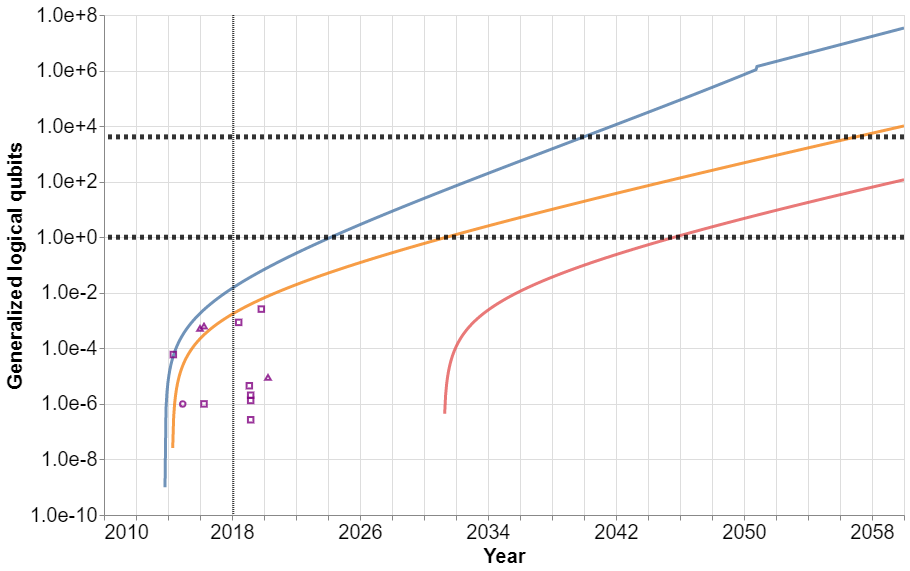}
		\caption{Extrapolated progress of generalized logical qubits based on $n=20$ data points corresponding to superconducting quantum computers developed between 2007 to 2018.}
		\label{fig:validation_2018}
\end{figure}
\twocolumngrid

\onecolumngrid

\definecolor{lightgreen}{RGB}{217,234,211}
\definecolor{lightred}{rgb}{0.96, 0.76, 0.76}
\begin{table}[H]
 			\centering  \setlength\extrarowheight{2pt}
\begin{tabular}{|>{\centering}p{0.27\textwidth}|>{\centering}p{0.13\textwidth}|>{\centering}p{0.13\textwidth}|>{\centering}p{0.13\textwidth}|>{\centering\arraybackslash}p{0.13\textwidth}|}
	\hline
	& \textbf{2019} & \textbf{2018} & \textbf{2017} & \textbf{2016} \\ \hline
	\textbf{Actual maximum} & 6.31e-03 & 4.29e-3 & 4.29e-3 & 1.44e-4 \\ \hline
	\textbf{2007-2018 data $n = 20$} & \cellcolor{lightgreen}1.29e-3; 1.94e-3; 7.99e-2 & N/A & N/A & N/A \\ \hline
	\textbf{2007-2017 data $n = 14$} & \cellcolor{lightgreen}0.; 6.21e-3; 2.69e-2 & \cellcolor{lightgreen}0.; 1.10e-2; 5.57e-2 & N/A & N/A \\ \hline
	\textbf{2007-2016 data $n = 7$} & \cellcolor{lightgreen}1.15e-4; 7.53e-3; 6.43e-2 & \cellcolor{lightgreen}5.82e-5; 4.44e-3; 3.02e-2 & \cellcolor{lightgreen}2.25e-5; 2.54e-3; 1-38e-2 & N/A \\ \hline
	\textbf{2007-2015 data $n=5$} & \cellcolor{lightgreen}3.97e-7; 5.84e-3; 1.57e-1 & \cellcolor{lightgreen}0.; 3.33e-3; 6.58e-2 & \cellcolor{lightgreen}0.; 1.18e-3; 2.55e-2 & \cellcolor{lightred}0.; 8.23e-4; 8.75e-3 \\ \hline
\end{tabular}
\caption{Actual and predicted records for GLQs in recent years, as predicted by models trained on different time spans, following the modelling choices explained in Section~\ref{extrapolating-current-trends} and Section~\ref{model-robustness}. Each entry contains the estimated 5\%, 50\%, and 95\% quantiles separated by semicolons. The confidence intervals where the actual value falls within the predicted 90\% confidence interval are shaded green, while the ones where it does not are shaded red. Note that when the predicted gate error rate is above the threshold where error correction is possible we consider the generalized logical qubits to be 0., hence the null entries.}
\label{tab:validation}
 \end{table}
\twocolumngrid



We proposed a single scalar figure of merit to measure progress towards large-scale fault-tolerant quantum computing, the generalized logical qubit (GLQ), which combines our two metrics. Using GLQs, we operationalized two milestone of fault-tolerant quantum computing: 

\begin{enumerate}
	\item Realization of a single GLQ, roughly corresponding to the beginning of \textit{scalable} quantum computation.

	\item Realization of 4100 GLQs, roughly corresponding to the arrival of computing power that has practical implications for real-world cryptographic systems.
\end{enumerate}

Ultimately, we extrapolated the dataset we had available to predict a less than $5\%$  chance that proof-of-concept fault-tolerant computation will be achieved before 2024, and less than $5\%$  chance that RSA-2048 Shor attacks will be feasible before 2039.  These predictions are dependent on disputable modeling choices we have made, although we generally feel our assumptions have been conservative in the sense that these are reasonable upper bounds on the rate of progress.

\subsection{Future work}
\label{future-work}
\addcontentsline{toc}{subsection}{5.2 Future work}
There is much room for future work to improve on our methods, especially as more data points appear in the coming years:

\begin{itemize}
	\item Other datasets, such as patents as in \citep{benson2015quantitative}.

	\item Modeling of other technical metrics like coherence time, qubit connectivity, and quantum volume, and incorporating other considerations (like the Steane code) into the primary figures of merit. 

	\item Substrates beyond superconducting qubits, especially ion traps.

	\item More sophisticated (but data-hungry) models like that of \citep{farmer2016how} or \citep{harvey1984time}.  

	\item More extensive validation. 

	\item Fitting the models to variables sensitive to global developments that may alter the effort put into quantum computing, such as cumulative inflation-adjustment capital investment in quantum computing research.

	\item Estimate when reversible computing, a necessary but not sufficient property of quantum computers, will be developed in order to maintain progress in \textit{classical} computing. (We briefly expand on this in Appendix 7.2.)
\end{itemize}

Our dataset is freely available \href{https://docs.google.com/spreadsheets/d/1pwb4gf0FxlxgfVhtXTaqEGS9b7FwsstsJ0v7Zb1naQ0/edit#gid=0}{\textcolor[HTML]{1155CC}{\ul{here}}} \footnote{Full URL: \texttt{https://\allowbreak{}docs.\allowbreak{}google.\allowbreak{}com/\allowbreak{}spreadsheets/\allowbreak{}d/\allowbreak{}1pwb4gf0F\allowbreak{}xlxgfVhtXTaq\allowbreak{}EGS9b7FwsstsJ0\allowbreak{}v7Zb1naQ0/\allowbreak{}edit\#\allowbreak{}gid=0}.}, and the code for our models is \href{https://\allowbreak{}colab.\allowbreak{}research.\allowbreak{}google.\allowbreak{}com/\allowbreak{}drive/\allowbreak{}1XkWcs\allowbreak{}Uy-Ci\allowbreak{}NDDJffPC3\allowbreak{}dNJbgvR6iDAqh#\allowbreak{}scrollTo=\allowbreak{}jV29mJwOiz\allowbreak{}9W&uniqifier=4}{\textcolor[HTML]{1155CC}{\ul{here}}} \footnote{Full URL: \texttt{https://\allowbreak{}colab.\allowbreak{}research.\allowbreak{}google.\allowbreak{}com/\allowbreak{}drive/\allowbreak{}1XkWcs\allowbreak{}Uy-Ci\allowbreak{}NDDJffPC3d\allowbreak{}NJbgvR6iD\allowbreak{}Aqh\#\allowbreak{}scrollTo=\allowbreak{}jV29mJw\allowbreak{}Oiz9W\&uniqifier=4}.}.

\subsection{Acknowledgements}
\label{acknowledgements}
\addcontentsline{toc}{subsection}{5.3 Acknowledgements}
We thank J. Doyne Farmer, Chris Ferrie, Steve Howard, Adrian Hutter, Francois Lafond, John-Clark Levin, Pablo Moreno, Florian Pein, Ehud Reiter, Daniel Sank, and Graeme Smith for thoughtful feedback and discussion.

This work is partly supported by a grant from the Center on Long Term Risk.  The Center on Long Term Risk and the employers of the authors did not have any role in the design of the study; in the collection, analysis and interpretation of data; in the writing of the article; or in the decision to submit the article for publication..



  \bibliography{QC-forecast}

\begin{thebibliography}{106}%
\makeatletter
\providecommand \@ifxundefined [1]{%
 \@ifx{#1\undefined}
}%
\providecommand \@ifnum [1]{%
 \ifnum #1\expandafter \@firstoftwo
 \else \expandafter \@secondoftwo
 \fi
}%
\providecommand \@ifx [1]{%
 \ifx #1\expandafter \@firstoftwo
 \else \expandafter \@secondoftwo
 \fi
}%
\providecommand \natexlab [1]{#1}%
\providecommand \enquote  [1]{``#1''}%
\providecommand \bibnamefont  [1]{#1}%
\providecommand \bibfnamefont [1]{#1}%
\providecommand \citenamefont [1]{#1}%
\providecommand \href@noop [0]{\@secondoftwo}%
\providecommand \href [0]{\begingroup \@sanitize@url \@href}%
\providecommand \@href[1]{\@@startlink{#1}\@@href}%
\providecommand \@@href[1]{\endgroup#1\@@endlink}%
\providecommand \@sanitize@url [0]{\catcode `\\12\catcode `\$12\catcode
  `\&12\catcode `\#12\catcode `\^12\catcode `\_12\catcode `\%12\relax}%
\providecommand \@@startlink[1]{}%
\providecommand \@@endlink[0]{}%
\providecommand \url  [0]{\begingroup\@sanitize@url \@url }%
\providecommand \@url [1]{\endgroup\@href {#1}{\urlprefix }}%
\providecommand \urlprefix  [0]{URL }%
\providecommand \Eprint [0]{\href }%
\providecommand \doibase [0]{http://dx.doi.org/}%
\providecommand \selectlanguage [0]{\@gobble}%
\providecommand \bibinfo  [0]{\@secondoftwo}%
\providecommand \bibfield  [0]{\@secondoftwo}%
\providecommand \translation [1]{[#1]}%
\providecommand \BibitemOpen [0]{}%
\providecommand \bibitemStop [0]{}%
\providecommand \bibitemNoStop [0]{.\EOS\space}%
\providecommand \EOS [0]{\spacefactor3000\relax}%
\providecommand \BibitemShut  [1]{\csname bibitem#1\endcsname}%
\let\auto@bib@innerbib\@empty
\bibitem [{\citenamefont {Grumbling}\ and\ \citenamefont
  {Horowitz}(2019)}]{grumbling2019quantum}%
  \BibitemOpen
  \bibinfo {editor} {\bibfnamefont {E.}~\bibnamefont {Grumbling}}\ and\
  \bibinfo {editor} {\bibfnamefont {M.}~\bibnamefont {Horowitz}},\ eds.,\
  \href@noop {} {\emph {\bibinfo {title} {Quantum Computing: Progress and
  Prospects}}}\ (\bibinfo  {publisher} {National Academies Press},\ \bibinfo
  {address} {Washington, {D.C.}},\ \bibinfo {year} {2019})\BibitemShut
  {NoStop}%
\bibitem [{Note1()}]{Note1}%
  \BibitemOpen
  \bibinfo {note} {Full URL: \protect \texttt {https://\protect \allowbreak
  {}docs.\protect \allowbreak {}google.\protect \allowbreak {}com/\protect
  \allowbreak {}spreadsheets/\protect \allowbreak {}d/\protect \allowbreak
  {}1pwb4gf0Fxl\protect \allowbreak {}xgfVhtXTaq\protect \allowbreak
  {}EGS9b7FwsstsJ\protect \allowbreak {}0v7Zb1naQ0/}.}\BibitemShut {Stop}%
\bibitem [{Note2()}]{Note2}%
  \BibitemOpen
  \bibinfo {note} {Full URL: \protect \texttt {https://\protect \allowbreak
  {}colab.\protect \allowbreak {}research.google.\protect \allowbreak
  {}com/\protect \allowbreak {}drive/\protect \allowbreak
  {}1XkWcsUy-CiND\protect \allowbreak {}DJffPC3d\protect \allowbreak
  NJbgvR6iDAqh}.}\BibitemShut {Stop}%
\bibitem [{\citenamefont {Aaronson}(2008)}]{aaronson2008limits}%
  \BibitemOpen
  \bibfield  {author} {\bibinfo {author} {\bibfnamefont {S.}~\bibnamefont
  {Aaronson}},\ }\href@noop {} {\bibfield  {journal} {\bibinfo  {journal}
  {Scientific American}\ }\textbf {\bibinfo {volume} {298}},\ \bibinfo {pages}
  {62} (\bibinfo {year} {2008})}\BibitemShut {NoStop}%
\bibitem [{\citenamefont {Aaronson}(2011)}]{aaronson2011quantum}%
  \BibitemOpen
  \bibfield  {author} {\bibinfo {author} {\bibfnamefont {S.}~\bibnamefont
  {Aaronson}},\ }\href@noop {} {\bibfield  {journal} {\bibinfo  {journal} {The
  New York Times}\ } (\bibinfo {year} {2011})}\BibitemShut {NoStop}%
\bibitem [{\citenamefont {Aaronson}(2013)}]{aaronson2013quantum}%
  \BibitemOpen
  \bibfield  {author} {\bibinfo {author} {\bibfnamefont {S.}~\bibnamefont
  {Aaronson}},\ }\href@noop {} {\emph {\bibinfo {title} {Quantum Computing
  Since Democritus}}}\ (\bibinfo  {publisher} {Cambridge University Press},\
  \bibinfo {year} {2013})\BibitemShut {NoStop}%
\bibitem [{\citenamefont {Perry}\ \emph {et~al.}(2019)\citenamefont {Perry},
  \citenamefont {Sun}, \citenamefont {Hughes}, \citenamefont {Isaacson},\ and\
  \citenamefont {Turner}}]{perry2019quantum}%
  \BibitemOpen
  \bibfield  {author} {\bibinfo {author} {\bibfnamefont {A.}~\bibnamefont
  {Perry}}, \bibinfo {author} {\bibfnamefont {R.}~\bibnamefont {Sun}}, \bibinfo
  {author} {\bibfnamefont {C.}~\bibnamefont {Hughes}}, \bibinfo {author}
  {\bibfnamefont {J.}~\bibnamefont {Isaacson}}, \ and\ \bibinfo {author}
  {\bibfnamefont {J.}~\bibnamefont {Turner}},\ }\href@noop {} {\bibfield
  {journal} {\bibinfo  {journal} {{arXiv:1905.00282}}\ } (\bibinfo {year}
  {2019})}\BibitemShut {NoStop}%
\bibitem [{\citenamefont {Rudolph}(2017)}]{rudolph2017q}%
  \BibitemOpen
  \bibfield  {author} {\bibinfo {author} {\bibfnamefont {T.}~\bibnamefont
  {Rudolph}},\ }\href@noop {} {\emph {\bibinfo {title} {Q is for quantum}}}\
  (\bibinfo  {publisher} {Terence Rudolph},\ \bibinfo {year}
  {2017})\BibitemShut {NoStop}%
\bibitem [{\citenamefont {Kitaev}\ \emph {et~al.}(2002)\citenamefont {Kitaev},
  \citenamefont {Shen}, \citenamefont {Vyalyi},\ and\ \citenamefont
  {Vyalyi}}]{kitaev2002classical}%
  \BibitemOpen
  \bibfield  {author} {\bibinfo {author} {\bibfnamefont {A.~Y.}\ \bibnamefont
  {Kitaev}}, \bibinfo {author} {\bibfnamefont {A.}~\bibnamefont {Shen}},
  \bibinfo {author} {\bibfnamefont {M.~N.}\ \bibnamefont {Vyalyi}}, \ and\
  \bibinfo {author} {\bibfnamefont {M.~N.}\ \bibnamefont {Vyalyi}},\
  }\href@noop {} {\emph {\bibinfo {title} {Classical and Quantum
  Computation}}}\ (\bibinfo  {publisher} {American Mathematical Society},\
  \bibinfo {year} {2002})\BibitemShut {NoStop}%
\bibitem [{\citenamefont {Matuschak}\ and\ \citenamefont
  {Nielsen}(2019)}]{matuschak2019quantum}%
  \BibitemOpen
  \bibfield  {author} {\bibinfo {author} {\bibfnamefont {A.}~\bibnamefont
  {Matuschak}}\ and\ \bibinfo {author} {\bibfnamefont {M.}~\bibnamefont
  {Nielsen}},\ }\href {https://quantum.country} {\enquote {\bibinfo {title}
  {Quantum country},}\ } (\bibinfo {year} {2019})\BibitemShut {NoStop}%
\bibitem [{\citenamefont {Mermin}(2007)}]{mermin2007quantum}%
  \BibitemOpen
  \bibfield  {author} {\bibinfo {author} {\bibfnamefont {N.~D.}\ \bibnamefont
  {Mermin}},\ }\href@noop {} {\emph {\bibinfo {title} {Quantum Computer
  Science: An Introduction}}}\ (\bibinfo  {publisher} {Cambridge University
  Press},\ \bibinfo {year} {2007})\BibitemShut {NoStop}%
\bibitem [{\citenamefont {Nielsen}\ and\ \citenamefont
  {Chuang}(2000)}]{nielsen2000quantum}%
  \BibitemOpen
  \bibfield  {author} {\bibinfo {author} {\bibfnamefont {M.~A.}\ \bibnamefont
  {Nielsen}}\ and\ \bibinfo {author} {\bibfnamefont {I.~L.}\ \bibnamefont
  {Chuang}},\ }\href@noop {} {\emph {\bibinfo {title} {Quantum Computation and
  Quantum Information}}}\ (\bibinfo  {publisher} {Cambridge University Press},\
  \bibinfo {address} {Cambridge ; New York},\ \bibinfo {year}
  {2000})\BibitemShut {NoStop}%
\bibitem [{\citenamefont {Ozols}\ and\ \citenamefont
  {Walter}(2018)}]{ozols2018quantum}%
  \BibitemOpen
  \bibfield  {author} {\bibinfo {author} {\bibfnamefont {M.}~\bibnamefont
  {Ozols}}\ and\ \bibinfo {author} {\bibfnamefont {M.}~\bibnamefont {Walter}},\
  }\href {www.quantum-quest.nl} {\enquote {\bibinfo {title} {The quantum
  quest},}\ } (\bibinfo {year} {2018})\BibitemShut {NoStop}%
\bibitem [{\citenamefont {de~Wolf}(2017)}]{dewolf2017potential}%
  \BibitemOpen
  \bibfield  {author} {\bibinfo {author} {\bibfnamefont {R.}~\bibnamefont
  {de~Wolf}},\ }\href@noop {} {\bibfield  {journal} {\bibinfo  {journal}
  {Ethics and Information Technology}\ ,\ \bibinfo {pages} {1}} (\bibinfo
  {year} {2017})}\BibitemShut {NoStop}%
\bibitem [{\citenamefont {Majot}\ and\ \citenamefont
  {Yampolskiy}(2015)}]{majot2015global}%
  \BibitemOpen
  \bibfield  {author} {\bibinfo {author} {\bibfnamefont {A.}~\bibnamefont
  {Majot}}\ and\ \bibinfo {author} {\bibfnamefont {R.}~\bibnamefont
  {Yampolskiy}},\ }\href@noop {} {\bibfield  {journal} {\bibinfo  {journal}
  {Futures}\ }\bibinfo {series} {Confronting Future Catastrophic Threats To
  Humanity},\ \textbf {\bibinfo {volume} {72}},\ \bibinfo {pages} {17}
  (\bibinfo {year} {2015})}\BibitemShut {NoStop}%
\bibitem [{\citenamefont {Möller}\ and\ \citenamefont
  {Vuik}(2017)}]{moeller2017impact}%
  \BibitemOpen
  \bibfield  {author} {\bibinfo {author} {\bibfnamefont {M.}~\bibnamefont
  {Möller}}\ and\ \bibinfo {author} {\bibfnamefont {C.}~\bibnamefont {Vuik}},\
  }\href@noop {} {\bibfield  {journal} {\bibinfo  {journal}
  {{arXiv:1705.07413}}\ } (\bibinfo {year} {2017})}\BibitemShut {NoStop}%
\bibitem [{Note3()}]{Note3}%
  \BibitemOpen
  \bibinfo {note} {These milestones include (1) experimental quantum annealers,
  (2) small (tens of qubits) computers, (3) gate-based quantum computers that
  demonstrates quantum supremacy, (4) quantum annealers that demonstrate
  quantum supremacy, (5) implementation of QEC for improved qubit quality, (6)
  commercially useful quantum computers and (7) large (\textgreater 1,000
  qubits), fault-tolerant, modular quantum computers.}\BibitemShut {Stop}%
\bibitem [{Note4()}]{Note4}%
  \BibitemOpen
  \bibinfo {note} {``Given the current state of quantum computing and recent
  rates of progress, it is highly unexpected that a quantum computer that can
  compromise RSA 2048 or comparable discrete logarithm-based public key
  cryptosystems will be built within the next decade.'' \protect \citep
  {grumbling2019quantum}.}\BibitemShut {Stop}%
\bibitem [{Note5()}]{Note5}%
  \BibitemOpen
  \bibinfo {note} {As one interesting point, the Academies report highlights
  Rock's law and contextualizes it as a warning that quantum computing may fail
  to achieve exponential progress unless it secures increasing funding:
  ``Moore's law is the result of a virtuous cycle, where improvements in
  integrated circuit manufacturing allow the manufacturer to reduce the price
  of their product, which in turn causes them to sell more products and
  increase their sales and profits. This increased revenue then enables them to
  improve the manufacturing process again, which is harder this time, since the
  easier changes have already been made.''}\BibitemShut {NoStop}%
\bibitem [{\citenamefont {Piani}\ and\ \citenamefont
  {Mosca}(2019)}]{piani2019quantum}%
  \BibitemOpen
  \bibfield  {author} {\bibinfo {author} {\bibfnamefont {M.}~\bibnamefont
  {Piani}}\ and\ \bibinfo {author} {\bibfnamefont {M.}~\bibnamefont {Mosca}},\
  }\href
  {https://globalriskinstitute.org/publications/quantum-threat-timeline/}
  {\emph {\bibinfo {title} {Quantum Threat Timeline Report}}},\ \bibinfo {type}
  {Tech. Rep.}\ (\bibinfo  {institution} {Global Risk Institute},\ \bibinfo
  {year} {2019})\BibitemShut {NoStop}%
\bibitem [{\citenamefont {Farmer}\ and\ \citenamefont
  {Lafond}(2016)}]{farmer2016how}%
  \BibitemOpen
  \bibfield  {author} {\bibinfo {author} {\bibfnamefont {J.~D.}\ \bibnamefont
  {Farmer}}\ and\ \bibinfo {author} {\bibfnamefont {F.}~\bibnamefont
  {Lafond}},\ }\href@noop {} {\bibfield  {journal} {\bibinfo  {journal}
  {Research Policy}\ }\textbf {\bibinfo {volume} {45}},\ \bibinfo {pages}
  {647–665} (\bibinfo {year} {2016})}\BibitemShut {NoStop}%
\bibitem [{\citenamefont {Grace}(2020)}]{grace2020discontinuous}%
  \BibitemOpen
  \bibfield  {author} {\bibinfo {author} {\bibfnamefont {K.}~\bibnamefont
  {Grace}},\ }\href
  {https://aiimpacts.org/discontinuous-progress-in-history-an-update/}
  {\enquote {\bibinfo {title} {Discontinuous progress in history: an update},}\
  } (\bibinfo {year} {2020})\BibitemShut {NoStop}%
\bibitem [{\citenamefont {Kott}(2019)}]{kott2019toward}%
  \BibitemOpen
  \bibfield  {author} {\bibinfo {author} {\bibfnamefont {A.}~\bibnamefont
  {Kott}},\ }\href@noop {} {\bibfield  {journal} {\bibinfo  {journal} {The
  Journal of Defense Modeling and Simulation: Applications, Methodology,
  Technology}\ ,\ \bibinfo {pages} {154851291987552}} (\bibinfo {year}
  {2019})}\BibitemShut {NoStop}%
\bibitem [{Note6()}]{Note6}%
  \BibitemOpen
  \bibinfo {note} {For example, one may expect that papers may only report the
  metrics they were explicitly trying to improve.}\BibitemShut {Stop}%
\bibitem [{Note7()}]{Note7}%
  \BibitemOpen
  \bibinfo {note} {There are important technical distinctions between error
  correction and fault tolerance, but we will not address them in this
  work.}\BibitemShut {Stop}%
\bibitem [{\citenamefont {Das}\ and\ \citenamefont
  {Chakrabarti}(2008)}]{das2008colloquium}%
  \BibitemOpen
  \bibfield  {author} {\bibinfo {author} {\bibfnamefont {A.}~\bibnamefont
  {Das}}\ and\ \bibinfo {author} {\bibfnamefont {B.~K.}\ \bibnamefont
  {Chakrabarti}},\ }\href@noop {} {\bibfield  {journal} {\bibinfo  {journal}
  {Reviews of Modern Physics}\ }\textbf {\bibinfo {volume} {80}},\ \bibinfo
  {pages} {1061} (\bibinfo {year} {2008})}\BibitemShut {NoStop}%
\bibitem [{\citenamefont {Yamamoto}\ \emph {et~al.}(2017)\citenamefont
  {Yamamoto}, \citenamefont {Aihara}, \citenamefont {Leleu}, \citenamefont
  {Kawarabayashi}, \citenamefont {Kako}, \citenamefont {Fejer}, \citenamefont
  {Inoue},\ and\ \citenamefont {Takesue}}]{yamamoto2017coherent}%
  \BibitemOpen
  \bibfield  {author} {\bibinfo {author} {\bibfnamefont {Y.}~\bibnamefont
  {Yamamoto}}, \bibinfo {author} {\bibfnamefont {K.}~\bibnamefont {Aihara}},
  \bibinfo {author} {\bibfnamefont {T.}~\bibnamefont {Leleu}}, \bibinfo
  {author} {\bibfnamefont {K.-i.}\ \bibnamefont {Kawarabayashi}}, \bibinfo
  {author} {\bibfnamefont {S.}~\bibnamefont {Kako}}, \bibinfo {author}
  {\bibfnamefont {M.}~\bibnamefont {Fejer}}, \bibinfo {author} {\bibfnamefont
  {K.}~\bibnamefont {Inoue}}, \ and\ \bibinfo {author} {\bibfnamefont
  {H.}~\bibnamefont {Takesue}},\ }\href@noop {} {\bibfield  {journal} {\bibinfo
   {journal} {npj Quantum Information}\ }\textbf {\bibinfo {volume} {3}},\
  \bibinfo {pages} {1–15} (\bibinfo {year} {2017})}\BibitemShut {NoStop}%
\bibitem [{Note8()}]{Note8}%
  \BibitemOpen
  \bibinfo {note} {Unfortunately, terminology is not completely standardized.
  Some authors take ``quantum supremacy'' and ``quantum advantage'' to be
  synonyms, while others distinguish them based on whether the problem could be
  solved by \protect \emph {existing} classical (super)computers vs. \protect
  \emph {any} feasible classical computer, or based on the economic value of
  the problem.}\BibitemShut {Stop}%
\bibitem [{Note9()}]{Note9}%
  \BibitemOpen
  \bibinfo {note} {It is worth emphasizing that the mathematical problem
  selected for demonstrating quantum supremacy is invariably contrived to be as
  easy as possible for the quantum devices and as hard as possible for the
  classical devices, i.e., the ``quantum home-field advantage'' has been
  maximized. In particular, quantum supremacy does not mean the quantum
  computer is better at all computational problems, or even a large class of
  them.}\BibitemShut {Stop}%
\bibitem [{\citenamefont {Karimi}\ \emph {et~al.}(2011)\citenamefont {Karimi},
  \citenamefont {Dickson}, \citenamefont {Hamze}, \citenamefont {Amin},
  \citenamefont {{Drew-Brook}}, \citenamefont {Chudak}, \citenamefont {Bunyk},
  \citenamefont {Macready},\ and\ \citenamefont
  {Rose}}]{karimi2011investigating}%
  \BibitemOpen
  \bibfield  {author} {\bibinfo {author} {\bibfnamefont {K.}~\bibnamefont
  {Karimi}}, \bibinfo {author} {\bibfnamefont {N.~G.}\ \bibnamefont {Dickson}},
  \bibinfo {author} {\bibfnamefont {F.}~\bibnamefont {Hamze}}, \bibinfo
  {author} {\bibfnamefont {M.~H.~S.}\ \bibnamefont {Amin}}, \bibinfo {author}
  {\bibfnamefont {M.}~\bibnamefont {{Drew-Brook}}}, \bibinfo {author}
  {\bibfnamefont {F.~A.}\ \bibnamefont {Chudak}}, \bibinfo {author}
  {\bibfnamefont {P.~I.}\ \bibnamefont {Bunyk}}, \bibinfo {author}
  {\bibfnamefont {W.~G.}\ \bibnamefont {Macready}}, \ and\ \bibinfo {author}
  {\bibfnamefont {G.}~\bibnamefont {Rose}},\ }\href@noop {} {\bibfield
  {journal} {\bibinfo  {journal} {{arXiv:1006.4147}}\ } (\bibinfo {year}
  {2011})}\BibitemShut {NoStop}%
\bibitem [{\citenamefont {Kelly}\ \emph {et~al.}(2015)\citenamefont {Kelly},
  \citenamefont {Barends}, \citenamefont {Fowler}, \citenamefont {Megrant},
  \citenamefont {Jeffrey}, \citenamefont {White}, \citenamefont {Sank},
  \citenamefont {Mutus}, \citenamefont {Campbell}, \citenamefont {Chen},
  \citenamefont {Chen}, \citenamefont {Chiaro}, \citenamefont {Dunsworth},
  \citenamefont {Hoi}, \citenamefont {Neill}, \citenamefont {{O’Malley}},
  \citenamefont {Quintana}, \citenamefont {Roushan}, \citenamefont
  {Vainsencher}, \citenamefont {Wenner}, \citenamefont {Cleland},\ and\
  \citenamefont {Martinis}}]{kelly2015state}%
  \BibitemOpen
  \bibfield  {author} {\bibinfo {author} {\bibfnamefont {J.}~\bibnamefont
  {Kelly}}, \bibinfo {author} {\bibfnamefont {R.}~\bibnamefont {Barends}},
  \bibinfo {author} {\bibfnamefont {A.~G.}\ \bibnamefont {Fowler}}, \bibinfo
  {author} {\bibfnamefont {A.}~\bibnamefont {Megrant}}, \bibinfo {author}
  {\bibfnamefont {E.}~\bibnamefont {Jeffrey}}, \bibinfo {author} {\bibfnamefont
  {T.~C.}\ \bibnamefont {White}}, \bibinfo {author} {\bibfnamefont
  {D.}~\bibnamefont {Sank}}, \bibinfo {author} {\bibfnamefont {J.~Y.}\
  \bibnamefont {Mutus}}, \bibinfo {author} {\bibfnamefont {B.}~\bibnamefont
  {Campbell}}, \bibinfo {author} {\bibfnamefont {Y.}~\bibnamefont {Chen}},
  \bibinfo {author} {\bibfnamefont {Z.}~\bibnamefont {Chen}}, \bibinfo {author}
  {\bibfnamefont {B.}~\bibnamefont {Chiaro}}, \bibinfo {author} {\bibfnamefont
  {A.}~\bibnamefont {Dunsworth}}, \bibinfo {author} {\bibfnamefont
  {I.}~\bibnamefont {Hoi}}, \bibinfo {author} {\bibfnamefont {C.}~\bibnamefont
  {Neill}}, \bibinfo {author} {\bibfnamefont {P.~J.~J.}\ \bibnamefont
  {{O’Malley}}}, \bibinfo {author} {\bibfnamefont {C.}~\bibnamefont
  {Quintana}}, \bibinfo {author} {\bibfnamefont {P.}~\bibnamefont {Roushan}},
  \bibinfo {author} {\bibfnamefont {A.}~\bibnamefont {Vainsencher}}, \bibinfo
  {author} {\bibfnamefont {J.}~\bibnamefont {Wenner}}, \bibinfo {author}
  {\bibfnamefont {A.~N.}\ \bibnamefont {Cleland}}, \ and\ \bibinfo {author}
  {\bibfnamefont {J.~M.}\ \bibnamefont {Martinis}},\ }\href@noop {} {\bibfield
  {journal} {\bibinfo  {journal} {Nature}\ }\textbf {\bibinfo {volume} {519}},\
  \bibinfo {pages} {66} (\bibinfo {year} {2015})}\BibitemShut {NoStop}%
\bibitem [{\citenamefont {Arute}\ \emph {et~al.}(2019)\citenamefont {Arute},
  \citenamefont {Arya}, \citenamefont {Babbush}, \citenamefont {Bacon},
  \citenamefont {Bardin}, \citenamefont {Barends}, \citenamefont {Biswas},
  \citenamefont {Boixo}, \citenamefont {Brandao}, \citenamefont {Buell},
  \citenamefont {Burkett}, \citenamefont {Chen}, \citenamefont {Chen},
  \citenamefont {Chiaro}, \citenamefont {Collins}, \citenamefont {Courtney},
  \citenamefont {Dunsworth}, \citenamefont {Farhi}, \citenamefont {Foxen},
  \citenamefont {Fowler}, \citenamefont {Gidney}, \citenamefont {Giustina},
  \citenamefont {Graff}, \citenamefont {Guerin}, \citenamefont {Habegger},
  \citenamefont {Harrigan}, \citenamefont {Hartmann}, \citenamefont {Ho},
  \citenamefont {Hoffmann}, \citenamefont {Huang}, \citenamefont {Humble},
  \citenamefont {Isakov}, \citenamefont {Jeffrey}, \citenamefont {Jiang},
  \citenamefont {Kafri}, \citenamefont {Kechedzhi}, \citenamefont {Kelly},
  \citenamefont {Klimov}, \citenamefont {Knysh}, \citenamefont {Korotkov},
  \citenamefont {Kostritsa}, \citenamefont {Landhuis}, \citenamefont
  {Lindmark}, \citenamefont {Lucero}, \citenamefont {Lyakh}, \citenamefont
  {Mandrà}, \citenamefont {{McClean}}, \citenamefont {{McEwen}}, \citenamefont
  {Megrant}, \citenamefont {Mi}, \citenamefont {Michielsen}, \citenamefont
  {Mohseni}, \citenamefont {Mutus}, \citenamefont {Naaman}, \citenamefont
  {Neeley}, \citenamefont {Neill}, \citenamefont {Niu}, \citenamefont {Ostby},
  \citenamefont {Petukhov}, \citenamefont {Platt}, \citenamefont {Quintana},
  \citenamefont {Rieffel}, \citenamefont {Roushan}, \citenamefont {Rubin},
  \citenamefont {Sank}, \citenamefont {Satzinger}, \citenamefont {Smelyanskiy},
  \citenamefont {Sung}, \citenamefont {Trevithick}, \citenamefont
  {Vainsencher}, \citenamefont {Villalonga}, \citenamefont {White},
  \citenamefont {Yao}, \citenamefont {Yeh}, \citenamefont {Zalcman},
  \citenamefont {Neven},\ and\ \citenamefont {Martinis}}]{arute2019quantum}%
  \BibitemOpen
  \bibfield  {author} {\bibinfo {author} {\bibfnamefont {F.}~\bibnamefont
  {Arute}}, \bibinfo {author} {\bibfnamefont {K.}~\bibnamefont {Arya}},
  \bibinfo {author} {\bibfnamefont {R.}~\bibnamefont {Babbush}}, \bibinfo
  {author} {\bibfnamefont {D.}~\bibnamefont {Bacon}}, \bibinfo {author}
  {\bibfnamefont {J.~C.}\ \bibnamefont {Bardin}}, \bibinfo {author}
  {\bibfnamefont {R.}~\bibnamefont {Barends}}, \bibinfo {author} {\bibfnamefont
  {R.}~\bibnamefont {Biswas}}, \bibinfo {author} {\bibfnamefont
  {S.}~\bibnamefont {Boixo}}, \bibinfo {author} {\bibfnamefont {F.~G. S.~L.}\
  \bibnamefont {Brandao}}, \bibinfo {author} {\bibfnamefont {D.~A.}\
  \bibnamefont {Buell}}, \bibinfo {author} {\bibfnamefont {B.}~\bibnamefont
  {Burkett}}, \bibinfo {author} {\bibfnamefont {Y.}~\bibnamefont {Chen}},
  \bibinfo {author} {\bibfnamefont {Z.}~\bibnamefont {Chen}}, \bibinfo {author}
  {\bibfnamefont {B.}~\bibnamefont {Chiaro}}, \bibinfo {author} {\bibfnamefont
  {R.}~\bibnamefont {Collins}}, \bibinfo {author} {\bibfnamefont
  {W.}~\bibnamefont {Courtney}}, \bibinfo {author} {\bibfnamefont
  {A.}~\bibnamefont {Dunsworth}}, \bibinfo {author} {\bibfnamefont
  {E.}~\bibnamefont {Farhi}}, \bibinfo {author} {\bibfnamefont
  {B.}~\bibnamefont {Foxen}}, \bibinfo {author} {\bibfnamefont
  {A.}~\bibnamefont {Fowler}}, \bibinfo {author} {\bibfnamefont
  {C.}~\bibnamefont {Gidney}}, \bibinfo {author} {\bibfnamefont
  {M.}~\bibnamefont {Giustina}}, \bibinfo {author} {\bibfnamefont
  {R.}~\bibnamefont {Graff}}, \bibinfo {author} {\bibfnamefont
  {K.}~\bibnamefont {Guerin}}, \bibinfo {author} {\bibfnamefont
  {S.}~\bibnamefont {Habegger}}, \bibinfo {author} {\bibfnamefont {M.~P.}\
  \bibnamefont {Harrigan}}, \bibinfo {author} {\bibfnamefont {M.~J.}\
  \bibnamefont {Hartmann}}, \bibinfo {author} {\bibfnamefont {A.}~\bibnamefont
  {Ho}}, \bibinfo {author} {\bibfnamefont {M.}~\bibnamefont {Hoffmann}},
  \bibinfo {author} {\bibfnamefont {T.}~\bibnamefont {Huang}}, \bibinfo
  {author} {\bibfnamefont {T.~S.}\ \bibnamefont {Humble}}, \bibinfo {author}
  {\bibfnamefont {S.~V.}\ \bibnamefont {Isakov}}, \bibinfo {author}
  {\bibfnamefont {E.}~\bibnamefont {Jeffrey}}, \bibinfo {author} {\bibfnamefont
  {Z.}~\bibnamefont {Jiang}}, \bibinfo {author} {\bibfnamefont
  {D.}~\bibnamefont {Kafri}}, \bibinfo {author} {\bibfnamefont
  {K.}~\bibnamefont {Kechedzhi}}, \bibinfo {author} {\bibfnamefont
  {J.}~\bibnamefont {Kelly}}, \bibinfo {author} {\bibfnamefont {P.~V.}\
  \bibnamefont {Klimov}}, \bibinfo {author} {\bibfnamefont {S.}~\bibnamefont
  {Knysh}}, \bibinfo {author} {\bibfnamefont {A.}~\bibnamefont {Korotkov}},
  \bibinfo {author} {\bibfnamefont {F.}~\bibnamefont {Kostritsa}}, \bibinfo
  {author} {\bibfnamefont {D.}~\bibnamefont {Landhuis}}, \bibinfo {author}
  {\bibfnamefont {M.}~\bibnamefont {Lindmark}}, \bibinfo {author}
  {\bibfnamefont {E.}~\bibnamefont {Lucero}}, \bibinfo {author} {\bibfnamefont
  {D.}~\bibnamefont {Lyakh}}, \bibinfo {author} {\bibfnamefont
  {S.}~\bibnamefont {Mandrà}}, \bibinfo {author} {\bibfnamefont {J.~R.}\
  \bibnamefont {{McClean}}}, \bibinfo {author} {\bibfnamefont {M.}~\bibnamefont
  {{McEwen}}}, \bibinfo {author} {\bibfnamefont {A.}~\bibnamefont {Megrant}},
  \bibinfo {author} {\bibfnamefont {X.}~\bibnamefont {Mi}}, \bibinfo {author}
  {\bibfnamefont {K.}~\bibnamefont {Michielsen}}, \bibinfo {author}
  {\bibfnamefont {M.}~\bibnamefont {Mohseni}}, \bibinfo {author} {\bibfnamefont
  {J.}~\bibnamefont {Mutus}}, \bibinfo {author} {\bibfnamefont
  {O.}~\bibnamefont {Naaman}}, \bibinfo {author} {\bibfnamefont
  {M.}~\bibnamefont {Neeley}}, \bibinfo {author} {\bibfnamefont
  {C.}~\bibnamefont {Neill}}, \bibinfo {author} {\bibfnamefont {M.~Y.}\
  \bibnamefont {Niu}}, \bibinfo {author} {\bibfnamefont {E.}~\bibnamefont
  {Ostby}}, \bibinfo {author} {\bibfnamefont {A.}~\bibnamefont {Petukhov}},
  \bibinfo {author} {\bibfnamefont {J.~C.}\ \bibnamefont {Platt}}, \bibinfo
  {author} {\bibfnamefont {C.}~\bibnamefont {Quintana}}, \bibinfo {author}
  {\bibfnamefont {E.~G.}\ \bibnamefont {Rieffel}}, \bibinfo {author}
  {\bibfnamefont {P.}~\bibnamefont {Roushan}}, \bibinfo {author} {\bibfnamefont
  {N.~C.}\ \bibnamefont {Rubin}}, \bibinfo {author} {\bibfnamefont
  {D.}~\bibnamefont {Sank}}, \bibinfo {author} {\bibfnamefont {K.~J.}\
  \bibnamefont {Satzinger}}, \bibinfo {author} {\bibfnamefont {V.}~\bibnamefont
  {Smelyanskiy}}, \bibinfo {author} {\bibfnamefont {K.~J.}\ \bibnamefont
  {Sung}}, \bibinfo {author} {\bibfnamefont {M.~D.}\ \bibnamefont
  {Trevithick}}, \bibinfo {author} {\bibfnamefont {A.}~\bibnamefont
  {Vainsencher}}, \bibinfo {author} {\bibfnamefont {B.}~\bibnamefont
  {Villalonga}}, \bibinfo {author} {\bibfnamefont {T.}~\bibnamefont {White}},
  \bibinfo {author} {\bibfnamefont {Z.~J.}\ \bibnamefont {Yao}}, \bibinfo
  {author} {\bibfnamefont {P.}~\bibnamefont {Yeh}}, \bibinfo {author}
  {\bibfnamefont {A.}~\bibnamefont {Zalcman}}, \bibinfo {author} {\bibfnamefont
  {H.}~\bibnamefont {Neven}}, \ and\ \bibinfo {author} {\bibfnamefont {J.~M.}\
  \bibnamefont {Martinis}},\ }\href@noop {} {\bibfield  {journal} {\bibinfo
  {journal} {Nature}\ }\textbf {\bibinfo {volume} {574}},\ \bibinfo {pages}
  {505} (\bibinfo {year} {2019})}\BibitemShut {NoStop}%
\bibitem [{\citenamefont {Pednault}\ \emph {et~al.}(2019)\citenamefont
  {Pednault}, \citenamefont {Gunnels},\ and\ \citenamefont
  {Gambetta}}]{pednault2019quantum}%
  \BibitemOpen
  \bibfield  {author} {\bibinfo {author} {\bibfnamefont {E.}~\bibnamefont
  {Pednault}}, \bibinfo {author} {\bibfnamefont {J.~A.}\ \bibnamefont
  {Gunnels}}, \ and\ \bibinfo {author} {\bibfnamefont {J.~M.}\ \bibnamefont
  {Gambetta}},\ }\href
  {https://www.ibm.com/blogs/research/2019/10/on-quantum-supremacy/} {\enquote
  {\bibinfo {title} {On {“Quantum} supremacy”},}\ } (\bibinfo {year}
  {2019})\BibitemShut {NoStop}%
\bibitem [{Note10()}]{Note10}%
  \BibitemOpen
  \bibinfo {note} {This is quantified by the average gate error rate, that we
  will explain in Section~\ref {metrics}. Google's average error rate is of the
  order of 1e3 \protect \citep {arute2019quantum}.}\BibitemShut {Stop}%
\bibitem [{\citenamefont {Preskill}(1998)}]{preskill1998quantum}%
  \BibitemOpen
  \bibfield  {author} {\bibinfo {author} {\bibfnamefont {J.}~\bibnamefont
  {Preskill}},\ }\href@noop {} {\bibfield  {journal} {\bibinfo  {journal}
  {Proceedings of the Royal Society of London A: Mathematical, Physical and
  Engineering Sciences}\ }\textbf {\bibinfo {volume} {454}},\ \bibinfo {pages}
  {469} (\bibinfo {year} {1998})}\BibitemShut {NoStop}%
\bibitem [{Note11()}]{Note11}%
  \BibitemOpen
  \bibinfo {note} {For a popular-level description, see \protect \citep
  {aaronson2017quantum}.}\BibitemShut {Stop}%
\bibitem [{\citenamefont {Acín}\ and\ \citenamefont
  {Masanes}(2016)}]{acin2016certified}%
  \BibitemOpen
  \bibfield  {author} {\bibinfo {author} {\bibfnamefont {A.}~\bibnamefont
  {Acín}}\ and\ \bibinfo {author} {\bibfnamefont {L.}~\bibnamefont
  {Masanes}},\ }\href@noop {} {\bibfield  {journal} {\bibinfo  {journal}
  {Nature}\ }\textbf {\bibinfo {volume} {540}},\ \bibinfo {pages} {213}
  (\bibinfo {year} {2016})}\BibitemShut {NoStop}%
\bibitem [{\citenamefont {King}\ \emph {et~al.}(2018)\citenamefont {King},
  \citenamefont {Carrasquilla}, \citenamefont {Ozfidan}, \citenamefont
  {Raymond}, \citenamefont {Andriyash}, \citenamefont {Berkley}, \citenamefont
  {Reis}, \citenamefont {Lanting}, \citenamefont {Harris}, \citenamefont
  {{Poulin-Lamarre}}, \citenamefont {Smirnov}, \citenamefont {Rich},
  \citenamefont {Altomare}, \citenamefont {Bunyk}, \citenamefont {Whittaker},
  \citenamefont {Swenson}, \citenamefont {Hoskinson}, \citenamefont {Sato},
  \citenamefont {Volkmann}, \citenamefont {Ladizinsky}, \citenamefont
  {Johnson}, \citenamefont {Hilton},\ and\ \citenamefont
  {Amin}}]{king2018observation}%
  \BibitemOpen
  \bibfield  {author} {\bibinfo {author} {\bibfnamefont {A.~D.}\ \bibnamefont
  {King}}, \bibinfo {author} {\bibfnamefont {J.}~\bibnamefont {Carrasquilla}},
  \bibinfo {author} {\bibfnamefont {I.}~\bibnamefont {Ozfidan}}, \bibinfo
  {author} {\bibfnamefont {J.}~\bibnamefont {Raymond}}, \bibinfo {author}
  {\bibfnamefont {E.}~\bibnamefont {Andriyash}}, \bibinfo {author}
  {\bibfnamefont {A.}~\bibnamefont {Berkley}}, \bibinfo {author} {\bibfnamefont
  {M.}~\bibnamefont {Reis}}, \bibinfo {author} {\bibfnamefont {T.~M.}\
  \bibnamefont {Lanting}}, \bibinfo {author} {\bibfnamefont {R.}~\bibnamefont
  {Harris}}, \bibinfo {author} {\bibfnamefont {G.}~\bibnamefont
  {{Poulin-Lamarre}}}, \bibinfo {author} {\bibfnamefont {A.~Y.}\ \bibnamefont
  {Smirnov}}, \bibinfo {author} {\bibfnamefont {C.}~\bibnamefont {Rich}},
  \bibinfo {author} {\bibfnamefont {F.}~\bibnamefont {Altomare}}, \bibinfo
  {author} {\bibfnamefont {P.}~\bibnamefont {Bunyk}}, \bibinfo {author}
  {\bibfnamefont {J.}~\bibnamefont {Whittaker}}, \bibinfo {author}
  {\bibfnamefont {L.}~\bibnamefont {Swenson}}, \bibinfo {author} {\bibfnamefont
  {E.}~\bibnamefont {Hoskinson}}, \bibinfo {author} {\bibfnamefont
  {Y.}~\bibnamefont {Sato}}, \bibinfo {author} {\bibfnamefont {M.}~\bibnamefont
  {Volkmann}}, \bibinfo {author} {\bibfnamefont {E.}~\bibnamefont
  {Ladizinsky}}, \bibinfo {author} {\bibfnamefont {M.}~\bibnamefont {Johnson}},
  \bibinfo {author} {\bibfnamefont {J.}~\bibnamefont {Hilton}}, \ and\ \bibinfo
  {author} {\bibfnamefont {M.~H.}\ \bibnamefont {Amin}},\ }\href@noop {}
  {\bibfield  {journal} {\bibinfo  {journal} {Nature}\ }\textbf {\bibinfo
  {volume} {560}},\ \bibinfo {pages} {456} (\bibinfo {year}
  {2018})}\BibitemShut {NoStop}%
\bibitem [{Note12()}]{Note12}%
  \BibitemOpen
  \bibinfo {note} {Here we use ``logic gate'' simply to distinguish this
  machine operation from other sorts of gates, like the ones that keep in
  horses. We are still referring to a gate operation on physical qubits. (In
  this paper, we will not need to discuss logical gates, which are a
  counterpart to logical qubits.)}\BibitemShut {NoStop}%
\bibitem [{Note13()}]{Note13}%
  \BibitemOpen
  \bibinfo {note} {We emphasize that qubit connectivity is essentially for
  fault tolerance and useful computations. (Trivially, $N$ separate and
  non-interacting quantum devices, each consisting of $2$ physical qubits
  interacting through a gate with error rate $r$, could be construed as a
  quantum ``computer'' with arbitrarily high physical qubits ($2N$) and
  arbitrarily low error rate ($r$).) The reasonableness of our model is based
  on the premise that, in practice, connectivity will to some extent track
  qubit counts and error rates within the field overall.}\BibitemShut {Stop}%
\bibitem [{\citenamefont {Cross}\ \emph {et~al.}(2019)\citenamefont {Cross},
  \citenamefont {Bishop}, \citenamefont {Sheldon}, \citenamefont {Nation},\
  and\ \citenamefont {Gambetta}}]{cross2019validating}%
  \BibitemOpen
  \bibfield  {author} {\bibinfo {author} {\bibfnamefont {A.~W.}\ \bibnamefont
  {Cross}}, \bibinfo {author} {\bibfnamefont {L.~S.}\ \bibnamefont {Bishop}},
  \bibinfo {author} {\bibfnamefont {S.}~\bibnamefont {Sheldon}}, \bibinfo
  {author} {\bibfnamefont {P.~D.}\ \bibnamefont {Nation}}, \ and\ \bibinfo
  {author} {\bibfnamefont {J.~M.}\ \bibnamefont {Gambetta}},\ }\href@noop {}
  {\bibfield  {journal} {\bibinfo  {journal} {Physical Review A}\ }\textbf
  {\bibinfo {volume} {100}},\ \bibinfo {pages} {032328} (\bibinfo {year}
  {2019})}\BibitemShut {NoStop}%
\bibitem [{\citenamefont {Barends}\ \emph {et~al.}(2014)\citenamefont
  {Barends}, \citenamefont {Kelly}, \citenamefont {Megrant}, \citenamefont
  {Veitia}, \citenamefont {Sank}, \citenamefont {Jeffrey}, \citenamefont
  {White}, \citenamefont {Mutus}, \citenamefont {Fowler}, \citenamefont
  {Campbell}, \citenamefont {Chen}, \citenamefont {Chen}, \citenamefont
  {Chiaro}, \citenamefont {Dunsworth}, \citenamefont {Neill}, \citenamefont
  {{O’Malley}}, \citenamefont {Roushan}, \citenamefont {Vainsencher},
  \citenamefont {Wenner}, \citenamefont {Korotkov}, \citenamefont {Cleland},\
  and\ \citenamefont {Martinis}}]{barends2014superconducting}%
  \BibitemOpen
  \bibfield  {author} {\bibinfo {author} {\bibfnamefont {R.}~\bibnamefont
  {Barends}}, \bibinfo {author} {\bibfnamefont {J.}~\bibnamefont {Kelly}},
  \bibinfo {author} {\bibfnamefont {A.}~\bibnamefont {Megrant}}, \bibinfo
  {author} {\bibfnamefont {A.}~\bibnamefont {Veitia}}, \bibinfo {author}
  {\bibfnamefont {D.}~\bibnamefont {Sank}}, \bibinfo {author} {\bibfnamefont
  {E.}~\bibnamefont {Jeffrey}}, \bibinfo {author} {\bibfnamefont {T.~C.}\
  \bibnamefont {White}}, \bibinfo {author} {\bibfnamefont {J.}~\bibnamefont
  {Mutus}}, \bibinfo {author} {\bibfnamefont {A.~G.}\ \bibnamefont {Fowler}},
  \bibinfo {author} {\bibfnamefont {B.}~\bibnamefont {Campbell}}, \bibinfo
  {author} {\bibfnamefont {Y.}~\bibnamefont {Chen}}, \bibinfo {author}
  {\bibfnamefont {Z.}~\bibnamefont {Chen}}, \bibinfo {author} {\bibfnamefont
  {B.}~\bibnamefont {Chiaro}}, \bibinfo {author} {\bibfnamefont
  {A.}~\bibnamefont {Dunsworth}}, \bibinfo {author} {\bibfnamefont
  {C.}~\bibnamefont {Neill}}, \bibinfo {author} {\bibfnamefont
  {P.}~\bibnamefont {{O’Malley}}}, \bibinfo {author} {\bibfnamefont
  {P.}~\bibnamefont {Roushan}}, \bibinfo {author} {\bibfnamefont
  {A.}~\bibnamefont {Vainsencher}}, \bibinfo {author} {\bibfnamefont
  {J.}~\bibnamefont {Wenner}}, \bibinfo {author} {\bibfnamefont {A.~N.}\
  \bibnamefont {Korotkov}}, \bibinfo {author} {\bibfnamefont {A.~N.}\
  \bibnamefont {Cleland}}, \ and\ \bibinfo {author} {\bibfnamefont {J.~M.}\
  \bibnamefont {Martinis}},\ }\href@noop {} {\bibfield  {journal} {\bibinfo
  {journal} {Nature}\ }\textbf {\bibinfo {volume} {508}},\ \bibinfo {pages}
  {500} (\bibinfo {year} {2014})}\BibitemShut {NoStop}%
\bibitem [{\citenamefont {Fowler}\ \emph {et~al.}(2011)\citenamefont {Fowler},
  \citenamefont {Wang},\ and\ \citenamefont {Hollenberg}}]{fowler2011surface}%
  \BibitemOpen
  \bibfield  {author} {\bibinfo {author} {\bibfnamefont {A.~G.}\ \bibnamefont
  {Fowler}}, \bibinfo {author} {\bibfnamefont {D.~S.}\ \bibnamefont {Wang}}, \
  and\ \bibinfo {author} {\bibfnamefont {L.~C.~L.}\ \bibnamefont
  {Hollenberg}},\ }\href@noop {} {\bibfield  {journal} {\bibinfo  {journal}
  {Quantum Information \& Computation}\ }\textbf {\bibinfo {volume} {11}},\
  \bibinfo {pages} {8–18} (\bibinfo {year} {2011})}\BibitemShut {NoStop}%
\bibitem [{\citenamefont {Fowler}\ \emph {et~al.}(2012)\citenamefont {Fowler},
  \citenamefont {Mariantoni}, \citenamefont {Martinis},\ and\ \citenamefont
  {Cleland}}]{fowler2012surface}%
  \BibitemOpen
  \bibfield  {author} {\bibinfo {author} {\bibfnamefont {A.~G.}\ \bibnamefont
  {Fowler}}, \bibinfo {author} {\bibfnamefont {M.}~\bibnamefont {Mariantoni}},
  \bibinfo {author} {\bibfnamefont {J.~M.}\ \bibnamefont {Martinis}}, \ and\
  \bibinfo {author} {\bibfnamefont {A.~N.}\ \bibnamefont {Cleland}},\
  }\href@noop {} {\bibfield  {journal} {\bibinfo  {journal} {Physical Review
  A}\ }\textbf {\bibinfo {volume} {86}} (\bibinfo {year} {2012})}\BibitemShut
  {NoStop}%
\bibitem [{\citenamefont {{Javadi-Abhari}}(2017)}]{javadi-abhari2017towards}%
  \BibitemOpen
  \bibfield  {author} {\bibinfo {author} {\bibfnamefont {A.}~\bibnamefont
  {{Javadi-Abhari}}},\ }\emph {\bibinfo {title} {Towards A Scalable Software
  Stack For Resource Estimation And Optimization In {General-Purpose} Quantum
  Computers}},\ \href@noop {} {Ph.D. thesis},\ \bibinfo  {school} {Princeton
  University} (\bibinfo {year} {2017})\BibitemShut {NoStop}%
\bibitem [{Note14()}]{Note14}%
  \BibitemOpen
  \bibinfo {note} {``Without QEC, it is unlikely that a complex quantum
  program, such as one that implements Shor's algorithm, would ever run
  correctly on a quantum computer.'' \protect \citep
  {grumbling2019quantum}.}\BibitemShut {Stop}%
\bibitem [{\citenamefont {Gidney}\ and\ \citenamefont
  {Ekerå}(2019)}]{gidney2019how}%
  \BibitemOpen
  \bibfield  {author} {\bibinfo {author} {\bibfnamefont {C.}~\bibnamefont
  {Gidney}}\ and\ \bibinfo {author} {\bibfnamefont {M.}~\bibnamefont
  {Ekerå}},\ }\href@noop {} {\bibfield  {journal} {\bibinfo  {journal}
  {{arXiv:1905.09749}}\ } (\bibinfo {year} {2019})}\BibitemShut {NoStop}%
\bibitem [{\citenamefont {Häner}\ \emph {et~al.}(2016)\citenamefont {Häner},
  \citenamefont {Roetteler},\ and\ \citenamefont
  {Svore}}]{haener2016factoring}%
  \BibitemOpen
  \bibfield  {author} {\bibinfo {author} {\bibfnamefont {T.}~\bibnamefont
  {Häner}}, \bibinfo {author} {\bibfnamefont {M.}~\bibnamefont {Roetteler}}, \
  and\ \bibinfo {author} {\bibfnamefont {K.~M.}\ \bibnamefont {Svore}},\
  }\href@noop {} {\bibfield  {journal} {\bibinfo  {journal}
  {{arXiv:1611.07995}}\ } (\bibinfo {year} {2016})}\BibitemShut {NoStop}%
\bibitem [{\citenamefont {Roetteler}\ \emph {et~al.}(2017)\citenamefont
  {Roetteler}, \citenamefont {Naehrig}, \citenamefont {Svore},\ and\
  \citenamefont {Lauter}}]{roetteler2017quantum}%
  \BibitemOpen
  \bibfield  {author} {\bibinfo {author} {\bibfnamefont {M.}~\bibnamefont
  {Roetteler}}, \bibinfo {author} {\bibfnamefont {M.}~\bibnamefont {Naehrig}},
  \bibinfo {author} {\bibfnamefont {K.~M.}\ \bibnamefont {Svore}}, \ and\
  \bibinfo {author} {\bibfnamefont {K.}~\bibnamefont {Lauter}},\ }\href@noop {}
  {\bibfield  {journal} {\bibinfo  {journal} {{arXiv:1706.06752}}\ } (\bibinfo
  {year} {2017})}\BibitemShut {NoStop}%
\bibitem [{Note15()}]{Note15}%
  \BibitemOpen
  \bibinfo {note} {Häner et al. describe a procedure for using $2(N+2)$ qubits
  to factor an $N$-bit number \protect \citep {haener2016factoring}. Recently,
  Gidney \& Ekerå have described a procedure that uses $ \sim $50$\%$ more
  qubits, $3N + 0.002 N \protect \qopname \relax o{log}_2 N$, but runs two
  orders of magnitude faster for these size numbers \protect \citep
  {gidney2019how}.}\BibitemShut {Stop}%
\bibitem [{\citenamefont {Finke}(2020{\natexlab{a}})}]{finke2020scorecards}%
  \BibitemOpen
  \bibfield  {author} {\bibinfo {author} {\bibfnamefont {D.}~\bibnamefont
  {Finke}},\ }\href {https://quantumcomputingreport.com/scorecards/} {\enquote
  {\bibinfo {title} {Scorecards},}\ } (\bibinfo {year}
  {2020}{\natexlab{a}})\BibitemShut {NoStop}%
\bibitem [{Note16()}]{Note16}%
  \BibitemOpen
  \bibinfo {note} {Full URL: \protect \texttt {https://\protect \allowbreak
  {}docs.\protect \allowbreak {}google.\protect \allowbreak {}com/\protect
  \allowbreak {}spreadsheets/\protect \allowbreak {}d/\protect \allowbreak
  {}1pwb4gf0\protect \allowbreak {}FxlxgfV\protect \allowbreak {}htXTaq\protect
  \allowbreak {}EGS9b7\protect \allowbreak {}FwsstsJ0v7Z\protect \allowbreak
  {}b1naQ0/\protect \allowbreak {}edit\#\protect \allowbreak
  {}gid=0}.}\BibitemShut {Stop}%
\bibitem [{\citenamefont {Intel}(2018)}]{intel2018}%
  \BibitemOpen
  \bibfield  {author} {\bibinfo {author} {\bibnamefont {Intel}},\ }\href@noop
  {} {\enquote {\bibinfo {title} {2018 {CES:} intel advances quantum and
  neuromorphic computing research},}\ } (\bibinfo {year} {2018})\BibitemShut
  {NoStop}%
\bibitem [{\citenamefont {Kelly}(2018)}]{kelly2018preview}%
  \BibitemOpen
  \bibfield  {author} {\bibinfo {author} {\bibfnamefont {J.}~\bibnamefont
  {Kelly}},\ }\href@noop {} {\enquote {\bibinfo {title} {A preview of
  bristlecone, google’s new quantum processor},}\ } (\bibinfo {year}
  {2018})\BibitemShut {NoStop}%
\bibitem [{\citenamefont {Gambetta}\ and\ \citenamefont
  {Sheldon}(2019)}]{gambetta2019cramming}%
  \BibitemOpen
  \bibfield  {author} {\bibinfo {author} {\bibfnamefont {J.~M.}\ \bibnamefont
  {Gambetta}}\ and\ \bibinfo {author} {\bibfnamefont {S.}~\bibnamefont
  {Sheldon}},\ }\href@noop {} {\enquote {\bibinfo {title} {Cramming more power
  into a quantum device},}\ } (\bibinfo {year} {2019})\BibitemShut {NoStop}%
\bibitem [{\citenamefont {Nay}(2019)}]{nay2019ibm}%
  \BibitemOpen
  \bibfield  {author} {\bibinfo {author} {\bibfnamefont {C.}~\bibnamefont
  {Nay}},\ }\href
  {https://newsroom.ibm.com/2019-09-18-IBM-Opens-Quantum-Computation-Center-in-New-York-Brings-Worlds-Largest-Fleet-of-Quantum-Computing-Systems-Online-Unveils-New-53-Qubit-Quantum-System-for-Broad-Use}
  {\enquote {\bibinfo {title} {{IBM} opens quantum computation center in new
  york; brings world's largest fleet of quantum computing systems online,
  unveils new {53-Qubit} quantum system for broad use},}\ } (\bibinfo {year}
  {2019})\BibitemShut {NoStop}%
\bibitem [{\citenamefont {{IBM}}(2020)}]{ibmquantumexperience2020ibm}%
  \BibitemOpen
  \bibfield  {author} {\bibinfo {author} {\bibnamefont {{IBM}}},\ }\href
  {https://quantum-computing.ibm.com} {\enquote {\bibinfo {title} {{IBM}
  quantum experience},}\ } (\bibinfo {year} {2020})\BibitemShut {NoStop}%
\bibitem [{\citenamefont {Intel}(2017)}]{intel2017intel}%
  \BibitemOpen
  \bibfield  {author} {\bibinfo {author} {\bibnamefont {Intel}},\ }\href
  {https://newsroom.intel.com/news/intel-delivers-17-qubit-superconducting-chip-advanced-packaging-qutech/}
  {\enquote {\bibinfo {title} {Intel delivers {17-Qubit} superconducting chip
  with advanced packaging to {QuTech}},}\ } (\bibinfo {year}
  {2017})\BibitemShut {NoStop}%
\bibitem [{\citenamefont {Amazon}(2019)}]{amazon2019amazon}%
  \BibitemOpen
  \bibfield  {author} {\bibinfo {author} {\bibnamefont {Amazon}},\ }\href
  {https://aws.amazon.com/braket/hardware-providers/} {\enquote {\bibinfo
  {title} {Amazon braket hardware providers},}\ } (\bibinfo {year}
  {2019})\BibitemShut {NoStop}%
\bibitem [{\citenamefont {{Rigetti
  Computing}}(2018{\natexlab{a}})}]{rigetticomputing2018quantum}%
  \BibitemOpen
  \bibfield  {author} {\bibinfo {author} {\bibnamefont {{Rigetti Computing}}},\
  }\href {http://docs.rigetti.com/en/1.9/qpu.html} {\enquote {\bibinfo {title}
  {The quantum processing unit {(QPU)} — {pyQuil} 1.9.1.dev0
  documentation},}\ } (\bibinfo {year} {2018}{\natexlab{a}})\BibitemShut
  {NoStop}%
\bibitem [{\citenamefont {Ballance}\ \emph {et~al.}(2016)\citenamefont
  {Ballance}, \citenamefont {Harty}, \citenamefont {Linke}, \citenamefont
  {Sepiol},\ and\ \citenamefont {Lucas}}]{ballance2016high-fidelity}%
  \BibitemOpen
  \bibfield  {author} {\bibinfo {author} {\bibfnamefont {C.~J.}\ \bibnamefont
  {Ballance}}, \bibinfo {author} {\bibfnamefont {T.~P.}\ \bibnamefont {Harty}},
  \bibinfo {author} {\bibfnamefont {N.~M.}\ \bibnamefont {Linke}}, \bibinfo
  {author} {\bibfnamefont {M.~A.}\ \bibnamefont {Sepiol}}, \ and\ \bibinfo
  {author} {\bibfnamefont {D.~M.}\ \bibnamefont {Lucas}},\ }\href@noop {}
  {\bibfield  {journal} {\bibinfo  {journal} {Physical Review Letters}\
  }\textbf {\bibinfo {volume} {117}},\ \bibinfo {pages} {060504} (\bibinfo
  {year} {2016})}\BibitemShut {NoStop}%
\bibitem [{\citenamefont {Barends}\ \emph {et~al.}(2016)\citenamefont
  {Barends}, \citenamefont {Shabani}, \citenamefont {Lamata}, \citenamefont
  {Kelly}, \citenamefont {Mezzacapo}, \citenamefont {Heras}, \citenamefont
  {Babbush}, \citenamefont {Fowler}, \citenamefont {Campbell}, \citenamefont
  {Chen}, \citenamefont {Chen}, \citenamefont {Chiaro}, \citenamefont
  {Dunsworth}, \citenamefont {Jeffrey}, \citenamefont {Lucero}, \citenamefont
  {Megrant}, \citenamefont {Mutus}, \citenamefont {Neeley}, \citenamefont
  {Neill}, \citenamefont {{O’Malley}}, \citenamefont {Quintana},
  \citenamefont {Roushan}, \citenamefont {Sank}, \citenamefont {Vainsencher},
  \citenamefont {Wenner}, \citenamefont {White}, \citenamefont {Solano},
  \citenamefont {Neven},\ and\ \citenamefont
  {Martinis}}]{barends2016digitized}%
  \BibitemOpen
  \bibfield  {author} {\bibinfo {author} {\bibfnamefont {R.}~\bibnamefont
  {Barends}}, \bibinfo {author} {\bibfnamefont {A.}~\bibnamefont {Shabani}},
  \bibinfo {author} {\bibfnamefont {L.}~\bibnamefont {Lamata}}, \bibinfo
  {author} {\bibfnamefont {J.}~\bibnamefont {Kelly}}, \bibinfo {author}
  {\bibfnamefont {A.}~\bibnamefont {Mezzacapo}}, \bibinfo {author}
  {\bibfnamefont {U.~L.}\ \bibnamefont {Heras}}, \bibinfo {author}
  {\bibfnamefont {R.}~\bibnamefont {Babbush}}, \bibinfo {author} {\bibfnamefont
  {A.~G.}\ \bibnamefont {Fowler}}, \bibinfo {author} {\bibfnamefont
  {B.}~\bibnamefont {Campbell}}, \bibinfo {author} {\bibfnamefont
  {Y.}~\bibnamefont {Chen}}, \bibinfo {author} {\bibfnamefont {Z.}~\bibnamefont
  {Chen}}, \bibinfo {author} {\bibfnamefont {B.}~\bibnamefont {Chiaro}},
  \bibinfo {author} {\bibfnamefont {A.}~\bibnamefont {Dunsworth}}, \bibinfo
  {author} {\bibfnamefont {E.}~\bibnamefont {Jeffrey}}, \bibinfo {author}
  {\bibfnamefont {E.}~\bibnamefont {Lucero}}, \bibinfo {author} {\bibfnamefont
  {A.}~\bibnamefont {Megrant}}, \bibinfo {author} {\bibfnamefont {J.~Y.}\
  \bibnamefont {Mutus}}, \bibinfo {author} {\bibfnamefont {M.}~\bibnamefont
  {Neeley}}, \bibinfo {author} {\bibfnamefont {C.}~\bibnamefont {Neill}},
  \bibinfo {author} {\bibfnamefont {P.~J.~J.}\ \bibnamefont {{O’Malley}}},
  \bibinfo {author} {\bibfnamefont {C.}~\bibnamefont {Quintana}}, \bibinfo
  {author} {\bibfnamefont {P.}~\bibnamefont {Roushan}}, \bibinfo {author}
  {\bibfnamefont {D.}~\bibnamefont {Sank}}, \bibinfo {author} {\bibfnamefont
  {A.}~\bibnamefont {Vainsencher}}, \bibinfo {author} {\bibfnamefont
  {J.}~\bibnamefont {Wenner}}, \bibinfo {author} {\bibfnamefont {T.~C.}\
  \bibnamefont {White}}, \bibinfo {author} {\bibfnamefont {E.}~\bibnamefont
  {Solano}}, \bibinfo {author} {\bibfnamefont {H.}~\bibnamefont {Neven}}, \
  and\ \bibinfo {author} {\bibfnamefont {J.~M.}\ \bibnamefont {Martinis}},\
  }\href@noop {} {\bibfield  {journal} {\bibinfo  {journal} {Nature}\ }\textbf
  {\bibinfo {volume} {534}},\ \bibinfo {pages} {222} (\bibinfo {year}
  {2016})}\BibitemShut {NoStop}%
\bibitem [{\citenamefont {Chow}\ \emph {et~al.}(2012)\citenamefont {Chow},
  \citenamefont {Gambetta}, \citenamefont {Corcoles}, \citenamefont {Merkel},
  \citenamefont {Smolin}, \citenamefont {Rigetti}, \citenamefont {Poletto},
  \citenamefont {Keefe}, \citenamefont {Rothwell}, \citenamefont {Rozen},
  \citenamefont {Ketchen},\ and\ \citenamefont {Steffen}}]{chow2012complete}%
  \BibitemOpen
  \bibfield  {author} {\bibinfo {author} {\bibfnamefont {J.~M.}\ \bibnamefont
  {Chow}}, \bibinfo {author} {\bibfnamefont {J.~M.}\ \bibnamefont {Gambetta}},
  \bibinfo {author} {\bibfnamefont {A.~D.}\ \bibnamefont {Corcoles}}, \bibinfo
  {author} {\bibfnamefont {S.~T.}\ \bibnamefont {Merkel}}, \bibinfo {author}
  {\bibfnamefont {J.~A.}\ \bibnamefont {Smolin}}, \bibinfo {author}
  {\bibfnamefont {C.}~\bibnamefont {Rigetti}}, \bibinfo {author} {\bibfnamefont
  {S.}~\bibnamefont {Poletto}}, \bibinfo {author} {\bibfnamefont {G.~A.}\
  \bibnamefont {Keefe}}, \bibinfo {author} {\bibfnamefont {M.~B.}\ \bibnamefont
  {Rothwell}}, \bibinfo {author} {\bibfnamefont {J.~R.}\ \bibnamefont {Rozen}},
  \bibinfo {author} {\bibfnamefont {M.~B.}\ \bibnamefont {Ketchen}}, \ and\
  \bibinfo {author} {\bibfnamefont {M.}~\bibnamefont {Steffen}},\ }\href@noop
  {} {\bibfield  {journal} {\bibinfo  {journal} {Physical Review Letters}\
  }\textbf {\bibinfo {volume} {109}},\ \bibinfo {pages} {060501} (\bibinfo
  {year} {2012})}\BibitemShut {NoStop}%
\bibitem [{\citenamefont {{Rigetti
  Computing}}(2018{\natexlab{b}})}]{rigetticomputing2018rigetti}%
  \BibitemOpen
  \bibfield  {author} {\bibinfo {author} {\bibnamefont {{Rigetti Computing}}},\
  }\href
  {https://medium.com/rigetti/the-rigetti-128-qubit-chip-and-what-it-means-for-quantum-df757d1b71ea}
  {\enquote {\bibinfo {title} {The rigetti 128-qubit chip and what it means for
  quantum},}\ } (\bibinfo {year} {2018}{\natexlab{b}})\BibitemShut {NoStop}%
\bibitem [{\citenamefont {Debnath}\ \emph {et~al.}(2016)\citenamefont
  {Debnath}, \citenamefont {Linke}, \citenamefont {Figgatt}, \citenamefont
  {Landsman}, \citenamefont {Wright},\ and\ \citenamefont
  {Monroe}}]{debnath2016demonstration}%
  \BibitemOpen
  \bibfield  {author} {\bibinfo {author} {\bibfnamefont {S.}~\bibnamefont
  {Debnath}}, \bibinfo {author} {\bibfnamefont {N.~M.}\ \bibnamefont {Linke}},
  \bibinfo {author} {\bibfnamefont {C.}~\bibnamefont {Figgatt}}, \bibinfo
  {author} {\bibfnamefont {K.~A.}\ \bibnamefont {Landsman}}, \bibinfo {author}
  {\bibfnamefont {K.}~\bibnamefont {Wright}}, \ and\ \bibinfo {author}
  {\bibfnamefont {C.}~\bibnamefont {Monroe}},\ }\href@noop {} {\bibfield
  {journal} {\bibinfo  {journal} {Nature}\ }\textbf {\bibinfo {volume} {536}},\
  \bibinfo {pages} {63} (\bibinfo {year} {2016})}\BibitemShut {NoStop}%
\bibitem [{\citenamefont {{DiCarlo}}\ \emph {et~al.}(2009)\citenamefont
  {{DiCarlo}}, \citenamefont {Chow}, \citenamefont {Gambetta}, \citenamefont
  {Bishop}, \citenamefont {Johnson}, \citenamefont {Schuster}, \citenamefont
  {Majer}, \citenamefont {Blais}, \citenamefont {Frunzio}, \citenamefont
  {Girvin},\ and\ \citenamefont {Schoelkopf}}]{dicarlo2009demonstration}%
  \BibitemOpen
  \bibfield  {author} {\bibinfo {author} {\bibfnamefont {L.}~\bibnamefont
  {{DiCarlo}}}, \bibinfo {author} {\bibfnamefont {J.~M.}\ \bibnamefont {Chow}},
  \bibinfo {author} {\bibfnamefont {J.~M.}\ \bibnamefont {Gambetta}}, \bibinfo
  {author} {\bibfnamefont {L.~S.}\ \bibnamefont {Bishop}}, \bibinfo {author}
  {\bibfnamefont {B.~R.}\ \bibnamefont {Johnson}}, \bibinfo {author}
  {\bibfnamefont {D.~I.}\ \bibnamefont {Schuster}}, \bibinfo {author}
  {\bibfnamefont {J.}~\bibnamefont {Majer}}, \bibinfo {author} {\bibfnamefont
  {A.}~\bibnamefont {Blais}}, \bibinfo {author} {\bibfnamefont
  {L.}~\bibnamefont {Frunzio}}, \bibinfo {author} {\bibfnamefont {S.~M.}\
  \bibnamefont {Girvin}}, \ and\ \bibinfo {author} {\bibfnamefont {R.~J.}\
  \bibnamefont {Schoelkopf}},\ }\href@noop {} {\bibfield  {journal} {\bibinfo
  {journal} {Nature}\ }\textbf {\bibinfo {volume} {460}},\ \bibinfo {pages}
  {240} (\bibinfo {year} {2009})}\BibitemShut {NoStop}%
\bibitem [{\citenamefont {Finke}(2020{\natexlab{b}})}]{finke2020qubit}%
  \BibitemOpen
  \bibfield  {author} {\bibinfo {author} {\bibfnamefont {D.}~\bibnamefont
  {Finke}},\ }\href
  {https://quantumcomputingreport.com/scorecards/qubit-quality/} {\enquote
  {\bibinfo {title} {Qubit quality},}\ } (\bibinfo {year}
  {2020}{\natexlab{b}})\BibitemShut {NoStop}%
\bibitem [{\citenamefont {Friis}\ \emph {et~al.}(2018)\citenamefont {Friis},
  \citenamefont {Marty}, \citenamefont {Maier}, \citenamefont {Hempel},
  \citenamefont {Holzäpfel}, \citenamefont {Jurcevic}, \citenamefont {Plenio},
  \citenamefont {Huber}, \citenamefont {Roos}, \citenamefont {Blatt},\ and\
  \citenamefont {Lanyon}}]{friis2018observation}%
  \BibitemOpen
  \bibfield  {author} {\bibinfo {author} {\bibfnamefont {N.}~\bibnamefont
  {Friis}}, \bibinfo {author} {\bibfnamefont {O.}~\bibnamefont {Marty}},
  \bibinfo {author} {\bibfnamefont {C.}~\bibnamefont {Maier}}, \bibinfo
  {author} {\bibfnamefont {C.}~\bibnamefont {Hempel}}, \bibinfo {author}
  {\bibfnamefont {M.}~\bibnamefont {Holzäpfel}}, \bibinfo {author}
  {\bibfnamefont {P.}~\bibnamefont {Jurcevic}}, \bibinfo {author}
  {\bibfnamefont {M.~B.}\ \bibnamefont {Plenio}}, \bibinfo {author}
  {\bibfnamefont {M.}~\bibnamefont {Huber}}, \bibinfo {author} {\bibfnamefont
  {C.}~\bibnamefont {Roos}}, \bibinfo {author} {\bibfnamefont {R.}~\bibnamefont
  {Blatt}}, \ and\ \bibinfo {author} {\bibfnamefont {B.}~\bibnamefont
  {Lanyon}},\ }\href@noop {} {\bibfield  {journal} {\bibinfo  {journal}
  {Physical Review X}\ }\textbf {\bibinfo {volume} {8}},\ \bibinfo {pages}
  {021012} (\bibinfo {year} {2018})}\BibitemShut {NoStop}%
\bibitem [{\citenamefont {Gaebler}\ \emph {et~al.}(2016)\citenamefont
  {Gaebler}, \citenamefont {Tan}, \citenamefont {Lin}, \citenamefont {Wan},
  \citenamefont {Bowler}, \citenamefont {Keith}, \citenamefont {Glancy},
  \citenamefont {Coakley}, \citenamefont {Knill}, \citenamefont {Leibfried},\
  and\ \citenamefont {Wineland}}]{gaebler2016high-fidelity}%
  \BibitemOpen
  \bibfield  {author} {\bibinfo {author} {\bibfnamefont {J.~P.}\ \bibnamefont
  {Gaebler}}, \bibinfo {author} {\bibfnamefont {T.~R.}\ \bibnamefont {Tan}},
  \bibinfo {author} {\bibfnamefont {Y.}~\bibnamefont {Lin}}, \bibinfo {author}
  {\bibfnamefont {Y.}~\bibnamefont {Wan}}, \bibinfo {author} {\bibfnamefont
  {R.}~\bibnamefont {Bowler}}, \bibinfo {author} {\bibfnamefont {A.~C.}\
  \bibnamefont {Keith}}, \bibinfo {author} {\bibfnamefont {S.}~\bibnamefont
  {Glancy}}, \bibinfo {author} {\bibfnamefont {K.}~\bibnamefont {Coakley}},
  \bibinfo {author} {\bibfnamefont {E.}~\bibnamefont {Knill}}, \bibinfo
  {author} {\bibfnamefont {D.}~\bibnamefont {Leibfried}}, \ and\ \bibinfo
  {author} {\bibfnamefont {D.~J.}\ \bibnamefont {Wineland}},\ }\href@noop {}
  {\bibfield  {journal} {\bibinfo  {journal} {Physical Review Letters}\
  }\textbf {\bibinfo {volume} {117}},\ \bibinfo {pages} {060505} (\bibinfo
  {year} {2016})}\BibitemShut {NoStop}%
\bibitem [{\citenamefont {Häffner}\ \emph {et~al.}(2005)\citenamefont
  {Häffner}, \citenamefont {Hänsel}, \citenamefont {Roos}, \citenamefont
  {Benhelm}, \citenamefont {Chek-al kar}, \citenamefont {Chwalla},
  \citenamefont {Körber}, \citenamefont {Rapol}, \citenamefont {Riebe},
  \citenamefont {Schmidt}, \citenamefont {Becher}, \citenamefont {Gühne},
  \citenamefont {Dür},\ and\ \citenamefont {Blatt}}]{haeffner2005scalable}%
  \BibitemOpen
  \bibfield  {author} {\bibinfo {author} {\bibfnamefont {H.}~\bibnamefont
  {Häffner}}, \bibinfo {author} {\bibfnamefont {W.}~\bibnamefont {Hänsel}},
  \bibinfo {author} {\bibfnamefont {C.~F.}\ \bibnamefont {Roos}}, \bibinfo
  {author} {\bibfnamefont {J.}~\bibnamefont {Benhelm}}, \bibinfo {author}
  {\bibfnamefont {D.}~\bibnamefont {Chek-al kar}}, \bibinfo {author}
  {\bibfnamefont {M.}~\bibnamefont {Chwalla}}, \bibinfo {author} {\bibfnamefont
  {T.}~\bibnamefont {Körber}}, \bibinfo {author} {\bibfnamefont {U.~D.}\
  \bibnamefont {Rapol}}, \bibinfo {author} {\bibfnamefont {M.}~\bibnamefont
  {Riebe}}, \bibinfo {author} {\bibfnamefont {P.~O.}\ \bibnamefont {Schmidt}},
  \bibinfo {author} {\bibfnamefont {C.}~\bibnamefont {Becher}}, \bibinfo
  {author} {\bibfnamefont {O.}~\bibnamefont {Gühne}}, \bibinfo {author}
  {\bibfnamefont {W.}~\bibnamefont {Dür}}, \ and\ \bibinfo {author}
  {\bibfnamefont {R.}~\bibnamefont {Blatt}},\ }\href@noop {} {\bibfield
  {journal} {\bibinfo  {journal} {Nature}\ }\textbf {\bibinfo {volume} {438}},\
  \bibinfo {pages} {643} (\bibinfo {year} {2005})}\BibitemShut {NoStop}%
\bibitem [{\citenamefont {Hong}\ \emph {et~al.}(2020)\citenamefont {Hong},
  \citenamefont {Papageorge}, \citenamefont {Sivarajah}, \citenamefont
  {Crossman}, \citenamefont {Didier}, \citenamefont {Polloreno}, \citenamefont
  {Sete}, \citenamefont {Turkowski}, \citenamefont {da~Silva},\ and\
  \citenamefont {Johnson}}]{hong2020demonstration}%
  \BibitemOpen
  \bibfield  {author} {\bibinfo {author} {\bibfnamefont {S.~S.}\ \bibnamefont
  {Hong}}, \bibinfo {author} {\bibfnamefont {A.~T.}\ \bibnamefont
  {Papageorge}}, \bibinfo {author} {\bibfnamefont {P.}~\bibnamefont
  {Sivarajah}}, \bibinfo {author} {\bibfnamefont {G.}~\bibnamefont {Crossman}},
  \bibinfo {author} {\bibfnamefont {N.}~\bibnamefont {Didier}}, \bibinfo
  {author} {\bibfnamefont {A.~M.}\ \bibnamefont {Polloreno}}, \bibinfo {author}
  {\bibfnamefont {E.~A.}\ \bibnamefont {Sete}}, \bibinfo {author}
  {\bibfnamefont {S.~W.}\ \bibnamefont {Turkowski}}, \bibinfo {author}
  {\bibfnamefont {M.~P.}\ \bibnamefont {da~Silva}}, \ and\ \bibinfo {author}
  {\bibfnamefont {B.~R.}\ \bibnamefont {Johnson}},\ }\href@noop {} {\bibfield
  {journal} {\bibinfo  {journal} {Physical Review A}\ }\textbf {\bibinfo
  {volume} {101}},\ \bibinfo {pages} {012302} (\bibinfo {year}
  {2020})}\BibitemShut {NoStop}%
\bibitem [{\citenamefont {Leibfried}\ \emph {et~al.}(2003)\citenamefont
  {Leibfried}, \citenamefont {{DeMarco}}, \citenamefont {Meyer}, \citenamefont
  {Lucas}, \citenamefont {Barrett}, \citenamefont {Britton}, \citenamefont
  {Itano}, \citenamefont {Jelenković}, \citenamefont {Langer}, \citenamefont
  {Rosenband},\ and\ \citenamefont {Wineland}}]{leibfried2003experimental}%
  \BibitemOpen
  \bibfield  {author} {\bibinfo {author} {\bibfnamefont {D.}~\bibnamefont
  {Leibfried}}, \bibinfo {author} {\bibfnamefont {B.}~\bibnamefont
  {{DeMarco}}}, \bibinfo {author} {\bibfnamefont {V.}~\bibnamefont {Meyer}},
  \bibinfo {author} {\bibfnamefont {D.}~\bibnamefont {Lucas}}, \bibinfo
  {author} {\bibfnamefont {M.}~\bibnamefont {Barrett}}, \bibinfo {author}
  {\bibfnamefont {J.}~\bibnamefont {Britton}}, \bibinfo {author} {\bibfnamefont
  {W.~M.}\ \bibnamefont {Itano}}, \bibinfo {author} {\bibfnamefont
  {B.}~\bibnamefont {Jelenković}}, \bibinfo {author} {\bibfnamefont
  {C.}~\bibnamefont {Langer}}, \bibinfo {author} {\bibfnamefont
  {T.}~\bibnamefont {Rosenband}}, \ and\ \bibinfo {author} {\bibfnamefont
  {D.~J.}\ \bibnamefont {Wineland}},\ }\href@noop {} {\bibfield  {journal}
  {\bibinfo  {journal} {Nature}\ }\textbf {\bibinfo {volume} {422}},\ \bibinfo
  {pages} {412} (\bibinfo {year} {2003})}\BibitemShut {NoStop}%
\bibitem [{\citenamefont {Nersisyan}\ \emph {et~al.}(2019)\citenamefont
  {Nersisyan}, \citenamefont {Poletto}, \citenamefont {Alidoust}, \citenamefont
  {Manenti}, \citenamefont {Renzas}, \citenamefont {Bui}, \citenamefont {Vu},
  \citenamefont {Whyland}, \citenamefont {Mohan}, \citenamefont {Sete},
  \citenamefont {Stanwyck}, \citenamefont {Bestwick},\ and\ \citenamefont
  {Reagor}}]{nersisyan2019manufacturing}%
  \BibitemOpen
  \bibfield  {author} {\bibinfo {author} {\bibfnamefont {A.}~\bibnamefont
  {Nersisyan}}, \bibinfo {author} {\bibfnamefont {S.}~\bibnamefont {Poletto}},
  \bibinfo {author} {\bibfnamefont {N.}~\bibnamefont {Alidoust}}, \bibinfo
  {author} {\bibfnamefont {R.}~\bibnamefont {Manenti}}, \bibinfo {author}
  {\bibfnamefont {R.}~\bibnamefont {Renzas}}, \bibinfo {author} {\bibfnamefont
  {C.}~\bibnamefont {Bui}}, \bibinfo {author} {\bibfnamefont {K.}~\bibnamefont
  {Vu}}, \bibinfo {author} {\bibfnamefont {T.}~\bibnamefont {Whyland}},
  \bibinfo {author} {\bibfnamefont {Y.}~\bibnamefont {Mohan}}, \bibinfo
  {author} {\bibfnamefont {E.~A.}\ \bibnamefont {Sete}}, \bibinfo {author}
  {\bibfnamefont {S.}~\bibnamefont {Stanwyck}}, \bibinfo {author}
  {\bibfnamefont {A.}~\bibnamefont {Bestwick}}, \ and\ \bibinfo {author}
  {\bibfnamefont {M.}~\bibnamefont {Reagor}},\ }\href@noop {} {\bibfield
  {journal} {\bibinfo  {journal} {{arXiv:1901.08042}}\ } (\bibinfo {year}
  {2019})}\BibitemShut {NoStop}%
\bibitem [{\citenamefont {Reagor}\ \emph {et~al.}(2018)\citenamefont {Reagor},
  \citenamefont {Osborn}, \citenamefont {Tezak}, \citenamefont {Staley},
  \citenamefont {Prawiroatmodjo}, \citenamefont {Scheer}, \citenamefont
  {Alidoust}, \citenamefont {Sete}, \citenamefont {Didier}, \citenamefont
  {da~Silva}, \citenamefont {Acala}, \citenamefont {Angeles}, \citenamefont
  {Bestwick}, \citenamefont {Block}, \citenamefont {Bloom}, \citenamefont
  {Bradley}, \citenamefont {Bui}, \citenamefont {Caldwell}, \citenamefont
  {Capelluto}, \citenamefont {Chilcott}, \citenamefont {Cordova}, \citenamefont
  {Crossman}, \citenamefont {Curtis}, \citenamefont {Deshpande}, \citenamefont
  {El~Bouayadi}, \citenamefont {Girshovich}, \citenamefont {Hong},
  \citenamefont {Hudson}, \citenamefont {Karalekas}, \citenamefont {Kuang},
  \citenamefont {Lenihan}, \citenamefont {Manenti}, \citenamefont {Manning},
  \citenamefont {Marshall}, \citenamefont {Mohan}, \citenamefont {{O’Brien}},
  \citenamefont {Otterbach}, \citenamefont {Papageorge}, \citenamefont
  {Paquette}, \citenamefont {Pelstring}, \citenamefont {Polloreno},
  \citenamefont {Rawat}, \citenamefont {Ryan}, \citenamefont {Renzas},
  \citenamefont {Rubin}, \citenamefont {Russel}, \citenamefont {Rust},
  \citenamefont {Scarabelli}, \citenamefont {Selvanayagam}, \citenamefont
  {Sinclair}, \citenamefont {Smith}, \citenamefont {Suska}, \citenamefont {To},
  \citenamefont {Vahidpour}, \citenamefont {Vodrahalli}, \citenamefont
  {Whyland}, \citenamefont {Yadav}, \citenamefont {Zeng},\ and\ \citenamefont
  {Rigetti}}]{reagor2018demonstration}%
  \BibitemOpen
  \bibfield  {author} {\bibinfo {author} {\bibfnamefont {M.}~\bibnamefont
  {Reagor}}, \bibinfo {author} {\bibfnamefont {C.~B.}\ \bibnamefont {Osborn}},
  \bibinfo {author} {\bibfnamefont {N.}~\bibnamefont {Tezak}}, \bibinfo
  {author} {\bibfnamefont {A.}~\bibnamefont {Staley}}, \bibinfo {author}
  {\bibfnamefont {G.}~\bibnamefont {Prawiroatmodjo}}, \bibinfo {author}
  {\bibfnamefont {M.}~\bibnamefont {Scheer}}, \bibinfo {author} {\bibfnamefont
  {N.}~\bibnamefont {Alidoust}}, \bibinfo {author} {\bibfnamefont {E.~A.}\
  \bibnamefont {Sete}}, \bibinfo {author} {\bibfnamefont {N.}~\bibnamefont
  {Didier}}, \bibinfo {author} {\bibfnamefont {M.~P.}\ \bibnamefont
  {da~Silva}}, \bibinfo {author} {\bibfnamefont {E.}~\bibnamefont {Acala}},
  \bibinfo {author} {\bibfnamefont {J.}~\bibnamefont {Angeles}}, \bibinfo
  {author} {\bibfnamefont {A.}~\bibnamefont {Bestwick}}, \bibinfo {author}
  {\bibfnamefont {M.}~\bibnamefont {Block}}, \bibinfo {author} {\bibfnamefont
  {B.}~\bibnamefont {Bloom}}, \bibinfo {author} {\bibfnamefont
  {A.}~\bibnamefont {Bradley}}, \bibinfo {author} {\bibfnamefont
  {C.}~\bibnamefont {Bui}}, \bibinfo {author} {\bibfnamefont {S.}~\bibnamefont
  {Caldwell}}, \bibinfo {author} {\bibfnamefont {L.}~\bibnamefont {Capelluto}},
  \bibinfo {author} {\bibfnamefont {R.}~\bibnamefont {Chilcott}}, \bibinfo
  {author} {\bibfnamefont {J.}~\bibnamefont {Cordova}}, \bibinfo {author}
  {\bibfnamefont {G.}~\bibnamefont {Crossman}}, \bibinfo {author}
  {\bibfnamefont {M.}~\bibnamefont {Curtis}}, \bibinfo {author} {\bibfnamefont
  {S.}~\bibnamefont {Deshpande}}, \bibinfo {author} {\bibfnamefont
  {T.}~\bibnamefont {El~Bouayadi}}, \bibinfo {author} {\bibfnamefont
  {D.}~\bibnamefont {Girshovich}}, \bibinfo {author} {\bibfnamefont
  {S.}~\bibnamefont {Hong}}, \bibinfo {author} {\bibfnamefont {A.}~\bibnamefont
  {Hudson}}, \bibinfo {author} {\bibfnamefont {P.}~\bibnamefont {Karalekas}},
  \bibinfo {author} {\bibfnamefont {K.}~\bibnamefont {Kuang}}, \bibinfo
  {author} {\bibfnamefont {M.}~\bibnamefont {Lenihan}}, \bibinfo {author}
  {\bibfnamefont {R.}~\bibnamefont {Manenti}}, \bibinfo {author} {\bibfnamefont
  {T.}~\bibnamefont {Manning}}, \bibinfo {author} {\bibfnamefont
  {J.}~\bibnamefont {Marshall}}, \bibinfo {author} {\bibfnamefont
  {Y.}~\bibnamefont {Mohan}}, \bibinfo {author} {\bibfnamefont
  {W.}~\bibnamefont {{O’Brien}}}, \bibinfo {author} {\bibfnamefont
  {J.}~\bibnamefont {Otterbach}}, \bibinfo {author} {\bibfnamefont
  {A.}~\bibnamefont {Papageorge}}, \bibinfo {author} {\bibfnamefont
  {J.}~\bibnamefont {Paquette}}, \bibinfo {author} {\bibfnamefont
  {M.}~\bibnamefont {Pelstring}}, \bibinfo {author} {\bibfnamefont
  {A.}~\bibnamefont {Polloreno}}, \bibinfo {author} {\bibfnamefont
  {V.}~\bibnamefont {Rawat}}, \bibinfo {author} {\bibfnamefont {C.~A.}\
  \bibnamefont {Ryan}}, \bibinfo {author} {\bibfnamefont {R.}~\bibnamefont
  {Renzas}}, \bibinfo {author} {\bibfnamefont {N.}~\bibnamefont {Rubin}},
  \bibinfo {author} {\bibfnamefont {D.}~\bibnamefont {Russel}}, \bibinfo
  {author} {\bibfnamefont {M.}~\bibnamefont {Rust}}, \bibinfo {author}
  {\bibfnamefont {D.}~\bibnamefont {Scarabelli}}, \bibinfo {author}
  {\bibfnamefont {M.}~\bibnamefont {Selvanayagam}}, \bibinfo {author}
  {\bibfnamefont {R.}~\bibnamefont {Sinclair}}, \bibinfo {author}
  {\bibfnamefont {R.}~\bibnamefont {Smith}}, \bibinfo {author} {\bibfnamefont
  {M.}~\bibnamefont {Suska}}, \bibinfo {author} {\bibfnamefont
  {T.}~\bibnamefont {To}}, \bibinfo {author} {\bibfnamefont {M.}~\bibnamefont
  {Vahidpour}}, \bibinfo {author} {\bibfnamefont {N.}~\bibnamefont
  {Vodrahalli}}, \bibinfo {author} {\bibfnamefont {T.}~\bibnamefont {Whyland}},
  \bibinfo {author} {\bibfnamefont {K.}~\bibnamefont {Yadav}}, \bibinfo
  {author} {\bibfnamefont {W.}~\bibnamefont {Zeng}}, \ and\ \bibinfo {author}
  {\bibfnamefont {C.~T.}\ \bibnamefont {Rigetti}},\ }\href@noop {} {\bibfield
  {journal} {\bibinfo  {journal} {Science Advances}\ }\textbf {\bibinfo
  {volume} {4}},\ \bibinfo {pages} {eaao3603} (\bibinfo {year}
  {2018})}\BibitemShut {NoStop}%
\bibitem [{\citenamefont {{Schmidt-Kaler}}\ \emph {et~al.}(2003)\citenamefont
  {{Schmidt-Kaler}}, \citenamefont {Häffner}, \citenamefont {Riebe},
  \citenamefont {Gulde}, \citenamefont {Lancaster}, \citenamefont {Deuschle},
  \citenamefont {Becher}, \citenamefont {Roos}, \citenamefont {Eschner},\ and\
  \citenamefont {Blatt}}]{schmidt-kaler2003realization}%
  \BibitemOpen
  \bibfield  {author} {\bibinfo {author} {\bibfnamefont {F.}~\bibnamefont
  {{Schmidt-Kaler}}}, \bibinfo {author} {\bibfnamefont {H.}~\bibnamefont
  {Häffner}}, \bibinfo {author} {\bibfnamefont {M.}~\bibnamefont {Riebe}},
  \bibinfo {author} {\bibfnamefont {S.}~\bibnamefont {Gulde}}, \bibinfo
  {author} {\bibfnamefont {G.~P.~T.}\ \bibnamefont {Lancaster}}, \bibinfo
  {author} {\bibfnamefont {T.}~\bibnamefont {Deuschle}}, \bibinfo {author}
  {\bibfnamefont {C.}~\bibnamefont {Becher}}, \bibinfo {author} {\bibfnamefont
  {C.~F.}\ \bibnamefont {Roos}}, \bibinfo {author} {\bibfnamefont
  {J.}~\bibnamefont {Eschner}}, \ and\ \bibinfo {author} {\bibfnamefont
  {R.}~\bibnamefont {Blatt}},\ }\href@noop {} {\bibfield  {journal} {\bibinfo
  {journal} {Nature}\ }\textbf {\bibinfo {volume} {422}},\ \bibinfo {pages}
  {408} (\bibinfo {year} {2003})}\BibitemShut {NoStop}%
\bibitem [{\citenamefont {Sheldon}\ \emph {et~al.}(2016)\citenamefont
  {Sheldon}, \citenamefont {Magesan}, \citenamefont {Chow},\ and\ \citenamefont
  {Gambetta}}]{sheldon2016procedure}%
  \BibitemOpen
  \bibfield  {author} {\bibinfo {author} {\bibfnamefont {S.}~\bibnamefont
  {Sheldon}}, \bibinfo {author} {\bibfnamefont {E.}~\bibnamefont {Magesan}},
  \bibinfo {author} {\bibfnamefont {J.~M.}\ \bibnamefont {Chow}}, \ and\
  \bibinfo {author} {\bibfnamefont {J.~M.}\ \bibnamefont {Gambetta}},\
  }\href@noop {} {\bibfield  {journal} {\bibinfo  {journal} {Physical Review
  A}\ }\textbf {\bibinfo {volume} {93}},\ \bibinfo {pages} {060302} (\bibinfo
  {year} {2016})}\BibitemShut {NoStop}%
\bibitem [{\citenamefont {Song}\ \emph {et~al.}(2017)\citenamefont {Song},
  \citenamefont {Xu}, \citenamefont {Liu}, \citenamefont {Yang}, \citenamefont
  {Zheng}, \citenamefont {Deng}, \citenamefont {Xie}, \citenamefont {Huang},
  \citenamefont {Guo}, \citenamefont {Zhang}, \citenamefont {Zhang},
  \citenamefont {Xu}, \citenamefont {Zheng}, \citenamefont {Zhu}, \citenamefont
  {Wang}, \citenamefont {Chen}, \citenamefont {Lu}, \citenamefont {Han},\ and\
  \citenamefont {Pan}}]{song2017-qubit}%
  \BibitemOpen
  \bibfield  {author} {\bibinfo {author} {\bibfnamefont {C.}~\bibnamefont
  {Song}}, \bibinfo {author} {\bibfnamefont {K.}~\bibnamefont {Xu}}, \bibinfo
  {author} {\bibfnamefont {W.}~\bibnamefont {Liu}}, \bibinfo {author}
  {\bibfnamefont {C.}~\bibnamefont {Yang}}, \bibinfo {author} {\bibfnamefont
  {S.}~\bibnamefont {Zheng}}, \bibinfo {author} {\bibfnamefont
  {H.}~\bibnamefont {Deng}}, \bibinfo {author} {\bibfnamefont {Q.}~\bibnamefont
  {Xie}}, \bibinfo {author} {\bibfnamefont {K.}~\bibnamefont {Huang}}, \bibinfo
  {author} {\bibfnamefont {Q.}~\bibnamefont {Guo}}, \bibinfo {author}
  {\bibfnamefont {L.}~\bibnamefont {Zhang}}, \bibinfo {author} {\bibfnamefont
  {P.}~\bibnamefont {Zhang}}, \bibinfo {author} {\bibfnamefont
  {D.}~\bibnamefont {Xu}}, \bibinfo {author} {\bibfnamefont {D.}~\bibnamefont
  {Zheng}}, \bibinfo {author} {\bibfnamefont {X.}~\bibnamefont {Zhu}}, \bibinfo
  {author} {\bibfnamefont {H.}~\bibnamefont {Wang}}, \bibinfo {author}
  {\bibfnamefont {Y.}~\bibnamefont {Chen}}, \bibinfo {author} {\bibfnamefont
  {C.}~\bibnamefont {Lu}}, \bibinfo {author} {\bibfnamefont {S.}~\bibnamefont
  {Han}}, \ and\ \bibinfo {author} {\bibfnamefont {J.}~\bibnamefont {Pan}},\
  }\href@noop {} {\bibfield  {journal} {\bibinfo  {journal} {Physical Review
  Letters}\ }\textbf {\bibinfo {volume} {119}},\ \bibinfo {pages} {180511}
  (\bibinfo {year} {2017})}\BibitemShut {NoStop}%
\bibitem [{\citenamefont {Steffen}\ \emph {et~al.}(2006)\citenamefont
  {Steffen}, \citenamefont {Ansmann}, \citenamefont {Bialczak}, \citenamefont
  {Katz}, \citenamefont {Lucero}, \citenamefont {{McDermott}}, \citenamefont
  {Neeley}, \citenamefont {Weig}, \citenamefont {Cleland},\ and\ \citenamefont
  {Martinis}}]{steffen2006measurement}%
  \BibitemOpen
  \bibfield  {author} {\bibinfo {author} {\bibfnamefont {M.}~\bibnamefont
  {Steffen}}, \bibinfo {author} {\bibfnamefont {M.}~\bibnamefont {Ansmann}},
  \bibinfo {author} {\bibfnamefont {R.~C.}\ \bibnamefont {Bialczak}}, \bibinfo
  {author} {\bibfnamefont {N.}~\bibnamefont {Katz}}, \bibinfo {author}
  {\bibfnamefont {E.}~\bibnamefont {Lucero}}, \bibinfo {author} {\bibfnamefont
  {R.}~\bibnamefont {{McDermott}}}, \bibinfo {author} {\bibfnamefont
  {M.}~\bibnamefont {Neeley}}, \bibinfo {author} {\bibfnamefont {E.~M.}\
  \bibnamefont {Weig}}, \bibinfo {author} {\bibfnamefont {A.~N.}\ \bibnamefont
  {Cleland}}, \ and\ \bibinfo {author} {\bibfnamefont {J.~M.}\ \bibnamefont
  {Martinis}},\ }\href@noop {} {\bibfield  {journal} {\bibinfo  {journal}
  {Science}\ }\textbf {\bibinfo {volume} {313}},\ \bibinfo {pages} {1423}
  (\bibinfo {year} {2006})}\BibitemShut {NoStop}%
\bibitem [{\citenamefont {Watson}(2017)}]{watson2017ibm}%
  \BibitemOpen
  \bibfield  {author} {\bibinfo {author} {\bibfnamefont {T.}~\bibnamefont
  {Watson}},\ }\href
  {https://developer.ibm.com/dwblog/2017/quantum-computing-16-qubit-processor/}
  {\enquote {\bibinfo {title} {{IBM} doubles compute power for quantum systems,
  developers execute {300K+} experiments on {IBM} quantum cloud},}\ } (\bibinfo
  {year} {2017})\BibitemShut {NoStop}%
\bibitem [{\citenamefont {Veldhorst}\ \emph {et~al.}(2015)\citenamefont
  {Veldhorst}, \citenamefont {Yang}, \citenamefont {Hwang}, \citenamefont
  {Huang}, \citenamefont {Dehollain}, \citenamefont {Muhonen}, \citenamefont
  {Simmons}, \citenamefont {Laucht}, \citenamefont {Hudson}, \citenamefont
  {Itoh}, \citenamefont {Morello},\ and\ \citenamefont
  {Dzurak}}]{veldhorst2015two}%
  \BibitemOpen
  \bibfield  {author} {\bibinfo {author} {\bibfnamefont {M.}~\bibnamefont
  {Veldhorst}}, \bibinfo {author} {\bibfnamefont {C.~H.}\ \bibnamefont {Yang}},
  \bibinfo {author} {\bibfnamefont {J.~C.~C.}\ \bibnamefont {Hwang}}, \bibinfo
  {author} {\bibfnamefont {W.}~\bibnamefont {Huang}}, \bibinfo {author}
  {\bibfnamefont {J.~P.}\ \bibnamefont {Dehollain}}, \bibinfo {author}
  {\bibfnamefont {J.~T.}\ \bibnamefont {Muhonen}}, \bibinfo {author}
  {\bibfnamefont {S.}~\bibnamefont {Simmons}}, \bibinfo {author} {\bibfnamefont
  {A.}~\bibnamefont {Laucht}}, \bibinfo {author} {\bibfnamefont {F.~E.}\
  \bibnamefont {Hudson}}, \bibinfo {author} {\bibfnamefont {K.~M.}\
  \bibnamefont {Itoh}}, \bibinfo {author} {\bibfnamefont {A.}~\bibnamefont
  {Morello}}, \ and\ \bibinfo {author} {\bibfnamefont {A.~S.}\ \bibnamefont
  {Dzurak}},\ }\href@noop {} {\bibfield  {journal} {\bibinfo  {journal}
  {Nature}\ }\textbf {\bibinfo {volume} {526}},\ \bibinfo {pages} {410}
  (\bibinfo {year} {2015})}\BibitemShut {NoStop}%
\bibitem [{\citenamefont {Wright}\ \emph {et~al.}(2019)\citenamefont {Wright},
  \citenamefont {Beck}, \citenamefont {Debnath}, \citenamefont {Amini},
  \citenamefont {Nam}, \citenamefont {Grzesiak}, \citenamefont {Chen},
  \citenamefont {Pisenti}, \citenamefont {Chmielewski}, \citenamefont
  {Collins}, \citenamefont {Hudek}, \citenamefont {Mizrahi}, \citenamefont
  {{Wong-Campos}}, \citenamefont {Allen}, \citenamefont {Apisdorf},
  \citenamefont {Solomon}, \citenamefont {Williams}, \citenamefont {Ducore},
  \citenamefont {Blinov}, \citenamefont {Kreikemeier}, \citenamefont {Chaplin},
  \citenamefont {Keesan}, \citenamefont {Monroe},\ and\ \citenamefont
  {Kim}}]{wright2019benchmarking}%
  \BibitemOpen
  \bibfield  {author} {\bibinfo {author} {\bibfnamefont {K.}~\bibnamefont
  {Wright}}, \bibinfo {author} {\bibfnamefont {K.~M.}\ \bibnamefont {Beck}},
  \bibinfo {author} {\bibfnamefont {S.}~\bibnamefont {Debnath}}, \bibinfo
  {author} {\bibfnamefont {J.~M.}\ \bibnamefont {Amini}}, \bibinfo {author}
  {\bibfnamefont {Y.}~\bibnamefont {Nam}}, \bibinfo {author} {\bibfnamefont
  {N.}~\bibnamefont {Grzesiak}}, \bibinfo {author} {\bibfnamefont
  {J.}~\bibnamefont {Chen}}, \bibinfo {author} {\bibfnamefont {N.~C.}\
  \bibnamefont {Pisenti}}, \bibinfo {author} {\bibfnamefont {M.}~\bibnamefont
  {Chmielewski}}, \bibinfo {author} {\bibfnamefont {C.}~\bibnamefont
  {Collins}}, \bibinfo {author} {\bibfnamefont {K.~M.}\ \bibnamefont {Hudek}},
  \bibinfo {author} {\bibfnamefont {J.}~\bibnamefont {Mizrahi}}, \bibinfo
  {author} {\bibfnamefont {J.~D.}\ \bibnamefont {{Wong-Campos}}}, \bibinfo
  {author} {\bibfnamefont {S.}~\bibnamefont {Allen}}, \bibinfo {author}
  {\bibfnamefont {J.}~\bibnamefont {Apisdorf}}, \bibinfo {author}
  {\bibfnamefont {P.}~\bibnamefont {Solomon}}, \bibinfo {author} {\bibfnamefont
  {M.}~\bibnamefont {Williams}}, \bibinfo {author} {\bibfnamefont {A.~M.}\
  \bibnamefont {Ducore}}, \bibinfo {author} {\bibfnamefont {A.}~\bibnamefont
  {Blinov}}, \bibinfo {author} {\bibfnamefont {S.~M.}\ \bibnamefont
  {Kreikemeier}}, \bibinfo {author} {\bibfnamefont {V.}~\bibnamefont
  {Chaplin}}, \bibinfo {author} {\bibfnamefont {M.}~\bibnamefont {Keesan}},
  \bibinfo {author} {\bibfnamefont {C.}~\bibnamefont {Monroe}}, \ and\ \bibinfo
  {author} {\bibfnamefont {J.}~\bibnamefont {Kim}},\ }\href@noop {} {\bibfield
  {journal} {\bibinfo  {journal} {Nature Communications}\ }\textbf {\bibinfo
  {volume} {10}},\ \bibinfo {pages} {5464} (\bibinfo {year}
  {2019})}\BibitemShut {NoStop}%
\bibitem [{Note17()}]{Note17}%
  \BibitemOpen
  \bibinfo {note} {When we only know the year of publication we imputed the
  date as 1 June that year.}\BibitemShut {Stop}%
\bibitem [{Note18()}]{Note18}%
  \BibitemOpen
  \bibinfo {note} {We could instead have used multiple linear regression where,
  for example, we predict the log error rate based on the year and log physical
  qubits. However this would require us to assume a linear dependency between
  the log error rate and log physical qubits, whereas the multivariate model is
  agnostic about the relation of the two metrics.}\BibitemShut {Stop}%
\bibitem [{\citenamefont {Vega}\ and\ \citenamefont
  {Rai}(2020)}]{vega2020multivariate}%
  \BibitemOpen
  \bibfield  {author} {\bibinfo {author} {\bibfnamefont {R.~D.~V.}\
  \bibnamefont {Vega}}\ and\ \bibinfo {author} {\bibfnamefont {A.~G.}\
  \bibnamefont {Rai}},\ }\href
  {https://brilliant.org/wiki/multivariate-regression/} {\enquote {\bibinfo
  {title} {Multivariate regression},}\ } (\bibinfo {year} {2020})\BibitemShut
  {NoStop}%
\bibitem [{\citenamefont {Izenman}(2008)}]{izenman2008modern}%
  \BibitemOpen
  \bibfield  {author} {\bibinfo {author} {\bibfnamefont {A.~J.}\ \bibnamefont
  {Izenman}},\ }\href@noop {} {\emph {\bibinfo {title} {Modern Multivariate
  Statistical Techniques: Regression, Classification, and Manifold
  Learning}}},\ Springer Texts in Statistics\ (\bibinfo  {publisher}
  {{Springer-Verlag}},\ \bibinfo {address} {New York},\ \bibinfo {year}
  {2008})\BibitemShut {NoStop}%
\bibitem [{Note19()}]{Note19}%
  \BibitemOpen
  \bibinfo {note} {For those unfamiliar with bootstrapping, we repeatedly
  sample n papers with replacements from our dataset and compute the estimator
  of the covariance $\sigma ^2_{12} = \protect \hat \Sigma _{1,0} $ over the
  resample. The quantiles 0.05 and 0.95 of this sampling procedure converge to
  a $90\%$ confidence interval for the covariance.}\BibitemShut {Stop}%
\bibitem [{\citenamefont {Efron}\ and\ \citenamefont
  {Hastie}(2016)}]{efron2016computer}%
  \BibitemOpen
  \bibfield  {author} {\bibinfo {author} {\bibfnamefont {B.}~\bibnamefont
  {Efron}}\ and\ \bibinfo {author} {\bibfnamefont {T.}~\bibnamefont {Hastie}},\
  }\href@noop {} {\emph {\bibinfo {title} {Computer Age Statistical Inference:
  Algorithms, Evidence, and Data Science}}}\ (\bibinfo  {publisher} {Cambridge
  University Press},\ \bibinfo {year} {2016})\BibitemShut {NoStop}%
\bibitem [{Note20()}]{Note20}%
  \BibitemOpen
  \bibinfo {note} {During bootstrapping, the median of ``record setting''
  physical qubits data points was 11, and the median of ``record setting''
  average two-qubit gate error rate data points was 6.}\BibitemShut {Stop}%
\bibitem [{Note21()}]{Note21}%
  \BibitemOpen
  \bibinfo {note} {Full URL: \protect \texttt {https://\protect \allowbreak
  {}colab.\protect \allowbreak {}research.\protect \allowbreak
  {}google.com/\protect \allowbreak {}drive/\protect \allowbreak
  {}1XkWcs\protect \allowbreak {}Uy-Ci\protect \allowbreak {}NDDJffPC3d\protect
  \allowbreak {}NJbgv\protect \allowbreak {}R6iDAqh\#\protect \allowbreak
  {}scrollTo=\protect \allowbreak {}jV29mJw\protect \allowbreak
  {}Oiz9W\&uniqifier\protect \allowbreak {}=3}.}\BibitemShut {Stop}%
\bibitem [{Note22()}]{Note22}%
  \BibitemOpen
  \bibinfo {note} {There are informal arguments for why error rates may fail to
  sustain exponential progress in the future even if the number of physical
  qubits grows exponentially.  Once the fault-tolerant threshold is achieved,
  there do not seem to be fundamental physical barriers to adding more physical
  qubits given sufficient effort and resources, but error rates refer to a
  single gate, and these will plausibly hit strongly diminishing returns with
  respect to how perfectly they can be engineered. Even if lower error rates
  are experimentally possible, there are reasons to think devoting engineering
  resources to more physical qubits may have a higher payoff in terms of the
  number of logical qubits. Unlike for traditional CPUs, superconducting qubits
  currently use parallel control channels.  Adding arbitrary numbers of
  physical qubits may require partially serializing this control, through
  shared control lines, which hasn't yet been attempted. We thank Daniel Sank
  for discussion on these points.}\BibitemShut {Stop}%
\bibitem [{Note23()}]{Note23}%
  \BibitemOpen
  \bibinfo {note} {Private communication with Farmer and Lafond.}\BibitemShut
  {Stop}%
\bibitem [{\citenamefont {Benson}\ and\ \citenamefont
  {Magee}(2015)}]{benson2015quantitative}%
  \BibitemOpen
  \bibfield  {author} {\bibinfo {author} {\bibfnamefont {C.~L.}\ \bibnamefont
  {Benson}}\ and\ \bibinfo {author} {\bibfnamefont {C.~L.}\ \bibnamefont
  {Magee}},\ }\href@noop {} {\bibfield  {journal} {\bibinfo  {journal} {{PLOS}
  {ONE}}\ }\textbf {\bibinfo {volume} {10}},\ \bibinfo {pages} {e0121635}
  (\bibinfo {year} {2015})}\BibitemShut {NoStop}%
\bibitem [{\citenamefont {Harvey}(1984)}]{harvey1984time}%
  \BibitemOpen
  \bibfield  {author} {\bibinfo {author} {\bibfnamefont {A.~C.}\ \bibnamefont
  {Harvey}},\ }\href@noop {} {\bibfield  {journal} {\bibinfo  {journal} {The
  Journal of the Operational Research Society}\ }\textbf {\bibinfo {volume}
  {35}},\ \bibinfo {pages} {641} (\bibinfo {year} {1984})}\BibitemShut
  {NoStop}%
\bibitem [{Note24()}]{Note24}%
  \BibitemOpen
  \bibinfo {note} {Full URL: \protect \texttt {https://\protect \allowbreak
  {}docs.\protect \allowbreak {}google.\protect \allowbreak {}com/\protect
  \allowbreak {}spreadsheets/\protect \allowbreak {}d/\protect \allowbreak
  {}1pwb4gf0F\protect \allowbreak {}xlxgfVhtXTaq\protect \allowbreak
  {}EGS9b7FwsstsJ0\protect \allowbreak {}v7Zb1naQ0/\protect \allowbreak
  {}edit\#\protect \allowbreak {}gid=0}.}\BibitemShut {Stop}%
\bibitem [{Note25()}]{Note25}%
  \BibitemOpen
  \bibinfo {note} {Full URL: \protect \texttt {https://\protect \allowbreak
  {}colab.\protect \allowbreak {}research.\protect \allowbreak
  {}google.\protect \allowbreak {}com/\protect \allowbreak {}drive/\protect
  \allowbreak {}1XkWcs\protect \allowbreak {}Uy-Ci\protect \allowbreak
  {}NDDJffPC3d\protect \allowbreak {}NJbgvR6iD\protect \allowbreak
  {}Aqh\#\protect \allowbreak {}scrollTo=\protect \allowbreak {}jV29mJw\protect
  \allowbreak {}Oiz9W\&uniqifier=4}.}\BibitemShut {Stop}%
\bibitem [{\citenamefont {Aaronson}(2017)}]{aaronson2017quantum}%
  \BibitemOpen
  \bibfield  {author} {\bibinfo {author} {\bibfnamefont {S.}~\bibnamefont
  {Aaronson}},\ }\href@noop {} {\bibfield  {journal} {\bibinfo  {journal}
  {American Scientist}\ }\textbf {\bibinfo {volume} {102}},\ \bibinfo {pages}
  {266} (\bibinfo {year} {2017})}\BibitemShut {NoStop}%
\bibitem [{Note26()}]{Note26}%
  \BibitemOpen
  \bibinfo {note} {In particular, note that once the two-gate error rate gets
  extremely low, below $10^{-6}$, it can become advantageous to use alternative
  protocols like the Steane concatenated code. This will somewhat change how
  fast we expect the available number of generalized logical qubits to increase
  with time (compared to the naive assumption that the surface code will always
  be used. However, such low error rates are very challenging to achieve,
  perhaps for a very long time, and for the foreseeable future something like
  the surface code is likely to dominate \protect \citep
  {javadi-abhari2017towards}.}\BibitemShut {Stop}%
\bibitem [{Note27()}]{Note27}%
  \BibitemOpen
  \bibinfo {note} {Earlier authors take a similar form but sometimes differ by
  a numerical constant due to different estimates for technical complications,
  e.g., a factor of $ \protect \sqrt {10}$ in \cite {fowler2011surface,
  fowler2012surface}.}\BibitemShut {Stop}%
\bibitem [{Note28()}]{Note28}%
  \BibitemOpen
  \bibinfo {note} {Indeed, we believe it would be misleading for any composite
  metric like GLQ to include the discrete structure associated with integer
  jumps in code distance since we are highly uncertain about the detailed
  features of future error correction schemes. (Any discrete structure would be
  ``fake detail''.) Rather, GLQ is intended only to capture the roughly scaling
  of the surface code.}\BibitemShut {Stop}%
\bibitem [{\citenamefont {Landauer}(1971)}]{landauer1971cooperative}%
  \BibitemOpen
  \bibfield  {author} {\bibinfo {author} {\bibfnamefont {R.}~\bibnamefont
  {Landauer}},\ }\href@noop {} {\bibfield  {journal} {\bibinfo  {journal}
  {Ferroelectrics}\ }\textbf {\bibinfo {volume} {2}},\ \bibinfo {pages} {47}
  (\bibinfo {year} {1971})}\BibitemShut {NoStop}%
\bibitem [{\citenamefont {Bennett}(2003)}]{bennett2003notes}%
  \BibitemOpen
  \bibfield  {author} {\bibinfo {author} {\bibfnamefont {C.~H.}\ \bibnamefont
  {Bennett}},\ }\href@noop {} {\bibfield  {journal} {\bibinfo  {journal}
  {Studies in History and Philosophy of Science Part B: Studies in History and
  Philosophy of Modern Physics}\ }\bibinfo {series} {Quantum Information and
  Computation},\ \textbf {\bibinfo {volume} {34}},\ \bibinfo {pages} {501}
  (\bibinfo {year} {2003})}\BibitemShut {NoStop}%
\bibitem [{\citenamefont {Bennett}(1973)}]{bennett1973logical}%
  \BibitemOpen
  \bibfield  {author} {\bibinfo {author} {\bibfnamefont {C.~H.}\ \bibnamefont
  {Bennett}},\ }\href@noop {} {\bibfield  {journal} {\bibinfo  {journal} {{IBM}
  Journal of Research and Development}\ }\textbf {\bibinfo {volume} {17}},\
  \bibinfo {pages} {525} (\bibinfo {year} {1973})}\BibitemShut {NoStop}%
\bibitem [{\citenamefont {{DeBenedictis}}\ \emph {et~al.}(2016)\citenamefont
  {{DeBenedictis}}, \citenamefont {Frank}, \citenamefont {Ganesh},\ and\
  \citenamefont {Anderson}}]{debenedictis2016path}%
  \BibitemOpen
  \bibfield  {author} {\bibinfo {author} {\bibfnamefont {E.~P.}\ \bibnamefont
  {{DeBenedictis}}}, \bibinfo {author} {\bibfnamefont {M.~P.}\ \bibnamefont
  {Frank}}, \bibinfo {author} {\bibfnamefont {N.}~\bibnamefont {Ganesh}}, \
  and\ \bibinfo {author} {\bibfnamefont {N.~G.}\ \bibnamefont {Anderson}},\
  }in\ \href@noop {} {\emph {\bibinfo {booktitle} {2016 {IEEE} International
  Conference on Rebooting Computing {(ICRC)}}}}\ (\bibinfo  {publisher}
  {{IEEE}},\ \bibinfo {address} {San Diego, {CA}},\ \bibinfo {year} {2016})\
  pp.\ \bibinfo {pages} {1--8}\BibitemShut {NoStop}%
\bibitem [{\citenamefont {{DeBenedictis}}\ \emph {et~al.}(2017)\citenamefont
  {{DeBenedictis}}, \citenamefont {Mee},\ and\ \citenamefont
  {Frank}}]{debenedictis2017opportunities}%
  \BibitemOpen
  \bibfield  {author} {\bibinfo {author} {\bibfnamefont {E.~P.}\ \bibnamefont
  {{DeBenedictis}}}, \bibinfo {author} {\bibfnamefont {J.~K.}\ \bibnamefont
  {Mee}}, \ and\ \bibinfo {author} {\bibfnamefont {M.~P.}\ \bibnamefont
  {Frank}},\ }\href@noop {} {\bibfield  {journal} {\bibinfo  {journal}
  {Computer}\ }\textbf {\bibinfo {volume} {50}},\ \bibinfo {pages} {76}
  (\bibinfo {year} {2017})}\BibitemShut {NoStop}%
\bibitem [{\citenamefont {{DeBenedictis}}\ and\ \citenamefont
  {Frank}(2018)}]{debenedictis2018national}%
  \BibitemOpen
  \bibfield  {author} {\bibinfo {author} {\bibfnamefont {E.~P.}\ \bibnamefont
  {{DeBenedictis}}}\ and\ \bibinfo {author} {\bibfnamefont {M.~P.}\
  \bibnamefont {Frank}},\ }\href@noop {} {\bibfield  {journal} {\bibinfo
  {journal} {Computer}\ }\textbf {\bibinfo {volume} {51}},\ \bibinfo {pages}
  {69} (\bibinfo {year} {2018})}\BibitemShut {NoStop}%
\bibitem [{\citenamefont {Rahaman}\ \emph {et~al.}(2012)\citenamefont
  {Rahaman}, \citenamefont {Chattopadhyay},\ and\ \citenamefont
  {Chattopadhyay}}]{rahaman2012progress}%
  \BibitemOpen
  \bibfield  {author} {\bibinfo {author} {\bibfnamefont {H.}~\bibnamefont
  {Rahaman}}, \bibinfo {author} {\bibfnamefont {S.}~\bibnamefont
  {Chattopadhyay}}, \ and\ \bibinfo {author} {\bibfnamefont {S.}~\bibnamefont
  {Chattopadhyay}},\ }\href@noop {} {\emph {\bibinfo {title} {Progress in
  {VLSI} Design and Test: Proceedings of the 16th International Symposium on
  {VSLI} Design and Test}}}\ (\bibinfo  {publisher} {Springer},\ \bibinfo
  {address} {Shipur, India},\ \bibinfo {year} {2012})\BibitemShut {NoStop}%
\end{thebibliography}%


\section{Appendix}
\addcontentsline{toc}{section}{7. Appendix}
This appendix is intended for readers with some technical familiarity with quantum computing.

\subsection{Generalized logical qubits}
\label{appendix:generalized-logical-qubits}
\addcontentsline{toc}{subsection}{7.1 Generalized logical qubits}
Our definition of the number of generalized logical qubits makes use of the scaling associated with the (planar) surface error correction code \citep{barends2014superconducting,fowler2011surface,fowler2012surface,javadi-abhari2017towards}.  Thus, when we project the arrival of a fault-tolerant QC for a particular year with this metric, we are implicitly assuming the device will use the surface code or something that makes similar trade-offs \footnote{In particular, note that once the two-gate error rate gets extremely low, below $10^{-6}$, it can become advantageous to use alternative protocols like the Steane concatenated code.  This will somewhat change how fast we expect the available number of generalized logical qubits to increase with time (compared to the naive assumption that the surface code will always be used.  However, such low error rates are very challenging to achieve, perhaps for a very long time, and for the foreseeable future something like the surface code is likely to dominate \citep{javadi-abhari2017towards}.}.

For the surface code, the number of physical qubits required to encode a single logical qubit is $\fQEC = (2 d - 1)^2$  where $d=1,3,5,7,\dots$ is the code distance, a measure of the size of the errors to which the code is robust. To a good approximation, this code satisfies \footnote{Earlier authors take a similar form but sometimes differ by a numerical constant due to different estimates for technical complications, e.g., a factor of $ \sqrt{10}$ in \cite{fowler2011surface, fowler2012surface}.}
\begin{align}
\pL \approx \sqrt{10}\, \pP \left(\frac{\pP}{\pth}\right)^{(d-1)/2}
\end{align}
where $\pP$ is the average two-qubit error rate on the raw physical qubits in the computer, $\pth \approx 10^{-2}$ is the approximate threshold error at which fault-tolerance becomes possible, and $\pL$ is the two-qubit error for the logical qubits, i.e., the \textit{effective} error rate of the mathematical computation \citep{javadi-abhari2017towards}. As long as $\pP$ is below $\pth$, the logical error rate can be driven arbitrarily low by choosing a sufficiently large code distance $d$ (at the expense of requiring more physical qubits to host the computation).

The inverse $1/\pL$ is the number of two-qubit gates that can typically be applied before an error occurs at the logical level, and that sets the maximum size of the computation.  The particular value we choose is not too important because our results depend only weakly (logarithmically) on this choice, which in principle is determined by the length of the particular computation one wants to do.  We follow \citep{grumbling2019quantum} by taking $\pL = 10^{-18}$, corresponding to computations with of order a trillion steps (not necessarily in serial).

We then ignore the discreteness \footnote{Indeed, we believe it would be misleading for any composite metric like GLQ to include the discrete structure associated with integer jumps in code distance since we are highly uncertain about the detailed features of future error correction schemes.  (Any discrete structure would be ``fake detail''.)  Rather, GLQ is intended only to capture the roughly scaling of the surface code.} of the code distance $d$, eliminate it from the equations, and derive ratio of physical qubits per logical qubits:
\begin{align}
    \fQEC = \left[4 \frac{\log \left(\sqrt{10}\, \pP/\pL\right)}{\log\left( \pth/\pP\right)} + 1 \right]^{-2}
\end{align}
We emphasize that the above formula is only an approximation, and in particular ignores the discreteness of the code distance $d$.

\subsection{Reversible computing}
\addcontentsline{toc}{subsection}{7.2 Reversible computing}
\label{reversible-computing}
Landaurer's limit \citep{landauer1971cooperative,bennett2003notes,bennett1973logical} is a fundamental thermodynamic barrier, imposing an energy cost on all computations that are irreversible (at the logical level).  It is known that it can be circumvented by adopting reversible computing, where all computations are modified to eliminate (almost all) irreversible steps \citep{bennett2003notes,bennett1973logical}.

Landaurer's limit is not yet a serious consideration for modern classical computers because current technology already wastes several orders of magnitude more energy, due to conventional engineering imperfections, than is required by Landaurer's limit.  But, as part of the progress associated with Moore's law, the amount of energy wasted has been decreasing exponentially each year, and eventually Landaurer's limit \emph{will} become important for continuing technological progress in classical computing \citep{debenedictis2016path,debenedictis2017opportunities}.  Once that happens, classical computers will likely need to become reversible (or partially reversible) to continue improving.

Reversible computing is a prerequisite for quantum computers in the sense that quantum computations are necessarily reversible and a subset of quantum operations (the classical operations with respect to some fixed computational basis) are sufficient to implement any classical computation reversibly \citep{debenedictis2018national}.  Indeed, our impression from discussion with experts is that achieving reversibility is a large component of the engineering difficulty of quantum computing.  It is unclear how much more difficult it would be to build a useful quantum computer once a useful reversible one is created.

Therefore, if we extrapolate current classical computation trends, the date at which classical computers become reversible is a good hint at when quantum computers may be available.  We have not performed a thorough search of the literature, but we did find a plot, Fig. 2 on page 386 of \citep{rahaman2012progress}, suggesting that achieving (partial) reversibility will become an important part of classical computing technological progress in the mid 2030's.  A quick skim of seminars and articles by DeBenedictis \& Frank seem broadly consistent with this \citep{debenedictis2016path,debenedictis2017opportunities,debenedictis2018national}.  

Constructing a more accurate projection and drawing implications for quantum computing is left for future work. Besides being valuable by virtue of being independent (and complementary to) approaches directly measuring research progress in quantum computing, the estimation approach based on reversible computing is notable because it naturally incorporates the economics of classical computing investments.



\end{document}